\documentclass[12pt,a4paper,ptm]{article}
\usepackage[utf8]{inputenc}
\usepackage[english]{babel}
\usepackage{amsmath}
\usepackage{amsfonts}
\usepackage{amssymb}
\usepackage{graphicx}
\usepackage{xcolor,tikz,datetime,multirow,lscape,multicol,float,soul,newtxtext,newtxmath,accents,extarrows}
\usepackage[outline]{contour}
\contourlength{.2pt}
\usepackage[round]{natbib}
\usepackage[affil-it]{authblk}
\usepackage[breaklinks=true,bookmarks=true,colorlinks=true,
            linkcolor=blue,citecolor=blue,urlcolor=blue]{hyperref}
\usepackage[toc,page]{appendix}
\usepackage[top=0.47in,bottom=0.51in,left=0.45in,right=0.45in]{geometry}
\newcommand{\figdir}{./}

\newcommand{\rev}[2]{{\color{black!} #2}}

\newif\ifcolmn
\colmnfalse
\newif\ifref
\reftrue
\newif\ifall
\alltrue
\newif\ifone
\newif\iftwo
\newif\ifthree
\newif\iffour
\newif\iffive
\newif\ifsix
\newif\ifseven
\newif\ifeight
\newif\ifnine
\newif\iften
\newif\ifeleven
\newif\iftwelve
\newif\ifthirteen
\newif\iffourteen
\newif\iffifteen
\newif\ifsixteen
\newif\ifseventeen
\newif\ifeighteen
\newif\ifnineteen
\newif\iftwenty
\newif\ifappenfig
\onefalse
\twofalse
\threefalse
\fourfalse
\fivefalse
\sixfalse
\sevenfalse
\eightfalse
\ninefalse
\tenfalse
\elevenfalse
\twelvefalse
\thirteenfalse
\fourteenfalse
\fifteenfalse
\sixteenfalse
\seventeenfalse
\eighteenfalse
\nineteenfalse
\twentyfalse
\appenfigfalse
\ifall
\onetrue
\twotrue
\threetrue
\fourtrue
\fivetrue
\sixtrue
\seventrue
\eighttrue
\ninetrue
\tentrue
\eleventrue
\twelvetrue
\thirteentrue
\fourteentrue
\fifteentrue
\sixteentrue
\seventeentrue
\eighteentrue
\nineteentrue
\twentytrue
\appenfigtrue
\fi
\sixtrue
\appenfigfalse
\date{September 2020}
\newcommand\emt[1]{\ensuremath{#1}}
\newcommand\xf[1]{\emt{  x_{#1}}}            
\newcommand\svf{\emt{    c}}                 
\newcommand\mob{\emt{    \Psi}}              
\newcommand\denss{\emt{  \varrho_s}}         
\newcommand\densf{\emt{  \varrho}}           
\newcommand\g{\emt{      g}}                 
\newcommand\vs{\emt{     v_s}}               
\newcommand\ds{\emt{     d}}                 
\newcommand\s{\emt{      s}}                 
\newcommand\U{\emt{      U_0}}               
\newcommand\del{\emt{    \delta}}            
\newcommand\om{\emt{     \omega}}            
\newcommand\TT{\emt{     T}}                 
\newcommand\Rdel{\emt{   R_\del}}            
\newcommand\Ga{\emt{     Ga}}                
\newcommand\Kc{\emt{     K_c}}               
\newcommand\beds{\emt{ x_{2,sf}}}            
\newcommand\beda{\emt{ \widehat{x}_{2,sf}}}  
\newcommand\salt{\emt{ \Delta_{sal}}}        
\newcommand\rbed{\emt{ x_{2,v0}}}            
\newcommand\xref{\emt{ \Delta x_{2,ref}}}    
\newcommand\cref{\emt{ c_{ref}}}             
\newcommand\vpx{\emt{ v_{1,p}}}              
\newcommand\vpy{\emt{ v_{2,p}}}              
\newcommand\vfx{\emt{ v_{1,f}}}              
\newcommand\shields{\emt{\theta}}            
\author[1]{Giovanna Vittori}
\author[1]{Paolo Blondeaux}
\author[1]{Marco Mazzuoli}
\author[2]{Julian Simeonov}
\author[2]{Joseph Calantoni}
\affil[1]{{\small Department of Civil, Chemical and
                  Environmental Engineering,
                  University of Genoa, Italy}}
\affil[2]{{\small Ocean Sciences Division, U.S. Naval Research Laboratory, 
                  Stennis Space Center, 
                  Mississipi, U.S.A.}}
\title{Sediment transport under oscillatory flows}
\begin{document}
\maketitle
\begin{abstract}
The results of Direct Numerical Simulations of the oscillatory flow over a cohesionless bed of spherical particles, mimicking sediment grains, are described. The flow around the sediment particles is explicitly computed by using the immersed boundary method, which allows the force and torque acting on the particles to be evaluated along with their dynamics. Different values of the Reynolds number and different values of the ratio between the grain size and the thickness of the boundary layer are considered such that the results are useful to quantify the sand transport generated by sea waves in the region offshore of the breaker line. Therefore, the results are used to test the capability of empirical sediment transport formulae to predict the sediment transport rate during the oscillatory cycle. %
\end{abstract}
\section{Introduction}

The amount of sediments transported by the water flowing over a cohesionless bottom depends on many factors such as the characteristics of the sediments (size, density, shape, ...) and those of the driving flow (steady/unsteady, laminar/turbulent, ...). %

\rev{%
Many sediment transport predictors do exist. Some of them provide an estimate of what is named ``bed load'', i.e. the sediment transport rate due to the sediment grains which roll, slide and saltate but move strictly interacting with the resting particles which make up the bottom \citep[][$\ldots$]{meyer1948,fernandez1976,vanrijn1984,nielsen1992}. %
}{%
The capability to evaluate the sediment transport rate is often requested by coastal and fluvial engineering applications and many sediment transport predictors do exist. %
At the scales relevant for practical purposes, both the fluid and the sediment can be reasonably assumed to be continua and the sediment transport can be considered an interface phenomenon. %
Therefore, the bottom is idealised by a surface and it is natural to associate the sediment transport with the shear stress measured at the bottom surface, in particular with the excess of shear stress above a threshold value. %
The threshold value is somehow related to the minimum shear stress necessary to entrain, namely set into motion, the sediments from the bottom surface. %

Then a distinction is made between the sediments moving in the vicinity of the bottom surface, where the effects of inter-particle contacts are significant, and the sediment motion sustained by the bulk flow away from the bottom \citep{bagnold1966}. %
On these grounds, the former transport mode, the so called \emph{bed load}, is assumed to depend on the average bottom shear stress and on its threshold value, as well as on the sediment properties, while the latter mode, i.e. the \emph{suspended load}, is characterised also by the settling velocity of the sediment particles and by the turbulence characteristics. %
Following a deterministic approach, some of the most known and currently used models to estimate the bed load transport were developed either in steady \citep[][...]{meyer1948,fernandez1976,vanrijn1984} or unsteady conditions \citep[e.g.][]{nielsen1992}. %

Obtaining accurate measurements of the shear stress close to the bottom is challenging. %
The strong vertical gradients which are present close to the bottom require a high measurement accuracy within a thin layer where the sediment no longer looks like a continuum. %
Moreover, it is impossible to measure simultaneously, the contributions to the shear stress associated with the viscous stress, the inter-particle contacts and the fluid sediment interactions. %
Therefore, the values of the shear stress are typically extrapolated from a convenient elevation towards the conventional mean bottom surface \citep[e.g.][located the reference bottom plane $0.25$ times the particle diameter below the top of the particles]{vanrijn1984}. %
This is partly justified by the fact that the total shear stress is linear in steady channel flow, but this approach might give erroneous results in oscillatory flows. %
Among others, \citet{dohmen2001} and \citet{liu2005} carried out measurements of the sediment transport in oscillatory flows and measured the excursion of the resting bottom elevation during the oscillation periods. %
When the sheet flow regime was attained, namely when ripples were washed out under high shear stress conditions and the bed is flat, they also measured the thickness of the sheet flow layer. %
However, accurate measurements of the bed load layer for small sediment flow rates are missing, since measuring the thickness of such a thin layer and estimating the dynamics of individual sediment grains can be extremely difficult to achieve experimentally. %
Indeed, in this case the bed load transport involves a layer of sediments, the thickness of which ranges between one to a few particle diameters. %

The bed load transport includes both particles that remain in contact with the bottom surface (rolling and sliding on the resting particles) and saltating particles, which perform small jumps. %
In particular, the saltating motion of particles reflects the instantaneous and local properties of the turbulent flow in the vicinity of the bed and %
inspired the formulation of bed-load transport predictors based on a stochastic (or ballistic) approach \citep{einstein1950,armanini2015}. %
Models based on the time-averaged characteristics of the fluid-solid interactions, fail in accounting for the force fluctuations typically encountered in turbulent flows, because the dynamics of sediment particles is significantly affected by turbulent vortices which can pick up the sediment grains from their resting position \citep{diplas2013}. %
Long lasting turbulent coherent structures, may impinge onto the particles laying on the bottom causing their entrainment or re-suspension \citep{sutherland1967}. %
The sediment particle dynamics is found to be dominated by the turbulent events that can develop a significant impulse of the hydrodynamic force \citep{diplas2008} or, similarly, produce significant work \citep{lee2012}. %
Therefore, the lifespan of the turbulent vortices and how long they interact with sediment particles determine the duration of force fluctuations and, consequently, the particle dynamics. %
\citet{cheng1998} assumed that the streamwise velocity fluctuations follow a Gaussian distribution and attempted to estimate the pick-up probability in steady conditions. %
\citet{wu2003} proposed an extension of \citet{cheng1998}'s model deriving also the rolling and lifting probabilities of sediment entrainment. %

On the other hand, models based on a stochastic approach do not suffer the uncertainty related to the determination of the initiation of sediment transport and explicitly account for the dynamics of sediment particles, but the high level of abstraction makes the values of the parameters difficult to be estimated, like the geometrical characteristics of the particle trajectories or the characteristic time scale of particle saltation \citep[e.g. the ``exchange time of bedload particle'' introduced by][]{einstein1950}. %

\rev{%
Other predictors allow an estimate of the \emph{total load}, i.e. the sediment transport rate that takes into account also the suspended load, namely sediments which are picked-up from the bottom and kept into suspension by the action of the turbulent vortices within which the sediments are trapped \citep[][$\ldots$]{enghan1972,ackers1973,vanrijn1984}. %
}{}%

Since turbulence dynamics in steady and oscillatory flows is remarkably different, in particular for moderate values of the Reynolds number such that the oscillatory flow re-laminarises every half-cycle, the effect of turbulent events on the sediment transport is also different. %
Nonetheless, almost all the sediment transport formulae currently available are elaborated on the basis of laboratory measurements carried out under steady currents and only a few of them are proposed to evaluate the sediment transport rate induced by the oscillatory flow generated by propagating sea waves close to the bottom. %

The irrotational flow generated by a monochromatic surface wave of small amplitude propagating over a flat sandy bottom provides a fair description of the actual flow that is observed in coastal environments seaward of the breaker zone. %
Close to the bottom, the flow turns into the oscillatory boundary layer which is induced by harmonic oscillations of a pressure gradient close to a wall. %
}%
A rough estimate of the Keulegan-Carpenter number of these oscillatory flows around sand grains, which is defined as %
\begin{equation}
\Kc=\dfrac{\U^* \TT^*}{\ds^*}
\:\:,
\label{eq_kc}
\end{equation}
shows that for field conditions this dimensionless parameter assumes large values. Hereinafter, $\U^*$ and $\TT^*$ denote the amplitude and period of the velocity oscillations induced close to the bottom by sea waves and $\ds^*$ is the mean grain size of the sediment grains. %
\rev{}{%
The asterisk superscript indicates dimensional quantities. %
}%
Therefore, it appears reasonable to assume that the sediments are driven by a sequence of steady flows characterized by a slowly varying amplitude even though, when the amplitude of the velocity oscillations is small, the pressure gradient and other effects due to the unsteadiness of the driving flow might be relevant. %

However, there are no experimental measurements showing that the instantaneous sediment transport rate induced by sea waves can be quantified by means of the sediment transport predictors proposed for steady flows and the empirical approaches that are used to evaluate the sediment transport rate in coastal environments are designed to provide only the sediment transport averaged over a semi-cycle of the surface wave \citep[][$\ldots$]{madsen1976,sleath1978,sleath1982,van1993,soulsby1997}. %

\rev{}{%
Early numerical investigations of the effect of the interaction between turbulent vortices and sediment particles in unsteady flows where carried out, among others, by \citet{finn2016a} and \citet{wu2017} using an Eulerian-Lagrangian approach, namely considering a two-way coupling point-particle method to simulate the motion of sediment grains. %
}%

Recently, \citet{mazzuoli2016a,mazzuoli2018a,mazzuoli2019a,mazzuoli2019b} have made direct numerical simulations (DNS) of the oscillatory flow generated by surface waves close to a bottom made up of spherical particles mimicking sand grains (both fixed \citep{mazzuoli2018a} and mobile  \citep{mazzuoli2016a,mazzuoli2019a,mazzuoli2019b} particles). %
In particular, the approach employed by \citet{mazzuoli2016a,mazzuoli2019a,mazzuoli2019b} evaluates the flow around the moving particles, the dynamics of which is explicitly evaluated by means of the numerical simulation of the Newton-Euler equations describing the translation and rotation of a rigid body. %
The results were supported by both qualitative and quantitative comparisons with laboratory data and numerical simulations (see \citealp{mazzuoli2016a,blondeaux2016} and \citealp{mazzuoli2019a}, respectively). %

In particular, the simulations carried out by \citet{mazzuoli2019b} show two important results. %
First, as previously discussed, for large values of the bottom shear stress, the sediment transport rate is found to be fairly described by the predictors proposed for steady flows. On the other hand, for the relatively small values of the bottom shear stress, which are present close to flow inversion, the sediment transport rate during the decelerating phase is different from that during the accelerating phase, even for the same value of the bottom shear stress, thus indicating that the dynamics of the sediment is controlled not only by the viscous stresses but also by the pressure gradient and the effects due to the self-interaction of the sediment grains and their interaction with the turbulent eddies. %
%
\rev{}{%
It is emphasized that the flow unsteadiness affects the sediment dynamics and, consequently, the sediment flow rate. %
It is expectable that unsteadiness effects manifest more markedly during the phases of the wave cycle close to the flow reversal. %
}%

In the following, the database provided by the DNS of \citet{mazzuoli2019b} and by further runs of the same code is used %
\rev{%
to analyse particle dynamics and to verify why, when and how much the sediment transport rate deviates from the predictors which provide its value for steady flows. %
A model based on an approach described in \citet{fredsoe1992} is adopted to estimate the bed load transport caused by an oscillatory flow over a bed of mono-dispersed spherical sediments. %
}{%
to evaluate the performance of a widely used bedload transport predictor. %
The considered predictor was developed for steady flows and is often used in coastal applications. %
Moreover, the numerical results allow for estimation of other quantities, such as the reference concentration and the vertical particle velocity, that are used in the modelling. %
In particular, the inputs for the model are obtained from the results of the DNS, which provide the precise and accurate values of the pressure and velocity field up to the surface of the individual grains. %
Moreover, the precise evaluation of the input quantities for the model, such as the thickness of bed load and saltation layers and the sediment flow rate during the wave-cycle, is obtained by means of DNS results in order to identify strengths and weaknesses of the predictor. %
}%

\section{Methods}
\label{Methods}

The flow within the boundary layer generated at the bottom of a propagating surface wave can be determined by considering a wave characterized by an amplitude much smaller than its wavelength and using the linear Stokes wave theory to describe the flow far from the bottom. Then, the flow close to the bottom can be evaluated by approximating it as the flow generated by an oscillating pressure gradient close to a horizontal plane. The pressure gradient is described by %
\begin{equation}
\label{pres}
\left( \frac{\partial p^*}{\partial x^*_1}, \ \frac{\partial p^*}{\partial x^*_2}, \frac{\partial p^*}{\partial x^*_3} \right) = 
\left( - \rho^* U^*_0 \omega^*\sin (\omega^*t^*),0,0\right) 
\end{equation}
where ($x_1^*, x_2^*,x_3^*$) is a Cartesian coordinate system with the $\xf{1}^*$-axis pointing in the direction of wave propagation and the $\xf{2}^*$-axis being vertical and pointing upwards. %
In (\ref{pres}), $\rho^*$ is the constant water density and $U^*_0$ and $\omega^*=2\pi/\TT^*$ are the amplitude and the angular frequency of the velocity oscillations induced by the surface wave close to the bottom ($T^*$ denotes the wave period). %
The pressure gradient drives not only the motion of the water but also the motion of spherical particles of density $\rho^*_s$ and diameter $d^*$, which represent sediment grains on the seabed. %
\rev{}{%
Hereinafter, wherever the asterisk superscript is omitted, length, time and velocity are assumed to be normalised by the reference quantities $\del^*=\sqrt{\nu^*\TT^*/\pi}$, $\TT^*/(2\pi)$ and $\U^*$, respectively, $\nu^*$ being the kinematic viscosity of sea water. %
Moreover, the time-averaged values of the physical quantities are denoted by an overline. %
}%

The results described in the following are obtained by the numerical simulation of Navier-Stokes and continuity equations to determine the velocity and pressure fields within the fluid, and the Newton-Euler equations to determine the dynamics of the spherical particles. %
%

\rev{}{%
The numerical solution of the incompressible Navier-Stokes equations is obtained with a second-order accurate finite-difference scheme which was previously used by \citet{kidanemariam2014a}, \citet{mazzuoli2016a}, \citet{mazzuoli2019a} and \citet{mazzuoli2019b}. %
The numerical approach consists of a semi-implicit fractional-step method based upon the combination of explicit (three-step Runge–Kutta) and implicit (Crank–Nicolson) discretisations of the nonlinear and viscous terms, respectively. %
The spatial operators are evaluated by standard centred second-order finite-difference approximations, written using a uniform (equispaced), staggered Cartesian grid. %
Periodic conditions are applied along the streamwise and spanwise directions, while free- and no-slip boundary conditions are enforced at the top and at the bottom of the computational domain, respectively. %
The no-slip condition at the sediment-fluid interface is enforced, using the immersed-boundary method proposed by \citet{uhlmann2005}, by means of a volume forcing term directly added to the momentum equations. %
The hydrodynamic force acting upon a particle is readily obtained by summing the additional volume forcing term over all discrete forcing points. %
An analogue procedure is applied for the computation of the hydrodynamic torque driving the angular particle motion. %
A soft-collision model is used to take into account the normal and tangential interactions between the solid particles. %
The model is based on a linear spring-dashpot system and was implemented into the code by \citet{kidanemariam2014b}. %
The interaction force depends on the relative distance and on the linear and angular velocities between the two approaching particles and it is enabled if the inter-space between the particles is smaller than one grid spacing. %
A detailed description of the numerical approach and of the tests carried out to support the reliability of the results is provided in \citet{mazzuoli2019a,mazzuoli2019b}. %
In the runs considered by \citet{mazzuoli2019b} and in the following, the sediment particles are characterised by the specific gravity $s=\denss^*/\densf^*$ equal to $2.65$, which is typical of silica sand, the Coulomb friction coefficient and the restitution coefficient characterising the inter-particle contacts are equal to $0.4$ and $0.9$, respectively, while the values of the dimensionless normal stiffness $k_n$ are indicated in table~\ref{tab0} along with the values of the other parameters. %
}%

\begin{figure}[t]
\begin{picture}(0,280)(0,0)
  \put(-20,0){\includegraphics[trim=5cm 1cm 0cm 0cm, clip, width=1\textwidth]{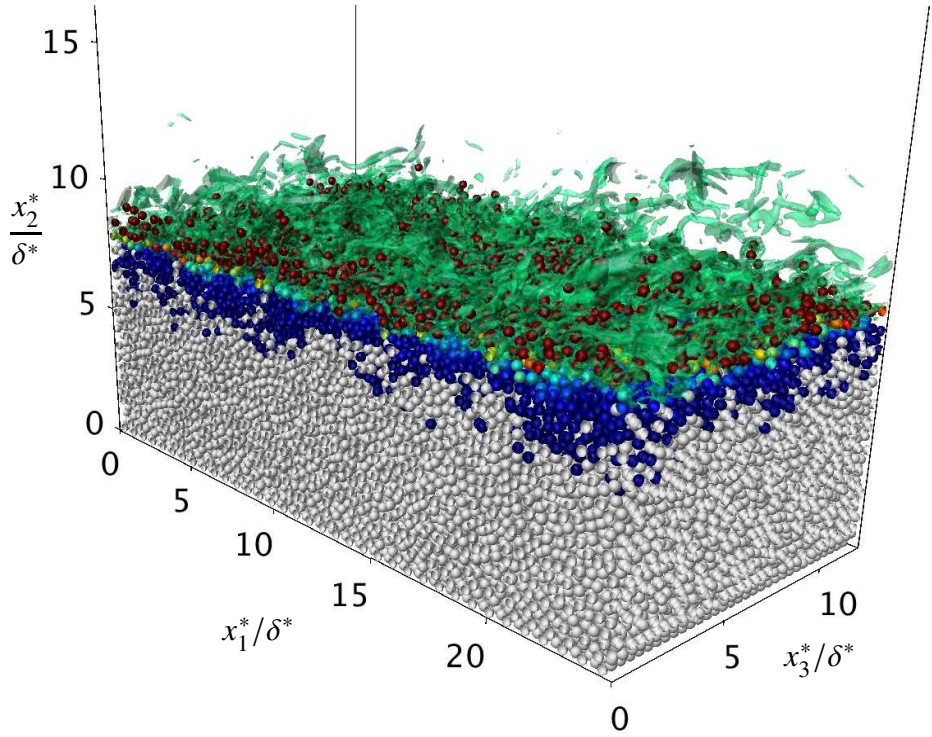}}
  \put(80,190){$\dfrac{\xf{2}^*}{\del^*}$}
  \put(160,40){$\xf{1}^*/\del^*$}
  \put(370,30){$\xf{3}^*/\del^*$}
\end{picture}
\caption{
Vortex structures generated within the oscillatory boundary layer above a cohesionless sediment bottom at $\om^*t^*=2.8 \pi$ for $\Rdel=1000$ and $\ds^*/\del^*=0.335$. %
The vortex structures are visualized by plotting the surfaces characterized by a constant values of $\lambda_2$, i.e. the second eigenvalue (the eigenvalues should be ordered
	in descending order) of the matrix given by the sum of the squares of the symmetric and antisymmetric parts of the gradient velocity tensor \citep{jeong1995}.
The sediment grains are coloured according to their velocity with grey particles at rest while the moving particles are coloured according to their increasing speed
(blue, light blue, yellow, orange, red).
}
\label{fig0}
\end{figure}

Figure \ref{fig0} shows an example of the results of the numerical simulations. %
In particular, detailed vortex structures are generated within the bottom boundary layer when turbulence appears at a particular phase of the cycle ($\om^*t^*=2.8 \pi)$ for $\Rdel=1000$ and $\ds^*/\del^*=0.335$. %
Hereinafter, the Reynolds number $\Rdel$ is defined by %
\begin{equation}
\label{Rdelta}
\Rdel 
= 
\frac{\U^* \del^*}{\nu^*}
\:\:.
\end{equation}
The Reynolds number of the boundary layer generated by propagating surface waves can be defined also by
\begin{equation}
\label{RE}
RE 
= 
\frac{\U^{*2}}{\om^*\nu^*}
\end{equation}
i.e., using $\U^*/\om^*$ as characteristic length scale of the phenomenon. It can be easily verified that $RE=\Rdel^2/2$. %

\begin{figure}[t]
\begin{picture}(0,280)(0,0)
  \put(0,-10){\includegraphics[trim=8cm 1cm 0cm 0cm, clip, width=1\textwidth]{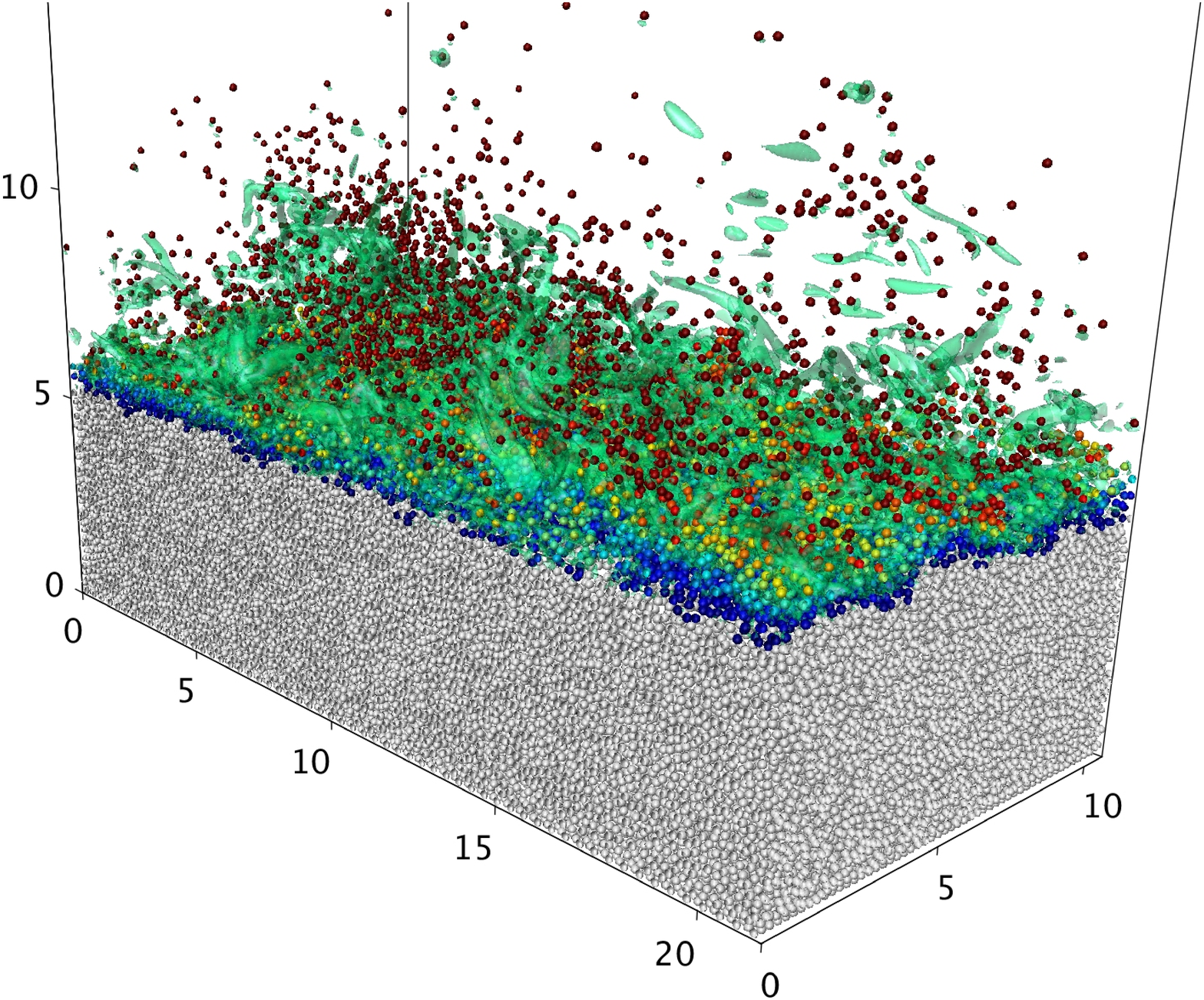}}
  \put(85,180){$\dfrac{\xf{2}^*}{\del^*}$}
  \put(190,35){$\xf{1}^*/\del^*$}
  \put(390,30){$\xf{3}^*/\del^*$}
\end{picture}
\caption{
Vortex structures generated within the oscillatory boundary layer above a cohesionless sediment bottom at $\om^*t^*=2.1 \pi$ for $\Rdel=1000$ and $\ds^*/\del^*=0.168$. %
\textit{Same as figure~\ref{fig0}.} %
}
\label{fig00}
\end{figure}
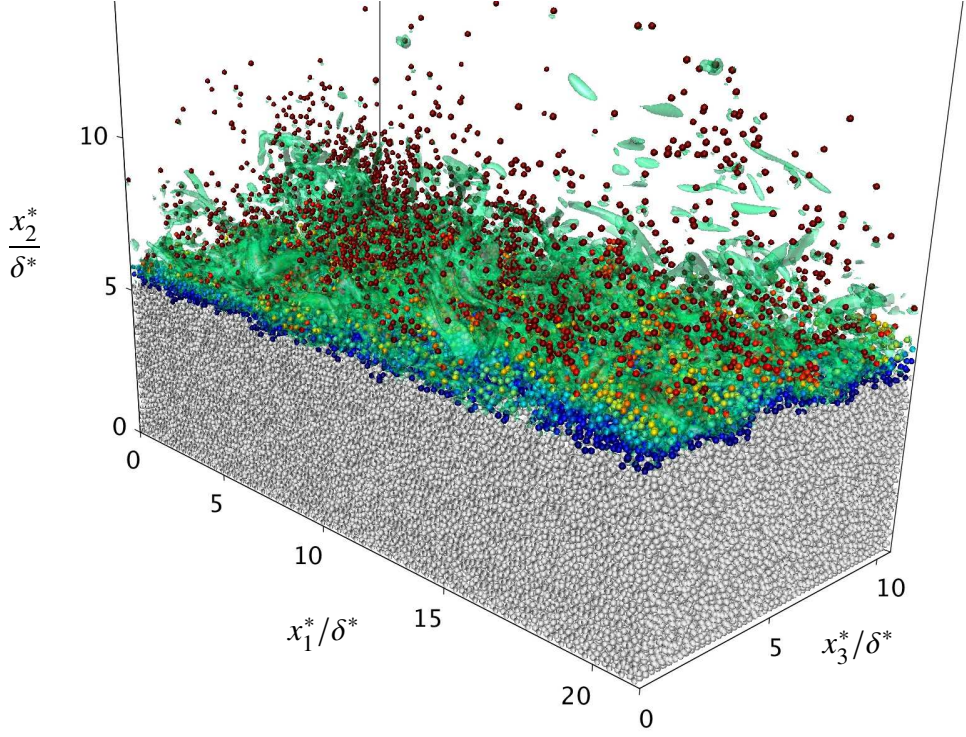

The results plotted in figure \ref{fig0} show that, for $\Rdel=1000$ and $\ds^*/\del^*=0.335$, turbulence is strong; turbulent vortex structures are able to pick-up a large number of sediment grains from the bottom and to set them into motion carrying some of them far from the resting particles. %
Larger values of the Reynolds number or smaller values of the grain size cause a larger number of particles to be carried into suspension as shown in figure \ref{fig00} where the results are plotted for $\Rdel=1000$ and $\ds^*/\del^*=0.168$. %
%
\rev{}{%
For the present values of the parameters, when the streamwise velocity is maximum, the sediment-turbulence interactions have a net dissipative effect on the turbulence which is characteristic of regime IV of the classification proposed by \citet{finn2016b}. %
\citet{finn2016b} classified the regime of particulate flows on the basis of values of the Shields parameter, the Galileo number, and the specific gravity. %
The Shields parameter is defined as
\begin{equation}
\label{theta1}
\shields 
= 
\frac{\tau_b^*}{\left( \denss^*-\densf^* \right) \g^*\ds^*} %
\:\:, 
\end{equation}
where $\tau_b^*$ is the shear stress acting on the bottom and $\g^*$ is the gravitational acceleration, %
while the Galileo number is %
\begin{equation}
Ga 
= 
\dfrac{\sqrt{(s-1)\g^*\ds^{*3}}}{\nu^*}=\dfrac{\Rdel}{\sqrt{\Psi}}\dfrac{\ds^*}{\del^*}
\:\:,
\label{gal}
\end{equation}
where $\mob$ denotes the mobility number, which is equal to $\U^{*2}/\vs^{*2}$, $\vs^*=\sqrt{(\s-1)\g^*\ds^*}$ indicating the characteristic velocity of sediment particles. %
In the regime IV, the particle motion is driven by vortices in the inertial range which are larger than the Kolmogorov length scale but smaller than the integral length scale of turbulence. %
In fact, \citet{mazzuoli2019b} found that, for the run~$2$, an effect of the particle motion was to increase the equivalent roughness and, consequently, the energy dissipation with respect to the same case with particles fixed at their initial positions. %
Instead, around the flow reversal, the particle dynamics falls in the regime II, i.e. it is dominated by gravitational forces. %
According to the picture proposed by \citet{finn2016b}, the sheet flow regime would be never attained in the present simulations. %
}%

\begin{table}[t]
        \begin{center}
                \begin{tabular}{l r c r c r c c r r r r r c}
                \hline
                  run  & $\Rdel$ & $\ds^*/\delta^*$ & \rev{}{$\Psi$} & \rev{$\denss^*/\densf^*$}{$s$} & \rev{}{$K_c/2\pi$} & 
                               \rev{}{$k_n$} & \rev{}{$\Delta x^*/\del^*$} & \rev{}{$N_{x1}$} & \rev{}{$N_{x2}$} & \rev{}{$N_{x3}$} & 
                               \rev{}{$\Delta t^*\om^*$} & \rev{}{${\cal T}^*/\TT^*$} \\
                \hline
			run 1 & $750$  & $0.335$ & $31$  & $2.65$ & $1119$ & $1.44\ 10^3$ & $2.39\ 10^{-2}$ & $1024$ & $1536$ & $512$ & $2.09\ 10^{-5}$ & $2.6$ \\
			run 2 & $1000$ & $0.335$ & $61$  & $2.65$ & $1493$ & $3.38\ 10^3$ & $2.39\ 10^{-2}$ & $1024$ & $1536$ & $512$ & $1.96\ 10^{-5}$ & $2.0$ \\
			run 3 & $1000$ & $0.168$ & $121$ & $2.65$ & $2976$ & $2.07\ 10^4$ & $2.09\ 10^{-2}$ & $1024$ & $1536$ & $512$ & $1.96\ 10^{-5}$ & $1.8$ \\
			run 4 & $1000$ & $0.670$ & $30$  & $2.65$ & $746$  & $4.23\ 10^2$ & $2.39\ 10^{-2}$ & $1024$ & $1536$ & $512$ & $1.96\ 10^{-5}$ & $2.0$ \\
			run 5 & $1500$ & $0.335$ & $124$ & $2.65$ & $2239$ & $4.18\ 10^3$ & $1.86\ 10^{-2}$ & $1280$ & $1536$ & $768$ & $7.85\ 10^{-6}$ & $1.5$ \\
                \hline
                \end{tabular}
        \end{center}
        \caption{
        Flow and sediment parameters of the numerical simulations. 
        From left to right: Reynolds number $\Rdel=U_0^*\del^*/\nu^*$, dimensionless grain size $\ds^*/\del^*$, %
        \rev{%
        specific gravity $\denss^*/\densf^*$ %
        }{%
        mobility number $\Psi=\U^{*2}/[(\s-1)\g^*\ds^*]$, specific gravity $s=\denss^*/\densf^*$, Keulegan-Carpenter number $\Kc/2\pi=\U^*/(\ds^*\om^*)$,         dimensionless stiffness of solid particles $k_n=(6/\pi)k_n^*\,\Delta x^*/(\ds^*\g^*\denss^*)$, grid spacing $\Delta x^*/\del^*$, number of grid points $N$ in the $x_1$-, $x_2$- and $x_3$-direction, fixed time step $\Delta t^*\om^*$ and number of periods that were simulated ${\cal T}^*/\TT^*$. %
        } 
        }
        \label{tab0}
\end{table}


\begin{figure}[t]
\begin{picture}(0,195)(-10,0)
  \put(85,-15){\includegraphics[trim=0cm 0cm 0cm 0cm, clip, width=.54\textwidth]{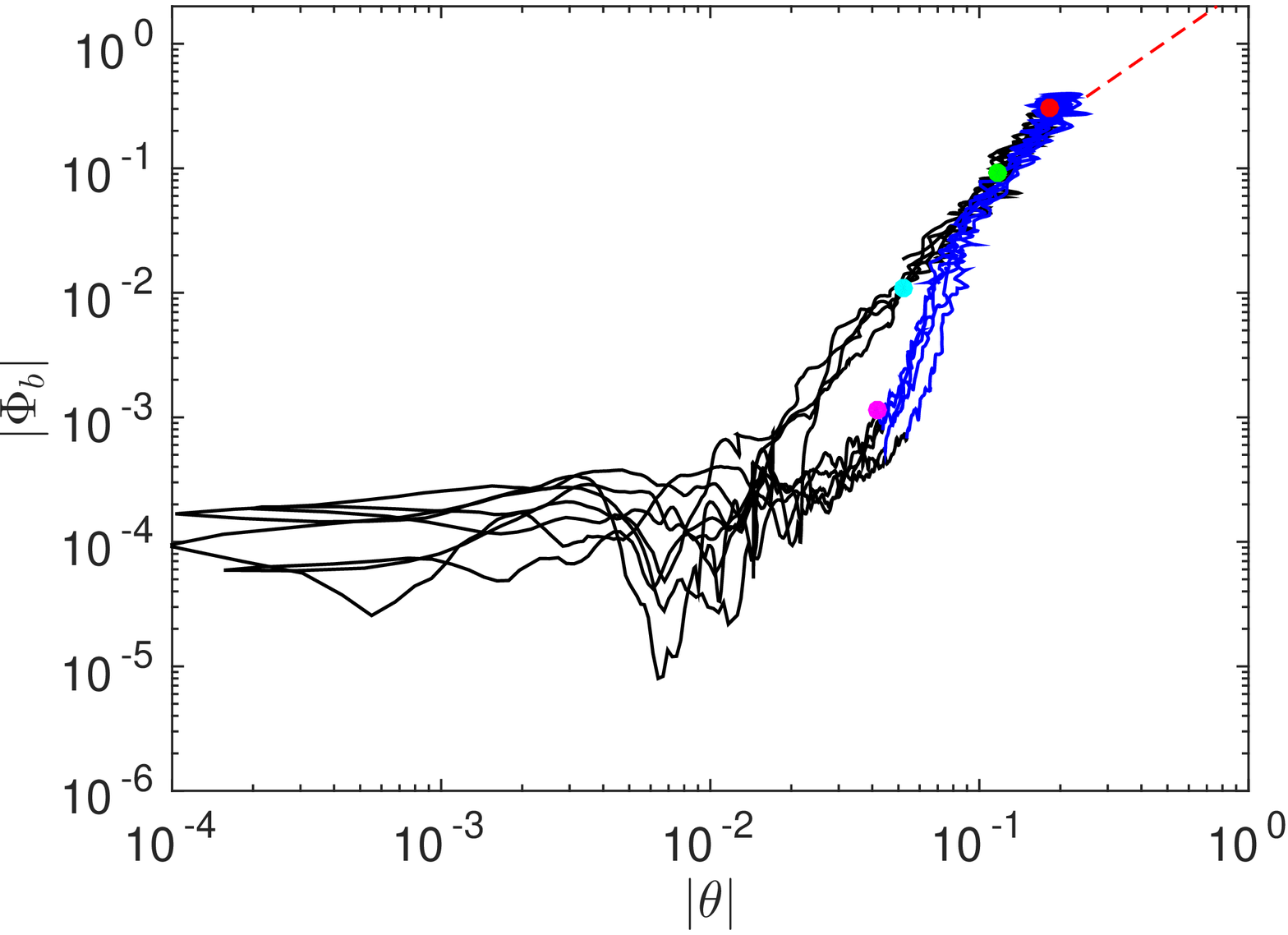}}
  \put(275,65){\vector(3,4){30}}
  \put(270,160){\vector(-1,-2){20}}
\end{picture}
\caption{
\rev{}{%
Dimensionless sediment flow rate $\vert\Phi_b\vert$ provided by the numerical simulation plotted versus 
the Shields parameter $\vert\shields\vert$ for $\Rdel=750$ and $\ds^*/\del^*=0.335$. %
The line is blue during the accelerating phases and black during the decelerating phases. %
The dashed red line indicates the values $\Phi_b\propto\shields^{3/2}$. %
Coloured markers refer to values obtained at: %
$[${\protect\tikz \protect\draw[red,fill=red] (0,0) circle (.5ex);}$]$ $t=2\pi$; 
$[${\protect\tikz \protect\draw[cyan,fill=cyan] (0,0) circle (.5ex);}$]$ $t=2.25\pi$; 
$[${\protect\tikz \protect\draw[magenta,fill=magenta] (0,0) circle (.5ex);}$]$ $t=2.5\pi$ and 
$[${\protect\tikz \protect\draw[green,fill=green] (0,0) circle (.5ex);}$]$ $t=2.75\pi$. %
}
}
\label{fig0a}
\end{figure}

In the following, we describe the results of the numerical simulations, and we use the database provided by previous DNSs to verify whether the assumptions usually introduced to quantify the value of sediment transport rate $q^*_b$ per unit width from the knowledge of the averaged flow quantities in steady flows are valid also for oscillatory flows. %
\rev{}{The simulations were carried out mostly on Marconi (CINECA, Italy) and required approximately $60$ million core hours spread over two years.}

Some of the numerical simulations are characterized by a value of the Reynolds number so that turbulence is generated twice during the wave cycle but the flow tends to recover a laminar like behaviour during the accelerating phases because of the rapid damping of turbulence. Such a flow regime is frequently encountered in the coastal region \citep[see][]{blondeaux2018}. %
Other runs are characterized by values of the Reynolds number and roughness large enough to generate a turbulent flow during the whole oscillatory cycle. %
Three different sediment grains sizes $d^*$ typical of medium and coarse sand were considered leading to values of $\ds^*/\del^*$ equal to $0.168, 0.335, 0.670$. 
Since the ratio between the amplitude of the fluid displacement oscillations and the particle size turns out to be
\begin{equation}
\label{KC}
\frac{\U^*}{\om^*\,d^*}=\frac{\Kc}{2 \pi}=\frac{\del^*}{\ds^*}\frac{\Rdel}{2}
\end{equation}
it can be verified that the runs are characterized by different values of the Keulegan-Carpenter number, $\Kc$, which in all cases are much larger than $1$. %

\section{Results}
\label{Results}

As a framework to analyse the sediment dynamics provided by the numerical simulations, let us follow \citet{ashida1972,fernandez1976,engelund1976} \citep[see also][]{fredsoe1992}, and let us consider moving sediment particles over an idealized bed of similar resting particles subject to a steady forcing flow. %
Since the Keulegan-Carpenter number of the phenomenon is large, the assumption of a steady flow should be a reasonable framework to analyse phase-by-phase the sediment dynamics except during the phases of the cycle close to flow inversion. %

It is usually assumed that the moving particles reduce the shear stress exerted by the fluid on the bottom because they exert a reaction force on the fluid which flows around them and the number of moving particles increases till the shear shear stress acting on the resting particles decreases to the critical value for the initiation of sediment motion. %
Using this hypothesis and employing slightly different relationships to quantify the number of moving particles and their speed, the Authors referenced above proposed sediment transport predictors which, for large values of the Shields parameter $\shields$, lead to values of the dimensionless sediment transport rate $\Phi_b$ per unit width (also known as Einstein bedload number) proportional to $\shields^{3/2}$. %
Hereinafter, the value of $\Phi_b$ is defined by %
\begin{equation}
\label{theta1}
\Phi_b=\frac{q^*_b}{\sqrt{(%
\rev{\denss^*/\densf^*}{\s}%
-1) \g^*\ds^{*\,3}}}
=
\dfrac{q^*_b}{\vs^*\ds^*}
\end{equation}
where $q^*_b$ is the volumetric sediment transport rate per unit width. %

The results of the numerical simulation carried out for $\Rdel=750$ and $\ds^*/\del^*=0.335$ are plotted in Figure~\ref{fig0a}, showing that for the largest values of the Shields parameter $\shields$ also an oscillatory flow generates a dimensionless sediment transport rate which tends to be proportional to $\shields^{3/2}$. %
\rev{%
However, as already pointed out in the introduction, when $\shields$ is close to $\shields_{cr}$, i.e. its critical value for the initiation of sediment motion, the sediment transport rate during the decelerating phases differs from that during the accelerating phases and $\Phi_b$ deviates from the values predicted by means of the empirical formulae proposed for steady flows. %
}{%
On the other hand, when $\shields$ is close to $\shields_{cr}$, i.e. its critical value for the initiation of sediment motion, the sediment transport rate is affected by the threshold parameter $\shields_{cr}$ and the values of $\Phi_b$ observed during the decelerating phases differ from those during the accelerating phases, thereby deviating from the values predicted by means of the empirical formulae proposed for steady flows. %
The blue lines in figure~\ref{fig0a} refer to the accelerating phases, i.e. when the flow far from the bottom accelerates. %
When the flow starts to decelerate (black line), the sediment transport rate does not recover the values attained during the accelerating phase for the same values of the Shields parameter. %
Thereafter, both $\Phi_b$ and $\shields$ reach the minimum and subsequently increase again earlier than the flow far from the bed reverses its direction, because of the relatively small flow inertia in the vicinity of the bottom. %
}%
To understand why the dynamics of the sediment for relatively small values of $\shields$ deviates from that predicted by the approaches previously mentioned, let us briefly summarize the main steps of the analyses that are based on average quantities and consider neither the unsteadiness of the average flow nor the random fluctuating components of the velocity generated by the turbulent eddies. %

The sediment particles are assumed to move under the action of a drag force $F^*_D$, which can be computed by means of %
\begin{equation}
\label{drag}
F^*_D=\frac{1}{2} \densf^* c_D \left( \vfx^*-\vpx^*\right)^2 \pi \frac{\ds^{*\,2}}{4} 
\end{equation}
and a resistance force, which can be evaluated by means of %
\begin{equation}
\label{resi}
F^*_R=\mu_d \left[ \densf^* \g^* (s-1) \pi \frac{\ds^{*\,3}}{6} - \frac{1}{2} \densf^* c_L \left( \vfx^*-\vpx^*\right)^2 \pi \frac{\ds^{*\,2}}{4} \right]
\end{equation}
where the term between the square brackets is the vertical component of the force on a sediment grain and $\mu_d$ is a dynamic friction coefficient, introduced to quantify the resistance force due to the interaction of the moving particles with the bottom. %
Sometimes the contribution due to the lift force is neglected without any justification. In (\ref{drag}) and (\ref{resi}), $\vpx^*$ and $\vfx^*$ indicate the average velocity of the sediment grains and the average velocity of the fluid which flows around the sediment grains, respectively. %
Moreover, %
\rev{%
$\s=\denss^*/\densf^*$ is the specific gravity of sand grains and %
}{}%
$c_D$ and $c_L$ are the drag and lift coefficients of the sediment grains, respectively. %

The fluid velocity close to the bed is assumed to be of the order of the shear velocity $u^*_{\tau}=\sqrt{\tau^*_b/\densf^*}$, i.e. $\vfx^*=\alpha u^*{_\tau}$ where $\alpha$ is a constant which is usually assumed to be of order $10$. %
Therefore, if sediment particles are supposed to move with a constant averaged velocity, the balance between the drag and resistance forces relates the particle velocity to the shear velocity %
\begin{equation}
\label{v_p}
\vpx^*= \alpha u^*_\tau \left[ 1- \sqrt{ \frac{\frac{4}{3}(\s-1) \g^* \ds^* \frac{\mu_d}{c_D+\mu_d c_L}}{\alpha^2 u^{*2}_\tau} } \right]=
\alpha u^*_\tau \left[ 1- \sqrt{ \frac{4 (\denss^*-\densf^*) \g^* \ds^* \mu_d}{3 \alpha^2 \tau^*_b (c_D+\mu_d c_L)}} \right]
\end{equation}

The condition of the incipient motion of the sediment grains is given by $F^*_D=F^*_R$ where $F^*_D$ and $F^*_R$ are provided by (\ref{drag}) and \eqref{resi} with $v^*_p$ equal to zero and the dynamic friction coefficient $\mu_d$ substituted by the static friction coefficient $\mu_s$. %
Hence, the critical value $\theta_{cr}$ of the Shields parameter which gives rise to sediment motion turns out to be %
\begin{equation}
\label{thetacr}
\shields_{cr} = \frac{4 \mu_s}{3 \alpha^2(c_D+\mu_s c_L)}
\:\:.
\end{equation}
At this stage, it is  worth pointing out that \eqref{thetacr} allows the value of $\alpha$ to be obtained if the critical value of the Shields parameter is known for example using the empirical predictors which are available in the literature and appropriate values of $c_D$, $c_L$ and $\mu_s$. %

Then, \eqref{v_p} leads to %
\begin{equation}
\label{v_p1}
\frac{\vpx^*}{u^*_\tau}= \alpha \left[ 1- c_1 \sqrt{ \frac{ \shields_{cr}}{\shields} } \right]
\end{equation}
where the value of $c_1$ is provided by %
\begin{equation}
\label{c_1}
c_1= \sqrt{\frac{\mu_d \left(c_D+\mu_s c_L \right)}{ \left(c_D+\mu_d c_L\right)\mu_s}}
\:\:.
\end{equation}
For values of $\shields$ larger than $\shields_{cr}$ but not too large, the bed load rate $q^*_b$ can be evaluated as a fraction $p$ of the number ($1/\ds^{*2}$) of sediment grains per unit area which are on the surface layer of the bottom times their volume and velocity. %
It follows that %
\begin{equation}
\label{Q_1}
q^*_b=  p \frac{1}{\ds^{*\,2}} \pi \frac{1}{6}\ds^{*\,3} \alpha u^*_\tau \left[ 1- c_1 \sqrt{ \frac{ \shields_{cr}}{\shields} } \right]
\:\:,
\end{equation}
which can be written in dimensionless form by introducing %
\rev{$\Phi_b=\frac{q^*_b}{\sqrt{(\s-1) \g^*\ds^{*3}}}$}{$\Phi_b$ defined by \eqref{theta1}} %
\begin{equation}
\label{Q_2}
\Phi_b%
\rev{=\frac{q^*_b}{\sqrt{(s-1) g^*d^{*3}}} }{}%
= \frac{\alpha \pi}{6} p \left[ \sqrt{\shields} - c_1 \sqrt{ \shields_{cr}} \right]\:\:.
\end{equation}

Finally, an estimate of $p$ can be obtained by assuming that only the skin friction part of the bottom shear stress tends to move the sediment particles while the residual part, which is transmitted to the bed by the collisions of the moving particles with the resting particles, does not contribute to set into motion the particles of the bed. %
Therefore, if $n^*$ denotes the number of moving particles per unit area, we have %
\begin{equation}
\label{p1}
\tau^*=\tau^*_{cr}+n^* F^*_D
\:\:.
\end{equation}
It follows %
\begin{equation}
\label{p2}
\tau^*=\tau^*_{cr}+n^* \frac{1}{2} \densf^* c_D \left( \vfx^*-\vpx^*\right)^2 \pi \frac{\ds^{*2}}{4} 
\:\:.
\end{equation}
Then, using (\ref{drag}) and (\ref{resi}) and taking into account that, for particles moving with a steady velocity, $F^*_D$ should be equal to $F^*_R$ %
\begin{equation}
\label{bal1}
\frac{1}{2} \densf^* c_D \left( \vfx^*-\vpx^*\right)^2 \pi \frac{\ds^{*\,2}}{4} =
\mu_d \densf^* \g^* (\s-1) \pi \frac{\ds^{*\,3} c_D}{6 \left( c_D+\mu_d c_L\right)}
\:\:.
\end{equation}
Hence %
\begin{equation}
\label{p3}
\shields-\shields_{cr}=n^* \mu_d \pi \frac{\ds^{*\,2} c_D}{ 6 \left( c_D+\mu_d c_L \right)} 
=p \mu_d \frac{\pi c_D}{ 6 \left( c_D+\mu_d c_L \right)}
\:\:,
\end{equation}
where $p=n^*\ds^{*2}$ is the ratio between the moving particles and the total number of particles in the surface layer. %
From (\ref{Q_2}) and (\ref{p3}), it is easy to obtain
\begin{equation}
\label{Q_3}
\Phi_b = 
\frac{\alpha \left( c_D+\mu_d c_L \right)}{ \mu_d c_D}
\left[\shields-\shields_{cr} \right]
\left[ \sqrt{\shields} - c_1 \sqrt{ \shields_{cr}} \right]
\:\:.
\end{equation}

To determine $\Phi_b$, it is necessary to evaluate the parameters appearing into the simplified approach previously summarized. Hereinafter, the critical value of the Shields parameter is evaluated by means of the relationship proposed by \citet{hanson2007} which is more suitable for coastal environments %
\begin{equation}
\label{theta_crit}
\shields_{cr}=0.08\left[1-\exp\left(-\frac{15}{\Ga^{2/3}}-0.02\,\Ga^{2/3}\right)\right]
\end{equation}
where $\Ga$ is the Galileo number (often referred to as sediment Reynolds number and indicated with $R_p$) defined by~\eqref{gal}. %

An estimate of the drag and lift coefficients can be obtained from the formulae of Schiller-Neumann \citep{clif1978} and \citet{takemura2003} %
\begin{equation}
\label{cd}
	c_D= \frac{24}{Re} \left( 1+0.15 Re^{0.687} \right)  
\end{equation}
\begin{equation}
\label{cl}
	c_L=c_{L0}\left( 1+0.6 Re^{0.5}-0.55 Re^{0.08}\right)^2 \left(\frac{L}{1.5}\right)^{-2\tanh\left(0.01 Re\right)} 
	\:\:,
\end{equation}
where the Reynolds number $Re$ is defined by %
\begin{equation}
\label{Re}
	Re= \frac{ \left(\vfx^*-\vpx^*\right)\ds^*}{\nu^*}
\end{equation}
and $L$ is the ratio $L^*/\ds^*$ between the average distance $L^*$ of the centre of the moving particles from the line which represents the idealized bottom surface and the diameter of the particles (cf. Appendix~\ref{appen1}). %
The value of $c_{L0}$ can be determined by means of %
\begin{equation}
\label{cl0}
	c_{L0}= \left( \frac{9}{8} +5.78 \times 10^{-6} \hat L^{4.58}\right) \beta^2 \exp\left(-0.292 \hat L\right)  \ \ \ \  \mbox{for} \ \ \ \  0<\hat L<10
\end{equation}
\[
	c_{L0}= 8.94 \beta^2 \hat L^{-2.09} \ \ \ \  \mbox{for} \ \ \ \ 10 \le \hat L < 300
\]
with $\hat L=L Re$ and $\beta=0.50698$. %

Reasonable values of $\mu_s$ and $\mu_d$ are 
\begin{equation}
\label{mu}
	\mu_s=\tan 32^\circ  \ \ \ \ \ \mu_d =  \tan 20^\circ
	\:\:.
\end{equation}

It is necessary to point out that the values of $c_D$ and $c_L$ provided by (\ref{cd}) and (\ref{cl}) might be rather different from the actual values because they were obtained under different conditions. %
Hence the values of $\Phi_b$ that are described in the following and are obtained using (\ref{cd}) and (\ref{cl}) should be considered just an estimate rather than a reliable quantitative prediction. %
Indeed these values are used simply to discuss qualitatively the results of the direct numerical simulations. %
Nonetheless, as discussed in the following, the main aspects of the average particle dynamics appear to be captured by the present model. %

\begin{figure}
\begin{picture}(0,155)(0,0)
  \put(0,-10){\includegraphics[trim=0cm 0cm 0cm 0cm, clip, width=.48\textwidth]{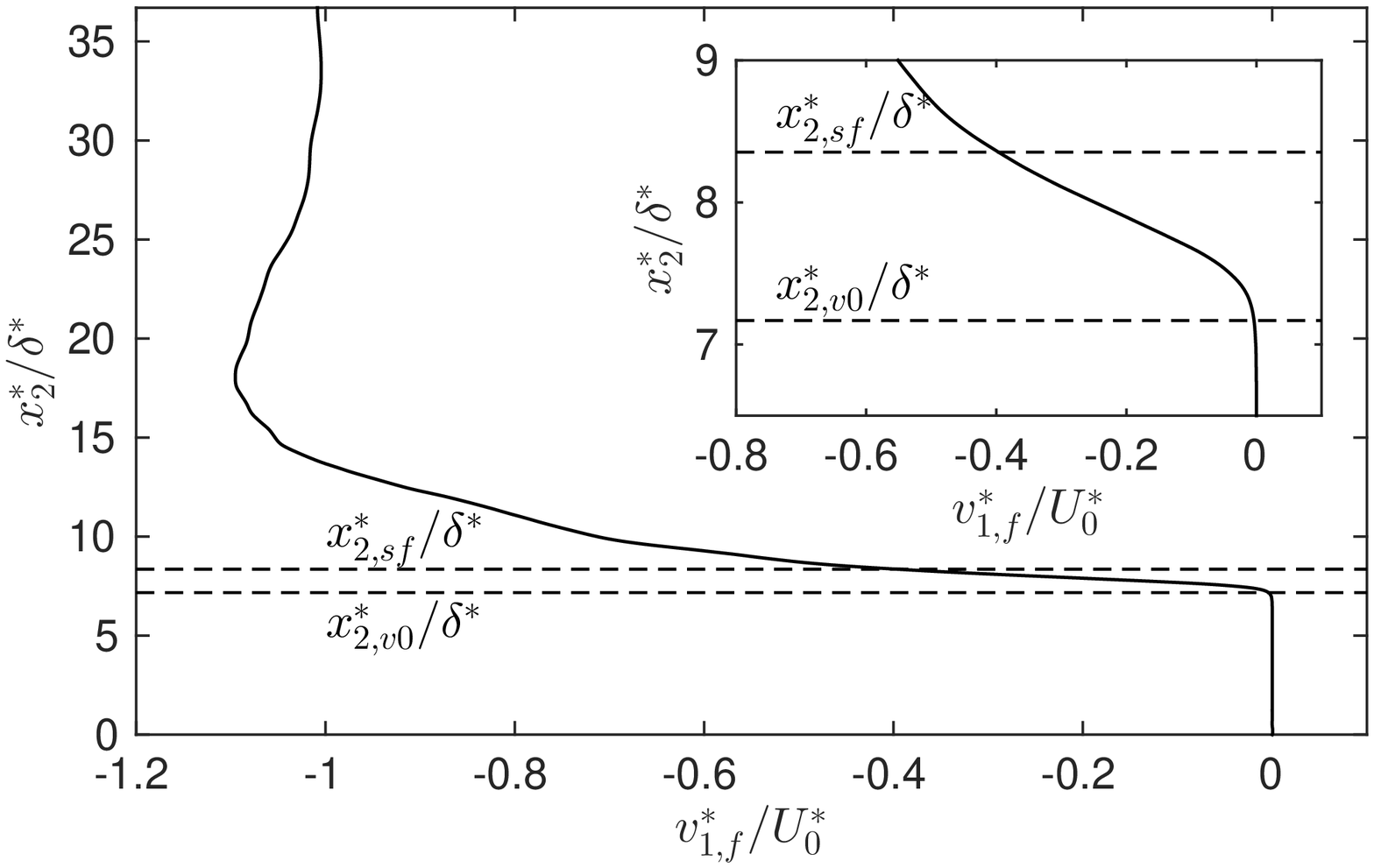}}
  \put(265,-10){\includegraphics[trim=0cm 0cm 0cm 0cm, clip, width=.48\textwidth]{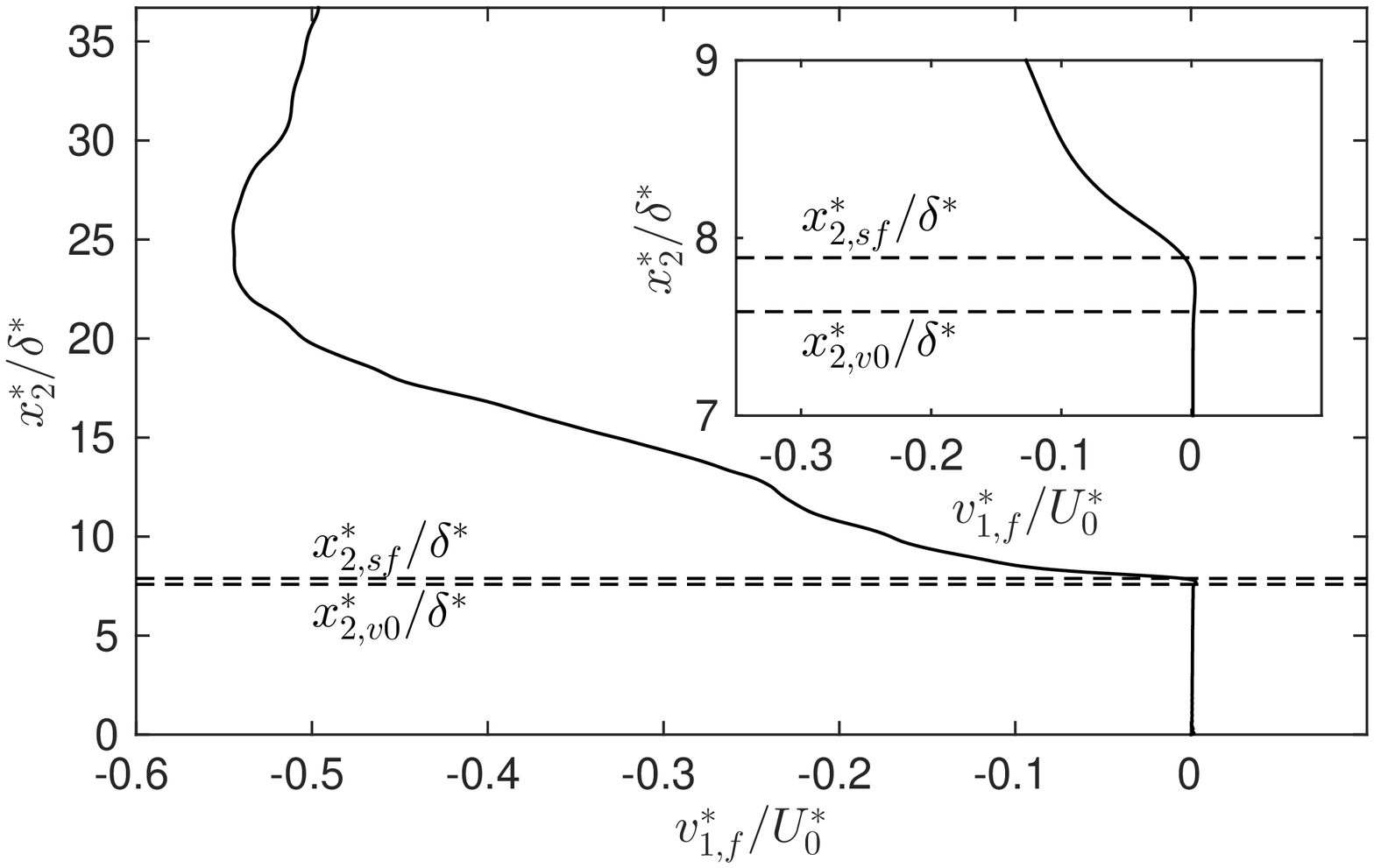}}
  \put(30,135){\small $(a)$}
  \put(295,135){\small $(b)$}
\end{picture}
\caption{%
\rev{Streamwise fluid-velocity component averaged over the horizontal plane plotted versus $\xf{2}$
for two different phases of the oscillatory cycle for $\ds^*/\del^*=0.335$ and $\Rdel=1000$ ($\om^*t^*= 4\,\pi $ (top panels) 
and $\om^*t^*=2.335\,\pi$ (right panel). The values of $\rbed$ and $\beds$ are indicated by a thin broken line. %
}{%
Streamwise velocity component averaged over the horizontal plane plotted versus $\xf{2}$ at $(a)$ $\om^*t^*= 4\,\pi $ and $(b)$ $\om^*t^*= 2.335\,\pi $, for $\ds^*/\del^*=0.335$ and $\Rdel=1000$. %
The values of $\rbed$ and $\beds$ are indicated by broken lines. %
Inset panels show the enlargement of the panels around the bottom surface elevation. %
}%
}%
\label{figcasa0}
\end{figure}

Since the approaches, which try to model sediment transport, consider a uniform flow over a horizontal bottom, the first point to be clarified is the position of the bottom. %
Let us introduce axiomatically two different ``bottom'' elevations $\rbed^*$ and $\beds^*$. %
The former is the largest value of $\xf{2}^*$ at which the solid volume fraction $c$ equals the value corresponding to the averaged value of the solid volume fraction within the sediment bulk. %
Hence, the particles do not move and this value is defined as the elevation of the ``resting bottom''. %
Even though the horizontal pressure gradient, which drives the flow, causes a slow filtration motion of the water within the resting sediments, from a practical point of view it can be assumed that the fluid velocity vanishes at $\xf{2}^*=\rbed^*$. %
Indeed the fluid velocity below this level turns out to be negligible with respect to $\U^*$ at any phase of the oscillatory cycle. %
The value of $\beds^*$ is the elevation at which the plane-averaged solid volume fraction $c$ is equal to $0.1$ and is named the elevation of the ``bottom surface'', since it approximates the surface dividing the water domain from the sediment domain. %
In figure \ref{figcasa0} the profiles of the streamwise velocity component $\vfx$ averaged along the $\xf{1}$- and $\xf{3}$-directions are plotted versus $\xf{2}$ at $\om^*t^*=4 \pi$ and $\om^*t^*=2.335\,\pi$ for $\ds^*/\del^*=0.335$ and $\Rdel=1000$. %
In the same figure the values of $\rbed$ and $\beds$ are also indicated and it clearly appears that below $\rbed$ the velocity practically vanishes at any phase while the velocity is significant also below the bottom surface ($\xf{2}=\beds$) but above the resting bed ($\xf{2}=\rbed$). %
%

\begin{figure}[t]
\begin{picture}(0,315)(0,0)
  \put(0,155){\includegraphics[trim=0cm 0cm 0cm 0cm, clip, width=.47\textwidth]{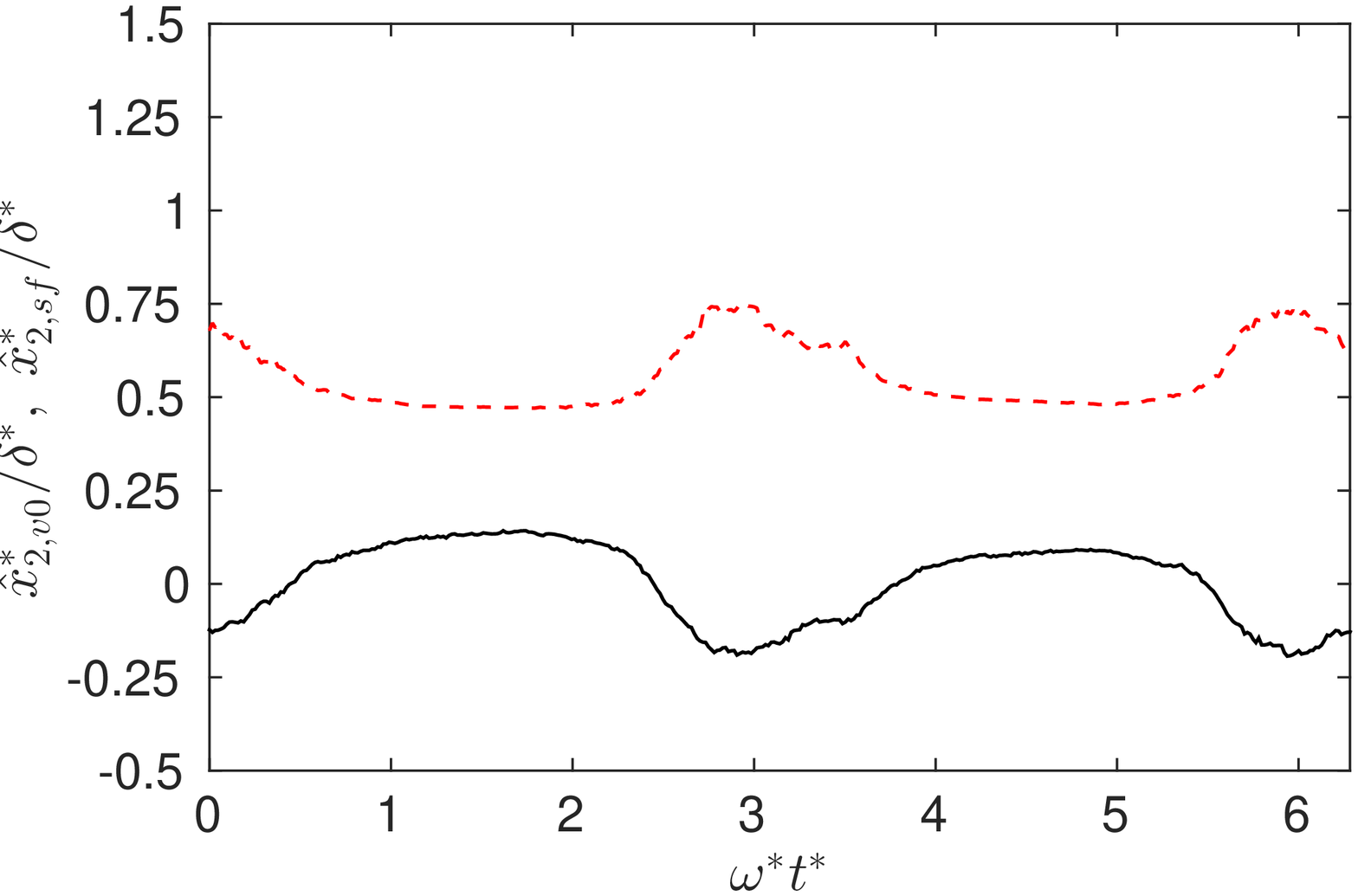}}
  \put(265,155){\includegraphics[trim=0cm 0cm 0cm 0cm, clip, width=.47\textwidth]{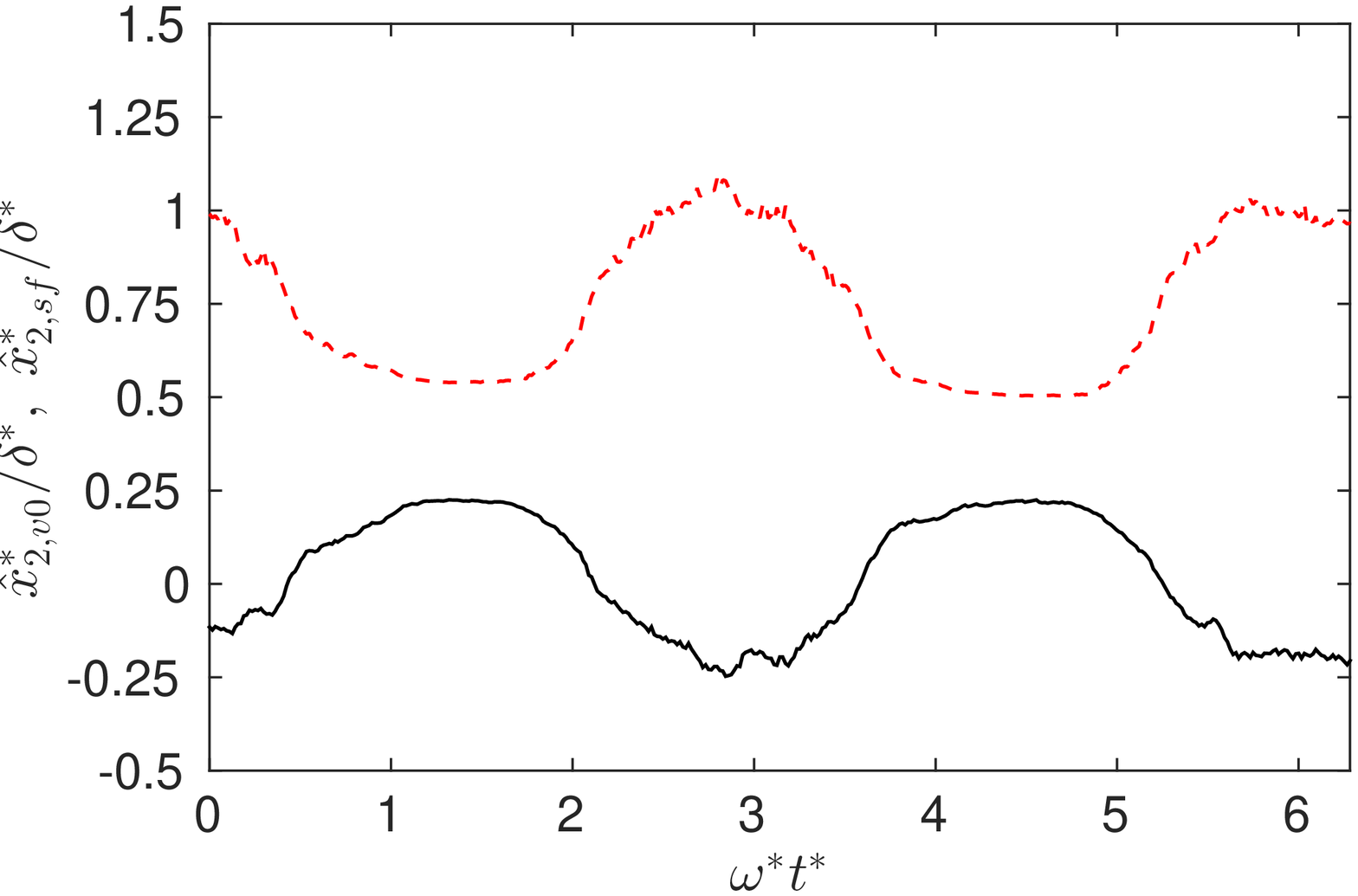}}
  \put(0,-10){\includegraphics[trim=0cm 0cm 0cm 0cm, clip, width=.47\textwidth]{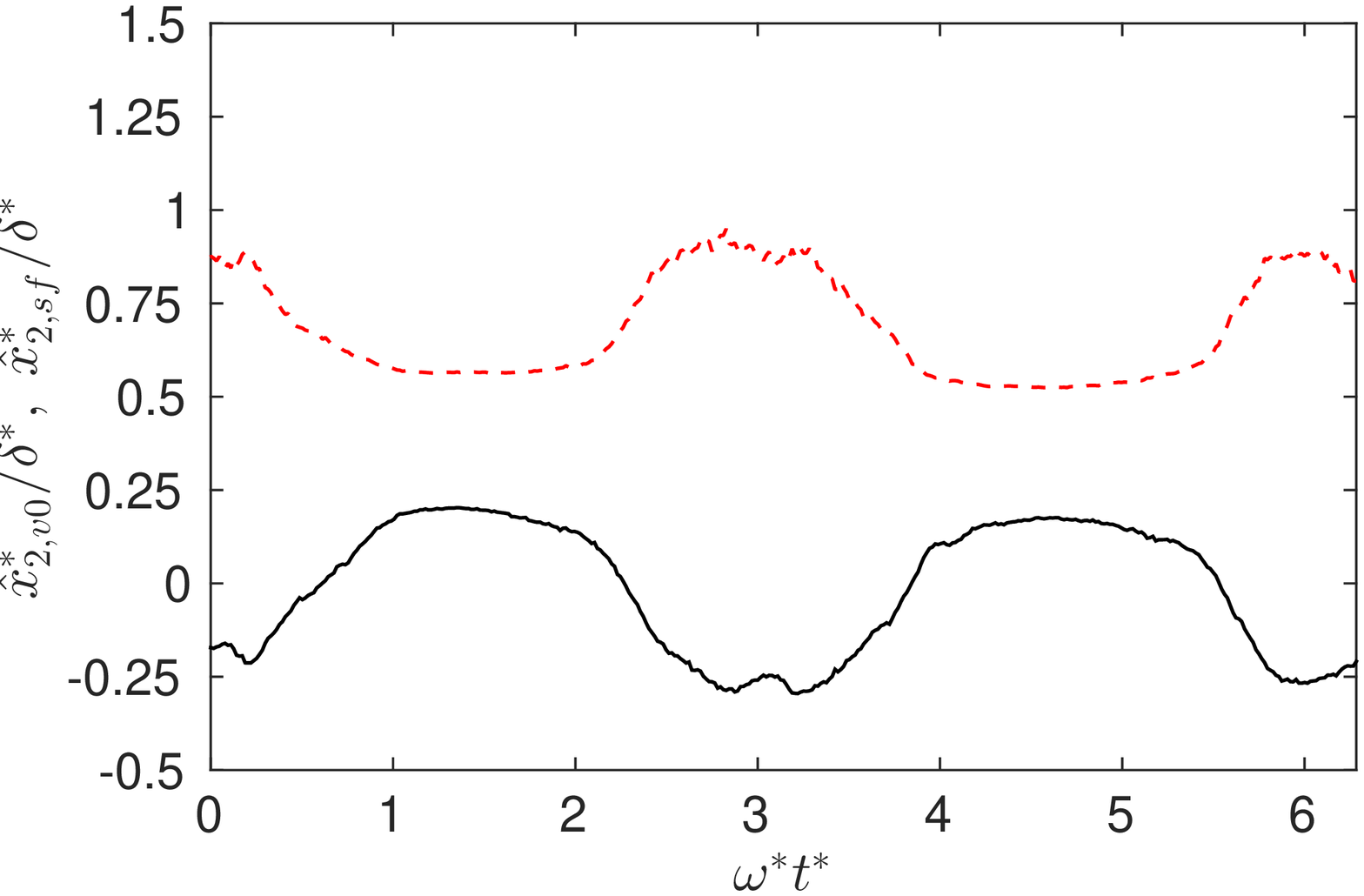}}
  \put(265,-10){\includegraphics[trim=0cm 0cm 0cm 0cm, clip, width=.46\textwidth]{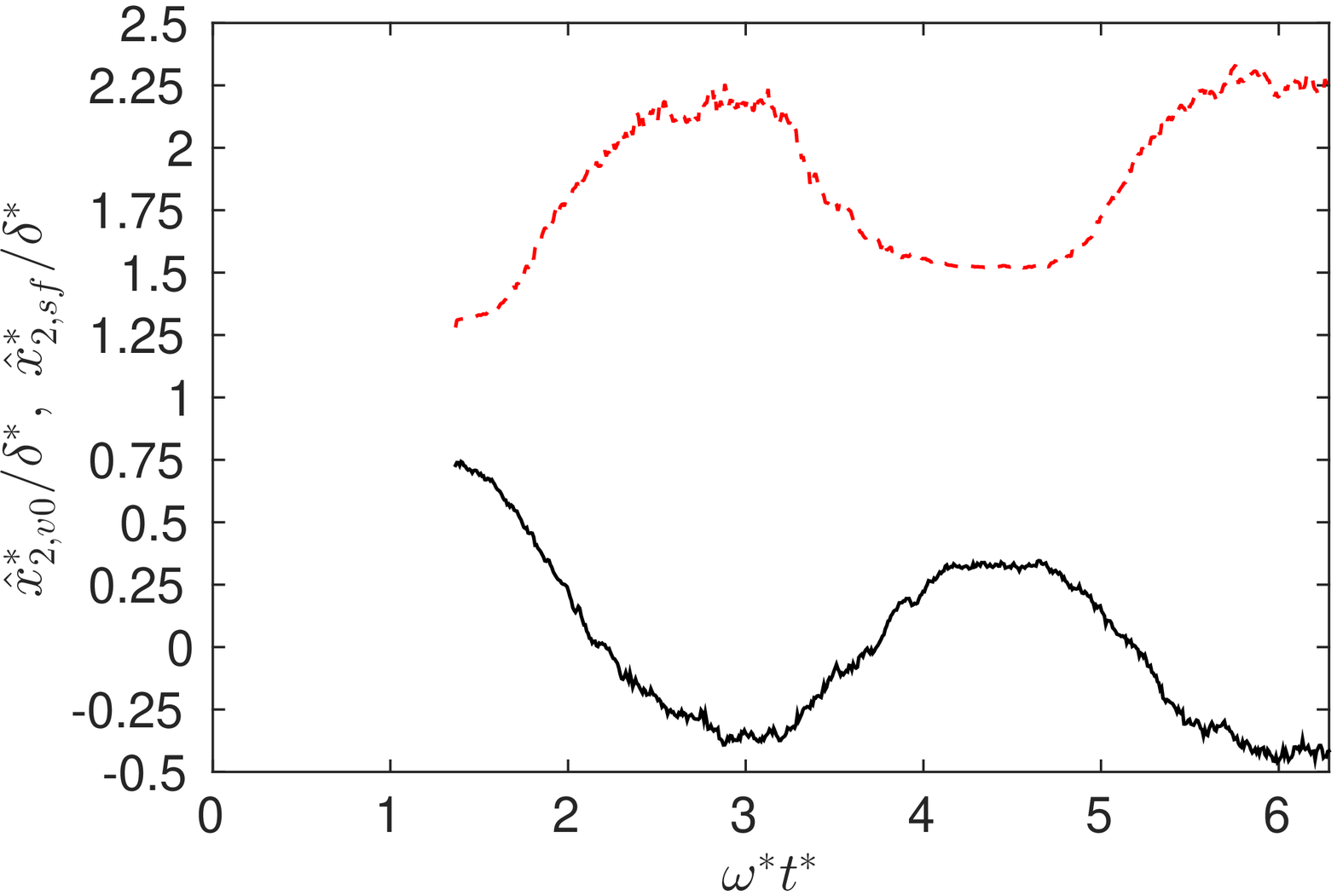}}
  \put(50,295){$(a)$}
  \put(315,295){$(b)$}
  \put(50,130){$(c)$}
  \put(315,130){$(d)$}
  \put(0,191){%
  \put(70,84){\small bottom surface}
  \put(90,80){\vector(0,-1){25}}
  \put(70,0){\small resting bottom surface}
  \put(90,9){\vector(0,1){22}}
  }%
\end{picture}
	\caption{\rev{%
	Values of $\beds$ (broken line) and $\rbed$ (solid line) as function of the phase within the wave cycle
	for $\ds^*/\del^*=0.335$ and $\Rdel=750$ (left panel) and $\Rdel=1000$ (right panel). %
	}{%
	Values of $\hat{x}_{2,sf}$ (broken line) and $\hat{x}_{2,v0}$ (solid line) as functions of the phase within the wave cycle
	for $(a)$ $\ds^*/\del^*=0.335$ and $\Rdel=750$, $(b)$ $\ds^*/\del^*=0.335$ and $\Rdel=1000$, 
	$(c)$ $\ds^*/\del^*=0.168$ and $\Rdel=1000$, $(d)$ $\ds^*/\del^*=0.335$ and $\Rdel=1500$. %
	}%
}
\label{figcasa1}
\end{figure}

\begin{figure}[t]
\begin{picture}(0,210)(0,0)
  \put(0,-10){\includegraphics[trim=0cm 0cm 0cm 0cm, clip, width=.48\textwidth]{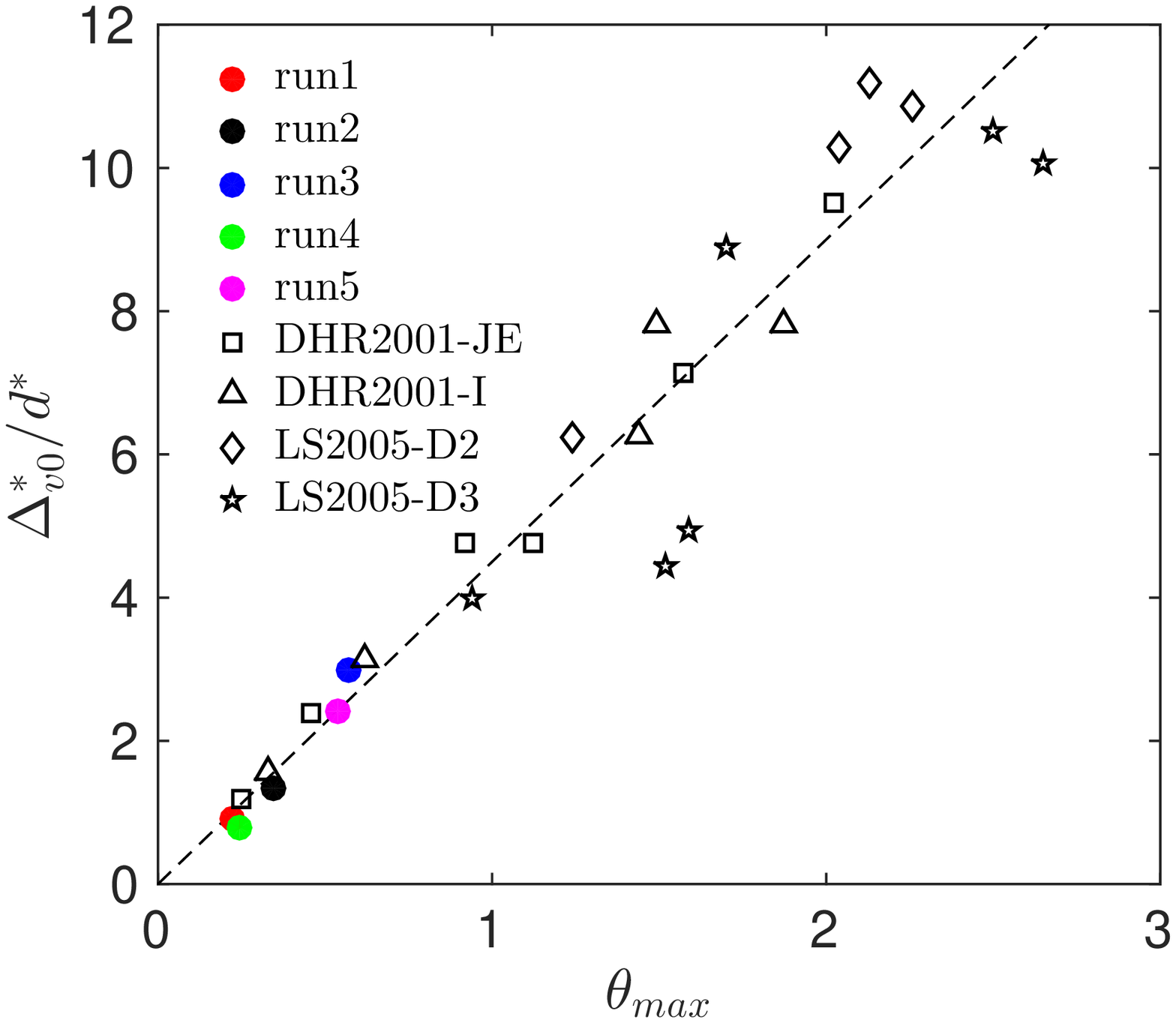}}
  \put(265,-10){\includegraphics[trim=0cm 0cm 0cm 0cm, clip, width=.5\textwidth]{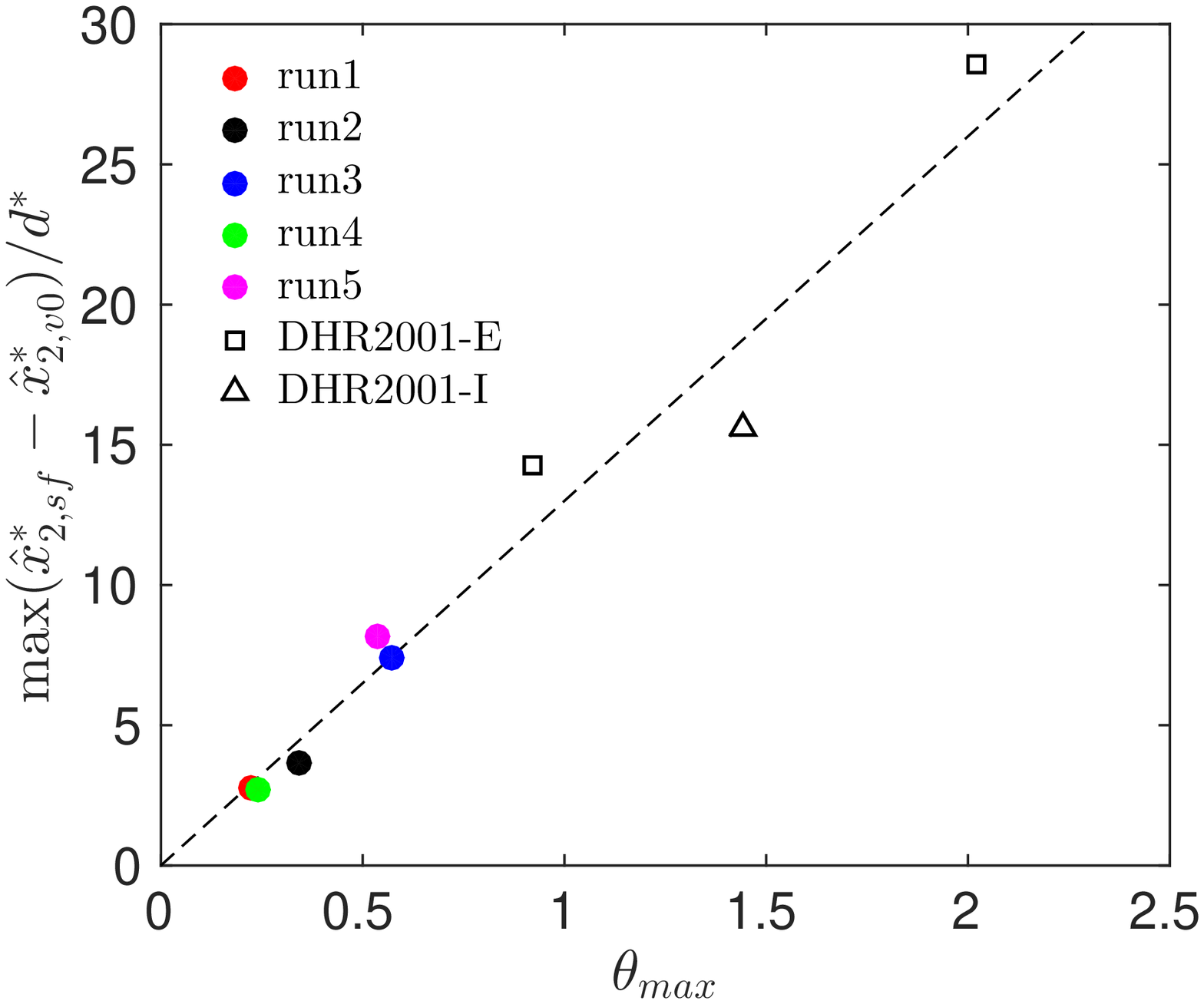}}
  \put(0,195){$(a)$}
  \put(265,195){$(b)$}
\end{picture}
	\caption{\rev{}{%
	$(a)$ Maximum vertical excursion $\Delta_{v0}^*$ of the resting bed elevation during the wave period ($\Delta_{v0}^*=\max(\rbed^*)-\min(\rbed^*)$ also referred to as \textit{erosion depth}) and $(b)$ maximum distance between the resting bed and the bed surface elevations plotted as functions of the maximum value $\shields_{max}$ of the Shields parameter. %
	Both quantities are normalized by $\ds^*$. %
	The experimental measurements of \citet{dohmen2001} and of \citet{liu2005} are indicated by the empty markers (DHR2001- and LS2005- followed by the experiment series letters, respectively. D2 and D3 stand for the diameter of sediments equal to $0.21$~mm and $0.3$~mm, respectively). %
	Broken lines are equal to $(a)$ $4.5\,\theta_{max}$ and $(b)$ $13\,\theta_{max}$, which were suggested by \citet{dohmen2001}. %
	}
}
\label{figcasa1a}
\end{figure}

In figure \ref{figcasa1} the values of $\rbed$ and $\beds$ are plotted versus the phase $\om^*t^*$ within the wave cycle for %
\rev{%
$\ds^*/\del^*=0.335$ and $\Rdel=750$ and $\Rdel=1000$ %
}{%
run~$1$, run~$2$, run~$3$ and run~$5$ %
}%
(in figure \ref{figcasa1}, the origin of the $x_2$-axis is fixed so that the time averaged value of $x_{2,v0}$ vanishes and just one oscillatory cycle is plotted). %
The results plotted in figure \ref{figcasa1} show that the bottom surface varies during the wave cycle because the bottom expands when the sediments are set into motion and modify their relative position. %
It follows that simultaneously the elevation of the resting bottom decreases. %
The periodic oscillations of both $\rbed$ and $\beds$ are relatively small, i.e. they are of the order of magnitude of the grain size as well as the difference between the values of $\rbed$ and $\beds$, which can be thought to be the order of magnitude of the thickness of the layer where the sediment particles roll and slide over the resting particles and collide among them with free paths that are of the order of magnitude of $\ds^*$. %
\rev{}{%
The vertical excursion of the resting bed elevation $\rbed^*$, hereafter referred to as \emph{erosion depth} and denoted by $\Delta_{v0}^*$, was measured by \citet{dibajnia2001} and \citet{liu2005} for values of the parameters comparable with those of the present simulations and also for cases where the sheet flow regime was observed. %
Figure~\ref{figcasa1a} shows that the present results are in good agreement with the experimental measurements of \citet{dibajnia2001} and \citet{liu2005}. %
Both the values of $\Delta_{v0}^*/\ds^*$ and of $\max(\beds^*-\rbed^*)/\ds^*$, plotted against the maximum value $\shields_{max}$ reached by the Shields parameter, lay on the lines that fit the experimental data of \citet{dibajnia2001}. %
Both quantities exhibit a good scaling once they are normalised by the particle diameter. %
The present results shown in figure~\ref{figcasa1a}b suggest that the thickness of the layer where particles roll and slide is proportional to the bed shear stress as much as the thickness of sheet flow layers. %
}%

When $\shields$ becomes large, a significant number of sediment grains are found above $\beds^*$. %
These sediment grains belong either to the so-called ``saltation layer'' or the so-called ``suspension layer''. %
In the saltation layer the sediments are picked-up by the strongest turbulent vortex structures and ``saltate'' making jumps much longer than their size but still of the order of magnitude of $\U^*/\om^*$, which is the characteristic length scale of the fluid motion. %
When the sediment particles are trapped within the turbulent vortex structures for time intervals much longer than $\TT^*$ and travel distances much longer than $\U^*/\om^*$ without interacting with the bed, it can be assumed that the sediment particles are set into suspension. %
%
\begin{figure}
\begin{picture}(0,170)(0,0)
  \put(0,-14){\includegraphics[trim=0cm 0cm 0cm 0cm, clip, width=.47\textwidth]{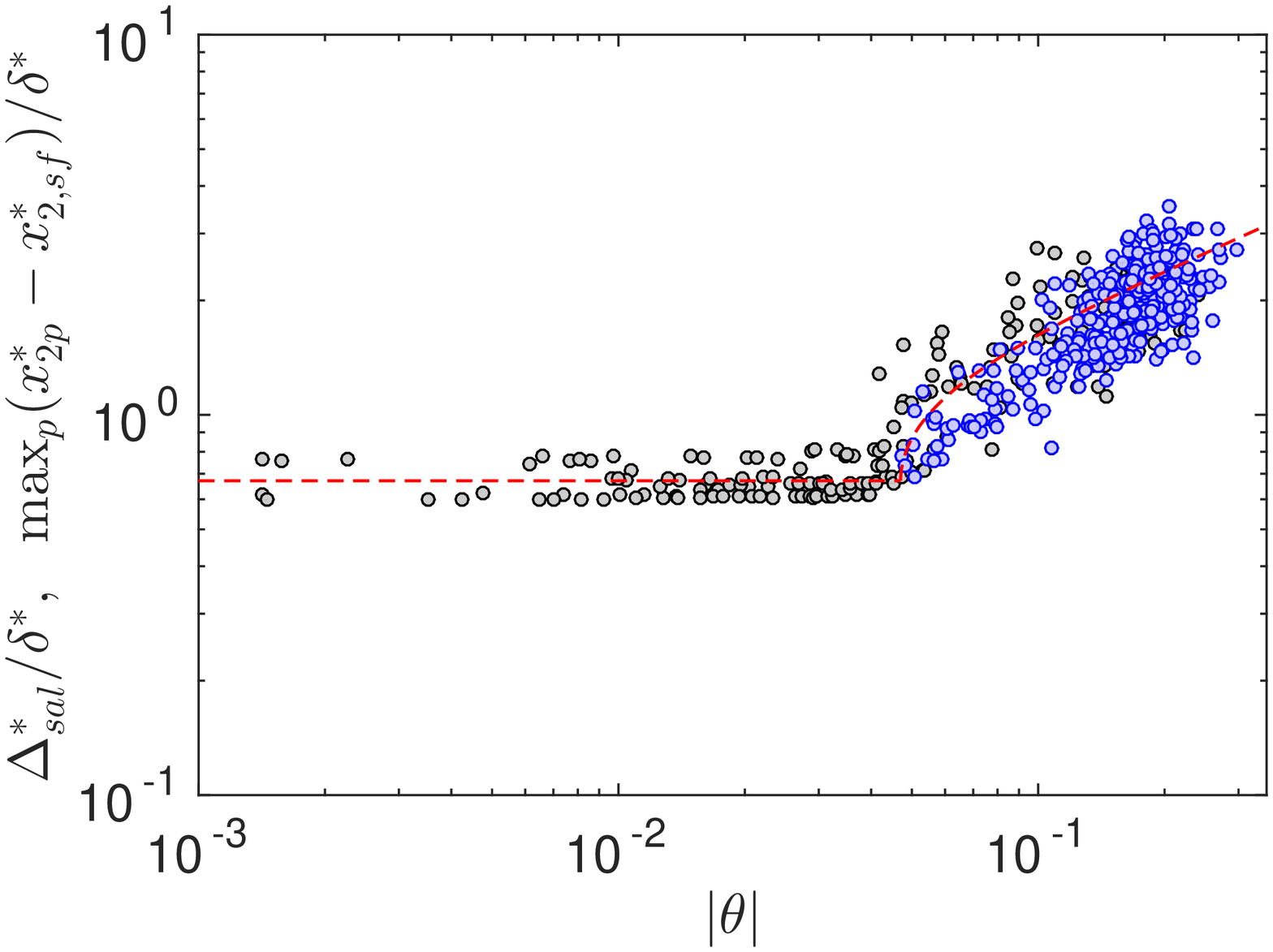}}
  \put(265,-14){\includegraphics[trim=0cm 0cm 0cm 0cm, clip, width=.47\textwidth]{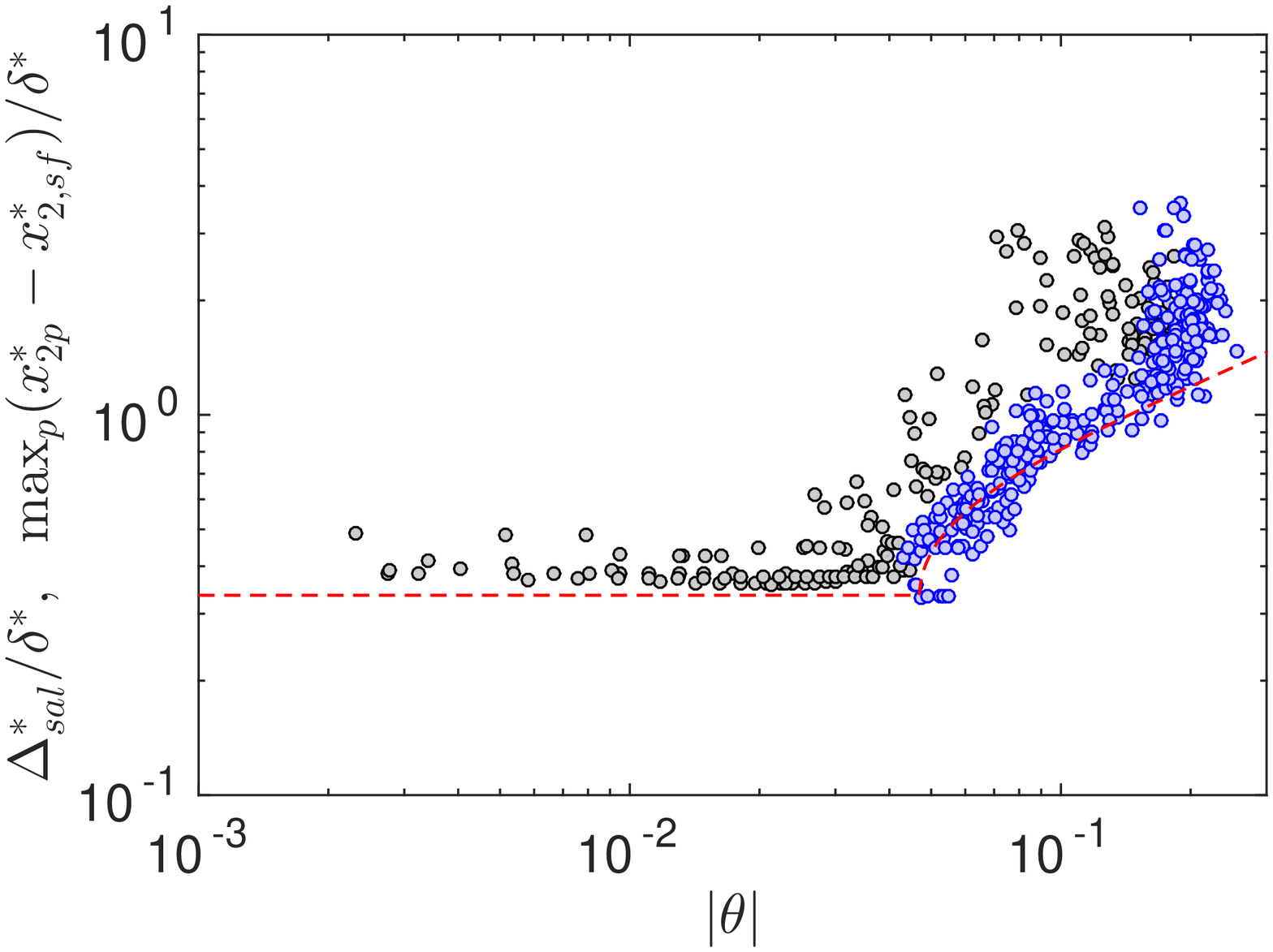}}
  \put(50,150){$(a)$}
  \put(310,150){$(b)$}
\end{picture}
\caption{Values of the maximum distance of particles from $\beds$ as a function of the Shields parameter
	for $(a)$ $\Rdel=1000$, $\ds^*/\del^*=0.67$ (run~$4$) and $(b)$ $\Rdel=750$, $\ds^*/\del^*=0.335$ (run~$1$). %
	Blue and black circles refer to accelerating and decelerating phases, respectively. %
	The red broken line indicates the thickness of the saltation layer evaluated by \eqref{mu}. %
}
\label{figcasa2}
\end{figure}
\begin{figure}
\begin{picture}(0,120)(0,0)
  \put(0,-10){\includegraphics[trim=0cm 0cm 0cm 0cm, clip, width=.31\textwidth]{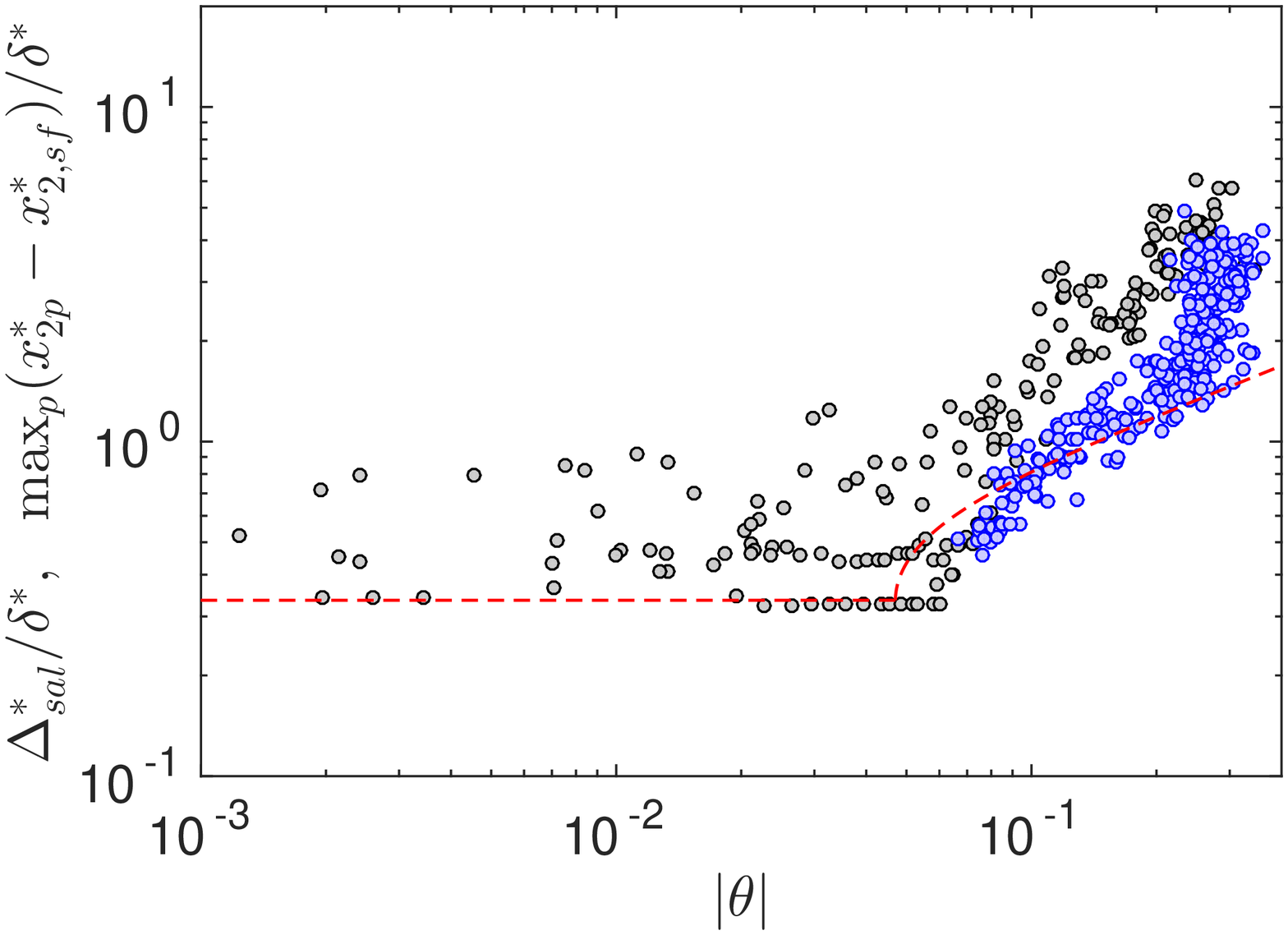}}
  \put(175,-10){\includegraphics[trim=0cm 0cm 0cm 0cm, clip, width=.31\textwidth]{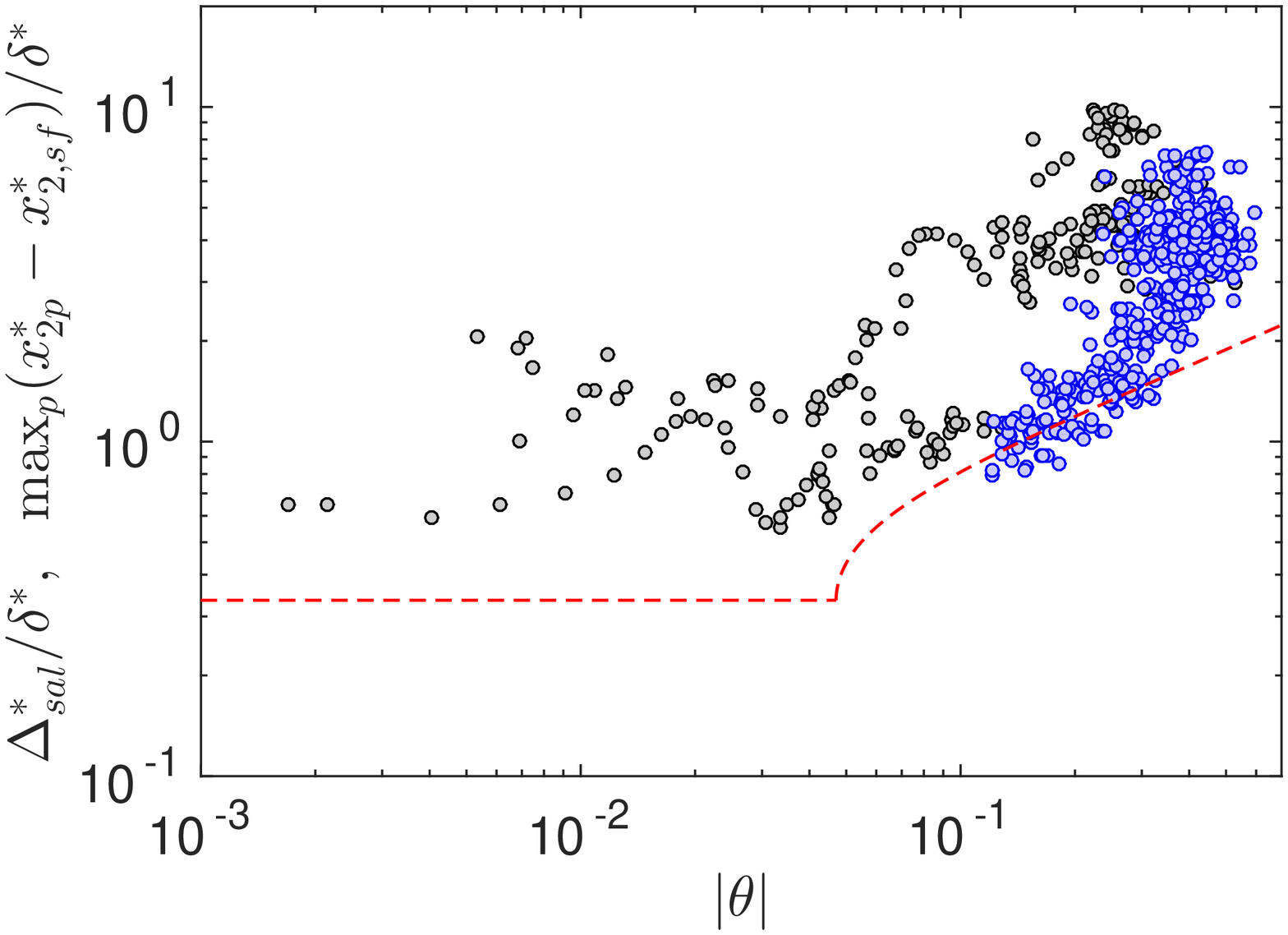}}
  \put(355,-10){\includegraphics[trim=0cm 0cm 0cm 0cm, clip, width=.31\textwidth]{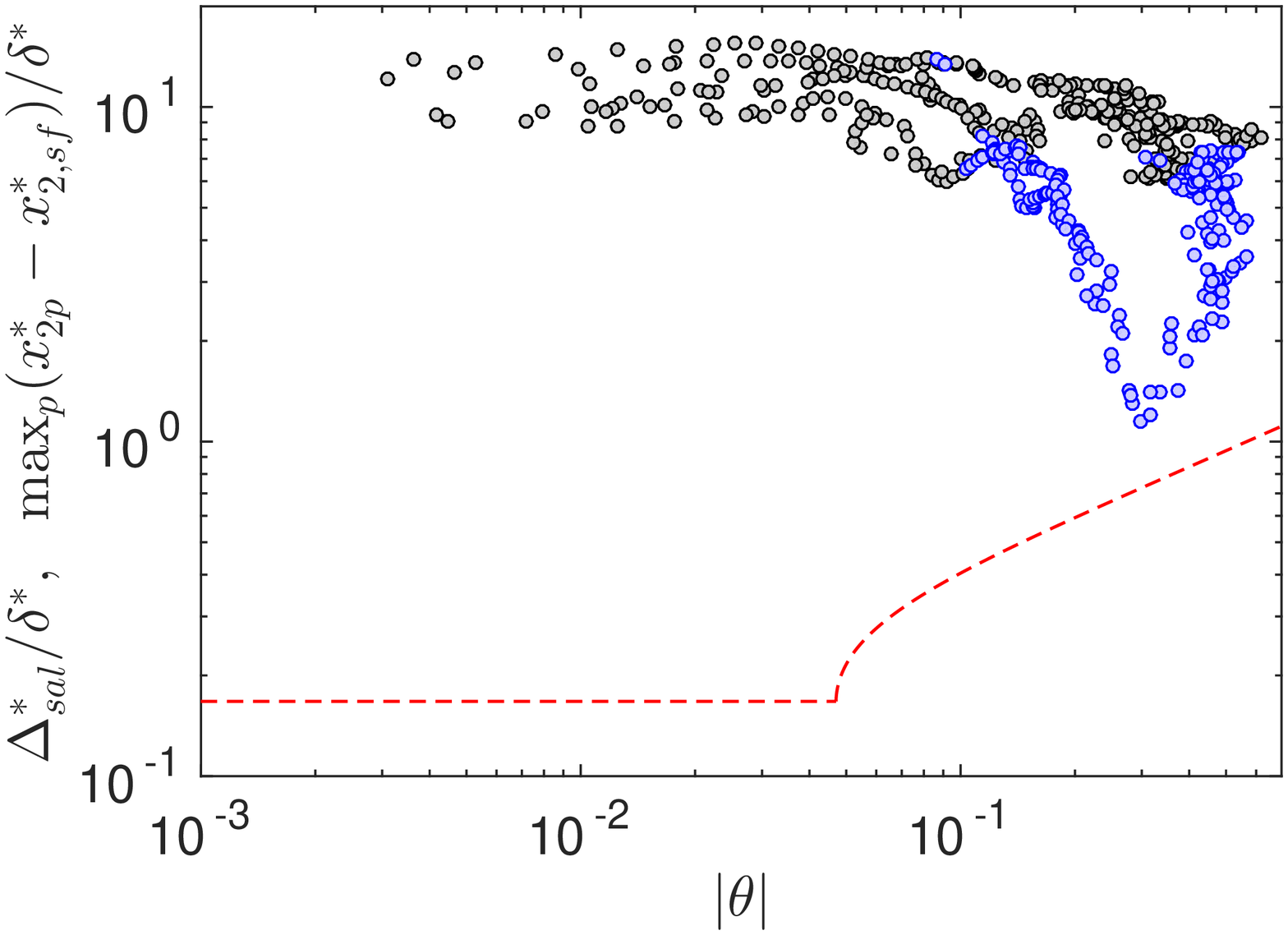}}
  \put(32,95){$(a)$}
  \put(205,95){$(b)$}
  \put(385,95){$(c)$}
\end{picture}
\caption{Values of the maximum distance of particles from $\beds$ as a function of the Shields parameter
	for $(a)$ $\Rdel=1000$, $\ds^*/\del^*=0.335$ (run~$2$), $(b)$ $\Rdel=1500$, $\ds^*/\del^*=0.335$ (run~$5$) and $(c)$ $\Rdel=1000$, $\ds^*/\del^*=0.168$ (run~$3$). %
	Blue and black circles refer to accelerating and decelerating phases, respectively. %
	The red broken line indicates the thickness of the saltation layer evaluated by \eqref{mu}. %
}
\label{figcasa4}
\end{figure}

Experimental observations show that the thickness $\salt^*$ of the layer where the sediment grains ``saltate'' depends on the sediment size and the Shields parameter. In particular the following relationship is found to fit reasonably the experimental data obtained in steady flow %
\begin{equation}
\label{mu}
	\salt^*=d^* \left[ 1 + A_b \left( \frac{\shields-\shields_{cs}}{\shields_{cs}}  \right)^{m} \right]
\:\:.
\end{equation}
As discussed in \citet{colombini2004}, a regression analysis on the experimental data of \citet{sekine1992} and \citet{lee1994} shows that the values of the constant $A_b$ and of the exponent $m$ in steady flows can be set equal to $1.33$ and to $0.55$, respectively, with $\shields_{cs}=0.047$ \citep[see also][]{lee2002}. %

Figure \ref{figcasa2}a shows the thickness of the layer above the bottom surface where sediments are found, as provided by the numerical simulations. %
The thickness of this layer is plotted as a function of the Shields parameter for $\Rdel=1000$ and $\ds^*/\del^*=0.67$. In the same figure, the value obtained by means of (\ref{mu}) is plotted as a continuous line. %
The results of the numerical simulations suggest that for such values of the parameters no sediment is put into suspension and the phenomenon can be approximated by a succession of steady flows \citep{mazzuoli2019b}. %
Similar results are found also for $\Rdel=750$ and $\ds^*/\del^*=0.335$ even though it appears that some particles move far from the bottom and the thickness of the layer where particles are found turns out to be slightly larger than that predicted by (\ref{mu}) (see figure \ref{figcasa2}b). %
\rev{}{%
Actually, figure~\ref{figcasa2} shows the results of the simulations where no suspended particles were observed during the wave cycle. %
}%
Larger values of the Reynolds number lead to a thickness of the layer, where sediments are found far from the bottom, that start to deviate from those predicted by (\ref{mu}) (see figure \ref{figcasa4}a) and a smaller particle size tends to trap the sediment particles within the turbulent eddies and the particles travel horizontal distances much larger than $U_0^*/\omega^*$ without interacting with the bottom (see figure \ref{figcasa4}b). %
In other words, for $d^*/\delta^*=0.168$ and $R_\delta=1000$, the sediments are suspended by the flow. %
Indeed, the value of the ratio between the maximum shear velocity and the fall velocity turns out to be larger than $1$, a value that some Authors assume
to be the limit for the appearance of the suspended load. %

\rev{%
When the sediment particles saltate, taking into account that the number of particles which do long and high jumps is significantly smaller than that of the particles doing short and low jumps, it is reasonable to consider values of $L^*$ significantly smaller than $0.5~\salt^*$. %
}{%
It is reasonable to relate the dimensionless parameter $L=L^*/\ds^*$, which appears in \eqref{cl}, to $\salt^*$. %
}%
The evaluation of the centre of gravity of the moving particles suggests $L^*=0.08~\salt^*$, which is approximately constant and equal to $0.3~\ds^*$ during the phases where sediment particles are saltating (see Appendix~\ref{appen1}). %
%

\begin{figure}
\begin{picture}(0,190)(0,10)
  \put(0,-8){\includegraphics[trim=0cm 0cm 0cm 0cm, clip, width=.48\textwidth]{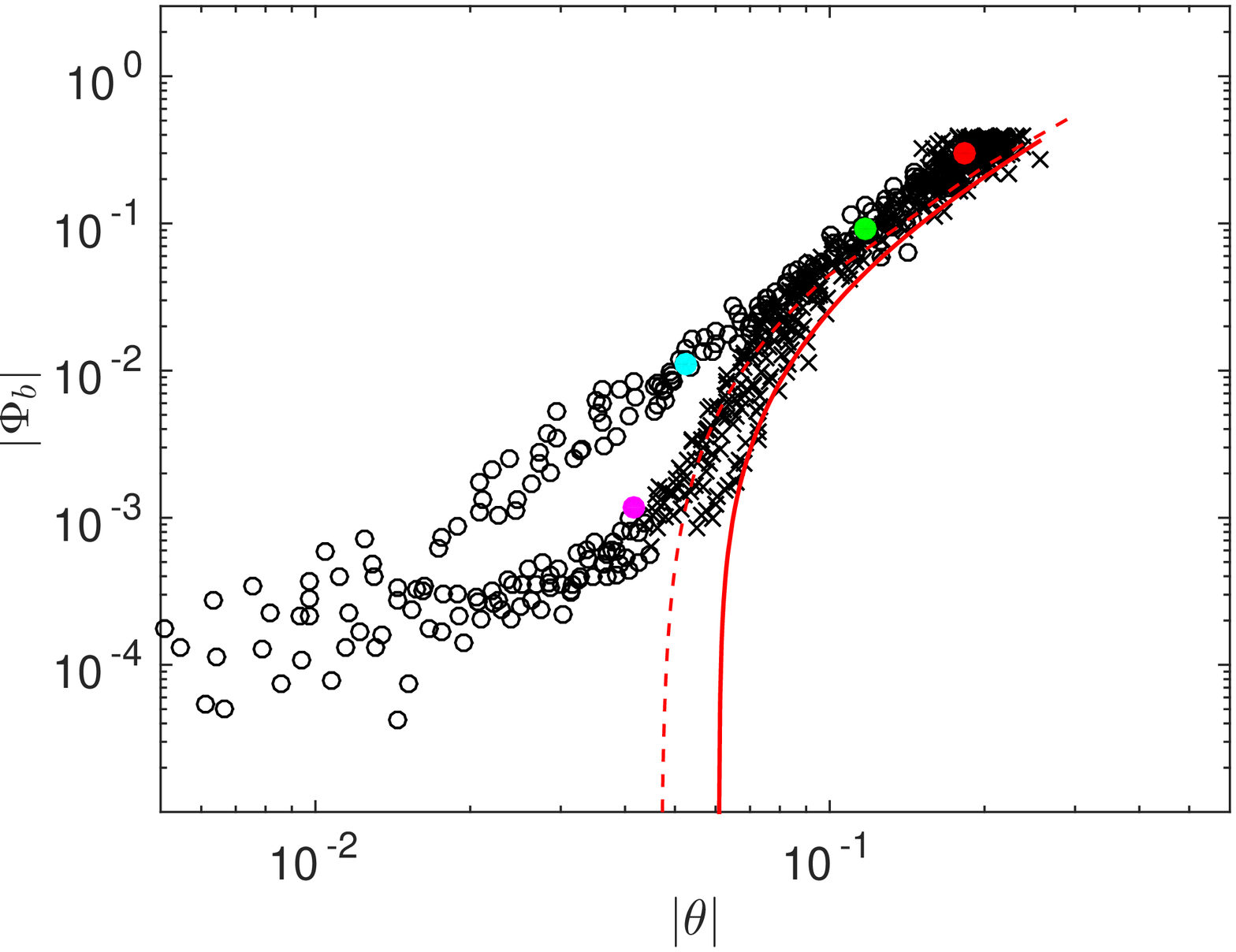}}
  \put(270,-8){\includegraphics[trim=0cm 0cm 0cm 0cm, clip, width=.48\textwidth]{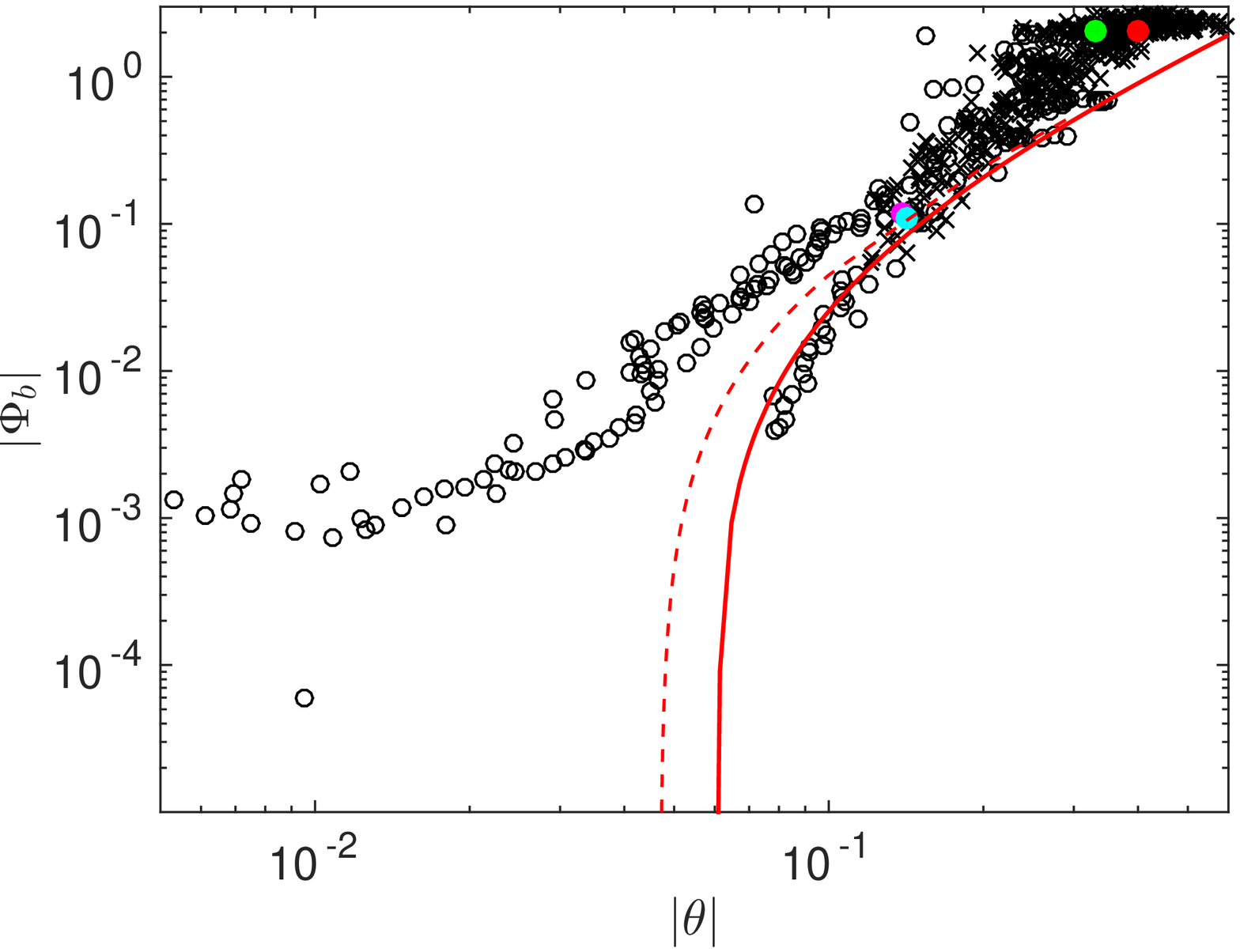}}
  \put(45,175){$(a)$}
  \put(320,175){$(b)$}
  \put(150,80){\vector(2,3){30}}
  \put(185,170){\vector(-3,-2){30}}
  \put(250,0){%
  \put(185,90){\vector(2,3){30}}
  \put(190,175){\vector(-3,-2){30}}
  }%
\end{picture}
\caption{
Dimensionless sediment flow rate $\Phi_b$ plotted versus the Shields parameter $\shields$ for $\ds^*/\del^*=0.335$ and $(a)$ $\Rdel=750$, $(b)$ $\Rdel=1500$. %
Results of the numerical simulations (crosses and circles refer to the accelerating and decelerating phases, respectively. %
\rev{}{%
The solid line indicates the values provided by \eqref{Q_3} and the broken red line indicates the values provided by \eqref{WP}. %
Arrows indicate the direction of the evolution during the wave-cycle. %
Coloured markers indicate values obtained at: %
$[${\protect\tikz \protect\draw[red,fill=red] (0,0) circle (.5ex);}$]$ $t=2\pi$; 
$[${\protect\tikz \protect\draw[cyan,fill=cyan] (0,0) circle (.5ex);}$]$ $t=2.25\pi$; 
$[${\protect\tikz \protect\draw[magenta,fill=magenta] (0,0) circle (.5ex);}$]$ $t=2.5\pi$ and 
$[${\protect\tikz \protect\draw[green,fill=green] (0,0) circle (.5ex);}$]$ $t=2.75\pi$. %
}
}%
\label{fig1a}
\end{figure}
\begin{figure}
\begin{picture}(0,125)(0,0)
  \put(0,-15){\includegraphics[trim=0cm 0cm 0cm 0cm, clip, width=.33\textwidth]{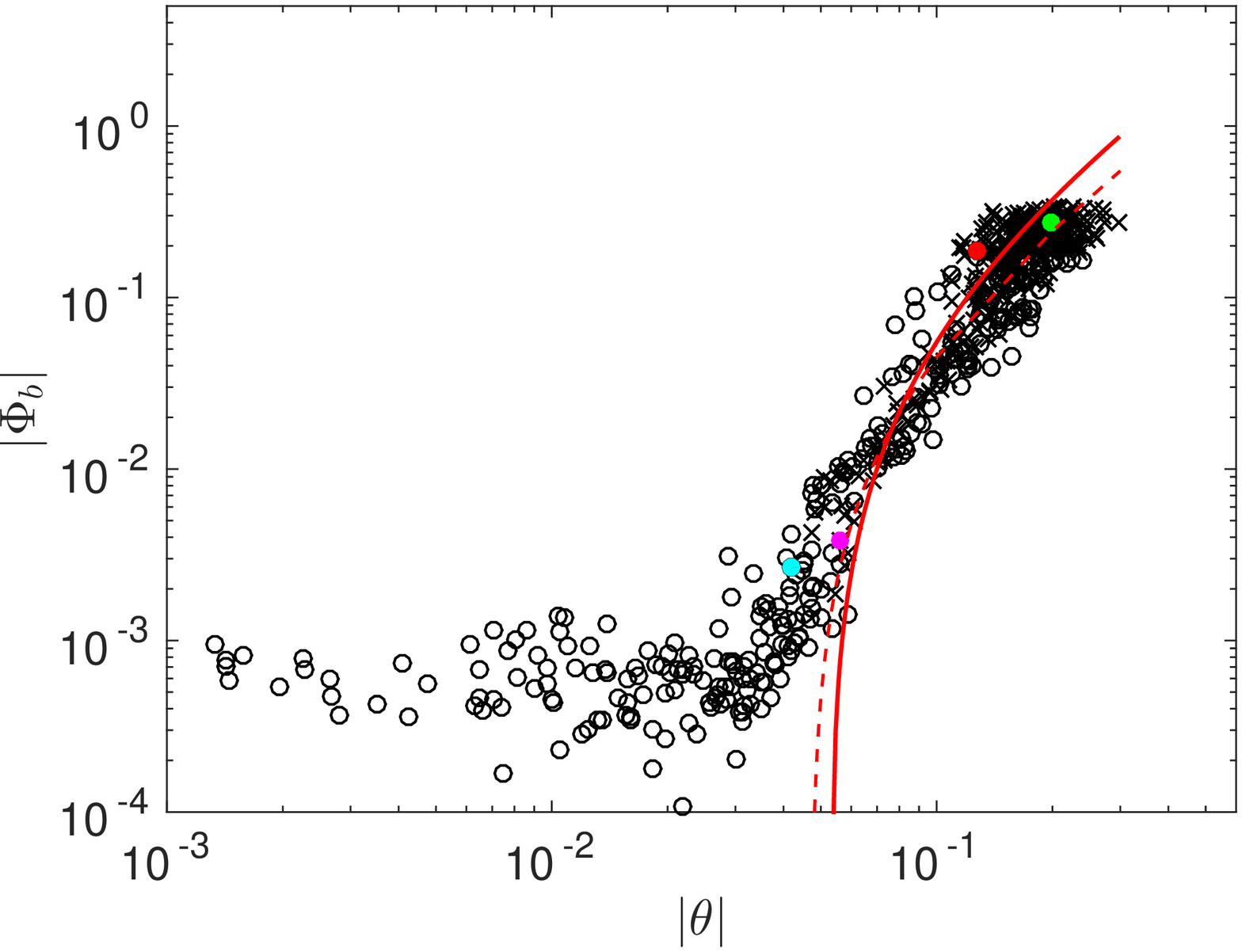}}
  \put(180,-15){\includegraphics[trim=0cm 0cm 0cm 0cm, clip, width=.33\textwidth]{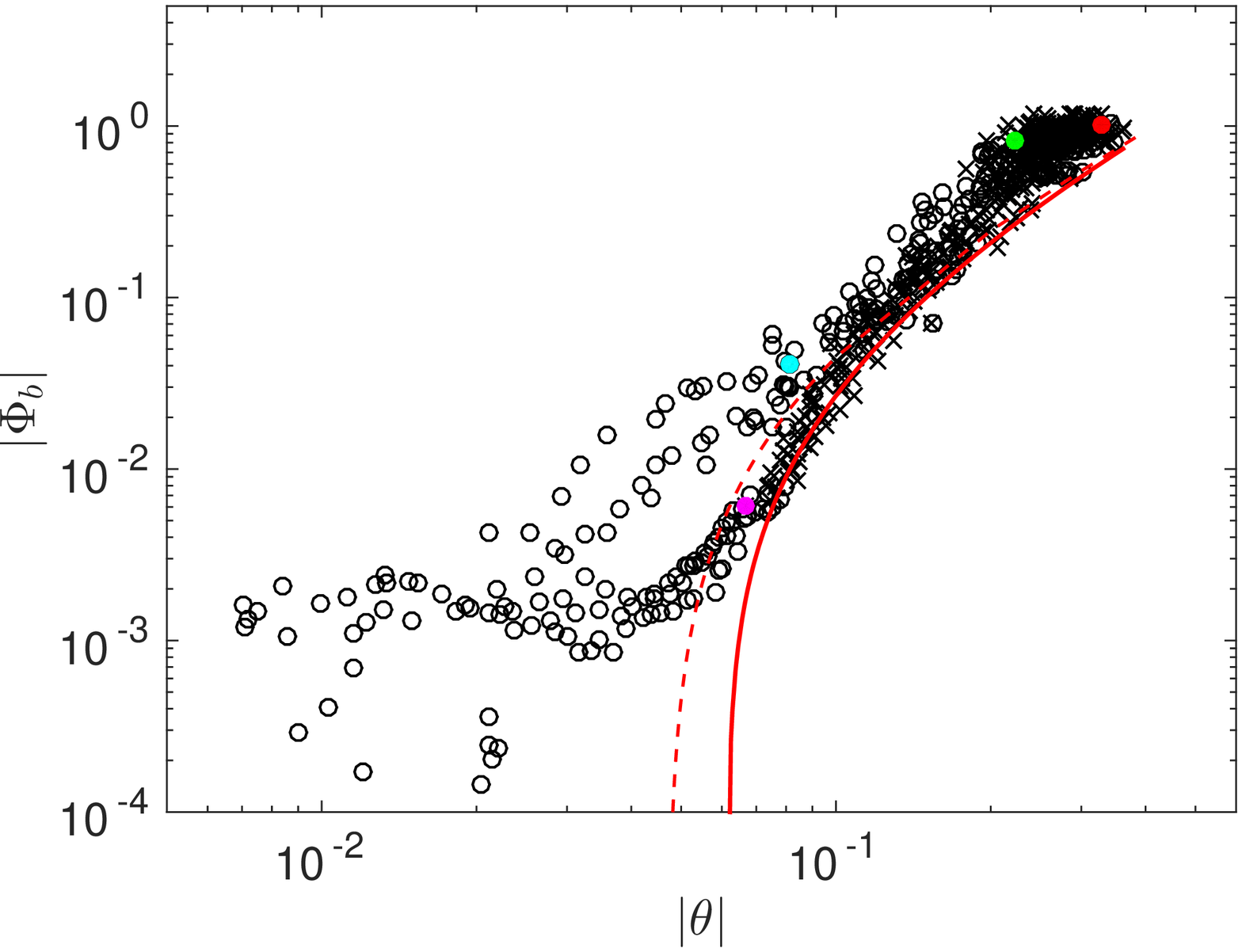}}
  \put(360,-15){\includegraphics[trim=0cm 0cm 0cm 0cm, clip, width=.33\textwidth]{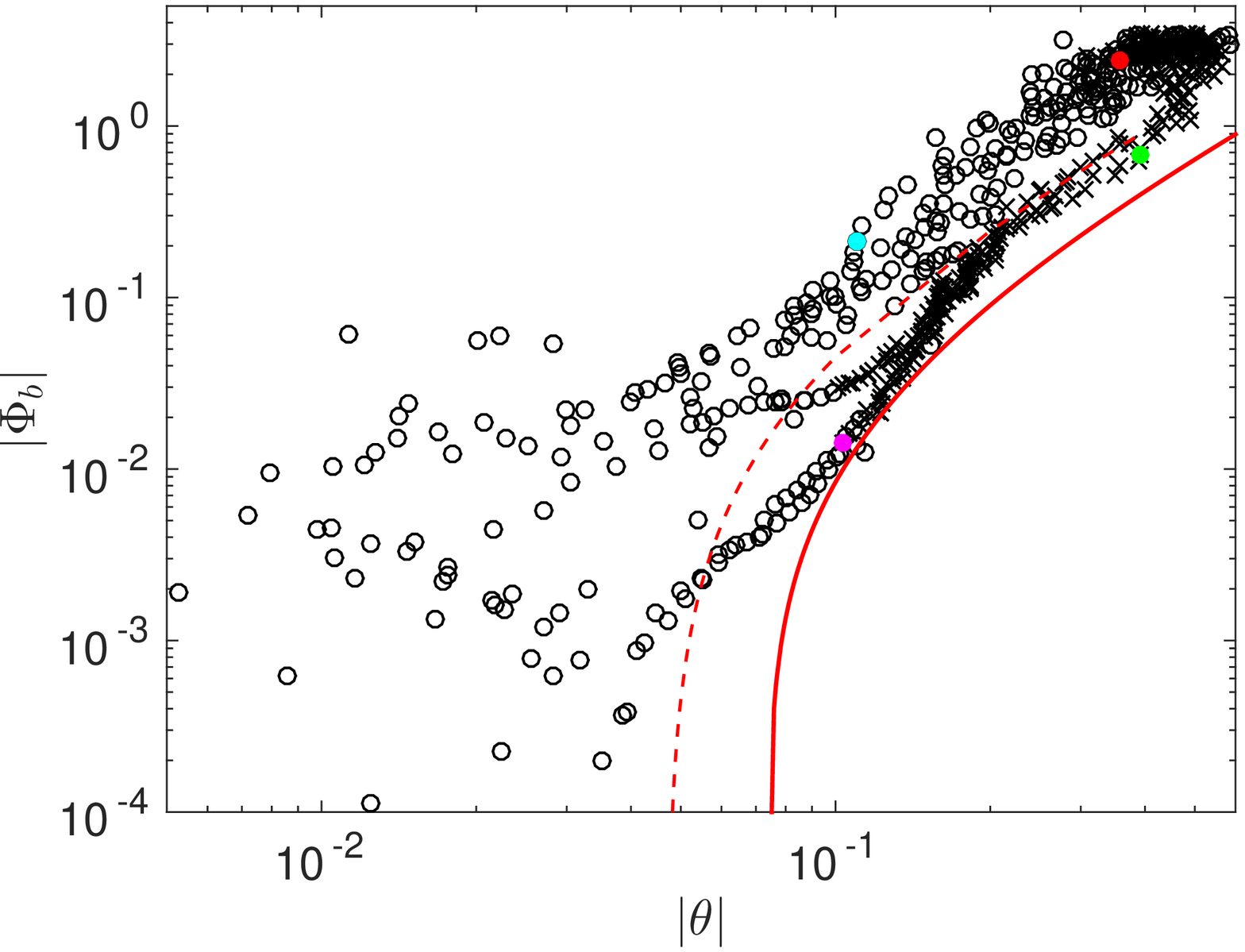}}
  \put(32,105){$(a)$}
  \put(210,105){$(b)$}
  \put(390,105){$(c)$}
\end{picture}
\caption{
\rev{}{%
Dimensionless sediment flow rate $\Phi_b$ plotted versus the Shields parameter $\shields$ for $\Rdel=1000$ and $\ds^*/\del^*$ equal to $(a)$ $0.670$, $(b)$ $0.335$, $(c)$ $0.168$. %
Results of the numerical simulations (crosses and circles refer to the accelerating and decelerating phases, respectively. %
The solid line indicates the values provided by equation~\eqref{Q_3} and the broken red line indicates the values provided by equation~\eqref{WP}. %
Coloured markers indicate values obtained at: %
$[${\protect\tikz \protect\draw[red,fill=red] (0,0) circle (.5ex);}$]$ $t=2\pi$; 
$[${\protect\tikz \protect\draw[cyan,fill=cyan] (0,0) circle (.5ex);}$]$ $t=2.25\pi$; 
$[${\protect\tikz \protect\draw[magenta,fill=magenta] (0,0) circle (.5ex);}$]$ $t=2.5\pi$ and 
$[${\protect\tikz \protect\draw[green,fill=green] (0,0) circle (.5ex);}$]$ $t=2.75\pi$. %
}
}
\label{fig2a}
\end{figure}
%
Figure~\ref{fig1a} a shows the values of $\Phi_b$ provided by the model (see relationship (\ref{Q_3})) and those provided by the numerical simulations for $d^*/\delta^*=0.335$ and $R_\delta=750$. %
As already pointed out, it appears that for rather small values of $\theta$, the sediment transport rate during the accelerating phases differs from that computed during the decelerating phases even if the value of $\theta$ is the same. %
\rev{}{%
In particular, two branches can be clearly distinguished in figure~\ref{fig1a} which tend to converge between the end of the accelerating phases and the early decelerating phases. %
}%
\rev{However}{In fact}, as soon as the Shields parameter and the sediment transport rate become significant, the values of $\Phi_b$ during the accelerating and decelerating phases are the same and \eqref{Q_3} provides reliable estimates of the sediment transport rate. %
The empirical formula \eqref{Q_3} provides a reasonable estimate of $\Phi_b$ also for larger values of $R_\delta$, as shown by figures~\ref{fig1a}b and \ref{fig2a}a,b, and smaller values of $\ds^*/\del^*$, as shown by figure~\ref{fig2a}c, even though the predicted values are slightly underestimated because a certain amount of particles start to be picked-up from the bottom and transported into suspension. %
As shown by figure \ref{figcasa4}a values of $\salt$ larger than those provided by \eqref{mu} are observed also for values of $\shields$ smaller than $\shields_{cr}$. %
The predictions obtained by means of \eqref{Q_3} largely deviate from the results of the numerical simulations when a significant number of particles is put into suspension, as shown by the results plotted in figures~\ref{figcasa4}b and \ref{fig2a}b where the value of $\Rdel$ is still equal to $1000$ but sediment particles smaller than those considered in figure~\ref{figcasa4}a are considered ($\ds^*/\del^*=0.168$). %

As expected, it can be concluded that the simplified approach which leads to \eqref{Q_3} can be used to obtain an estimate of the sediment transport rate only when the Shields parameter is not far from its critical value for the initiation of sediment motion and smaller than the threshold value $\shields_{cr,susp}$ above which the sediments are carried into suspension. %
Even for $\shields$ smaller than $\shields_{cr,susp}$, the values of $\Phi_b$ provided by \eqref{Q_3} should be used with caution because the average flow field is modified when a large number of particles move and make high jumps. Moreover, the uncertainty in the evaluation of $c_D$ and $c_L$  certainly affects the results. %

For example, the empirical predictor of \citet{wong2006}, which was considered in the earlier analysis by \citet{mazzuoli2019b} %
\begin{equation}
\label{WP}
	\Phi_b= 4.93 \left(\theta-0.047\right)^{1.6},
\end{equation}
provides values of $\Phi_b$ close to those obtained by means of \eqref{Q_3} for $R_\delta=750$ and $d^*/\delta^*=0.335$ (see figure \ref{fig1a}a) but it appears to provide better results for $R_\delta=1000$ and $d^*/\delta^*=0.67$ (see figure \ref{fig2a}a). %
This finding is not surprising  since the value of $p$ which is obtained from (\ref{p3}) implies that only the particles on the surface layer can be dragged by the flow and the accuracy of (\ref{Q_3}) decreases as soon as $\theta-\theta_{cr}$ increases and more layers of particles are set into motion by the oscillatory flow. %
In the \textit{supplementary material}, it is possible to see three videos showing the position of the sediment particles (shadowed by their velocity) during the oscillatory cycle. %
Video~$1$ shows simultaneously the visualisations for $\ds^*/\del^*=0.335$, $\Rdel=450,\,750,\,1000$, and for $\ds^*/\del^*=0.67$, in order to compare different modes of bedload sediment transport. %
In video~$2$ and video~$3$, which are obtained for $\ds^*/\del^*=0.168$, $\Rdel=1000$ and for $\ds^*/\del^*=0.335$, $\Rdel=1500$ respectively, a significant amount of particles are put into suspension by the turbulent vortices. %
In all the videos, when the bottom shear stress is close to its critical value, only a few particles roll and slide over the resting particles. %
When large values of the bottom shear stress are generated by the oscillatory flow, the largest particles start to make jumps but the interaction with the resting particles is still an element which controls sediment dynamics. %
On the other hand, for large values of the bottom shear stress, the smallest particles start to be trapped by the turbulent eddies and put into suspension and their interaction with the bottom is weak. %

\begin{figure}
\begin{picture}(0,120)(0,10)
  \put(-2,-10){\includegraphics[trim=0cm 0cm 0cm 0cm, clip, width=.32\textwidth]{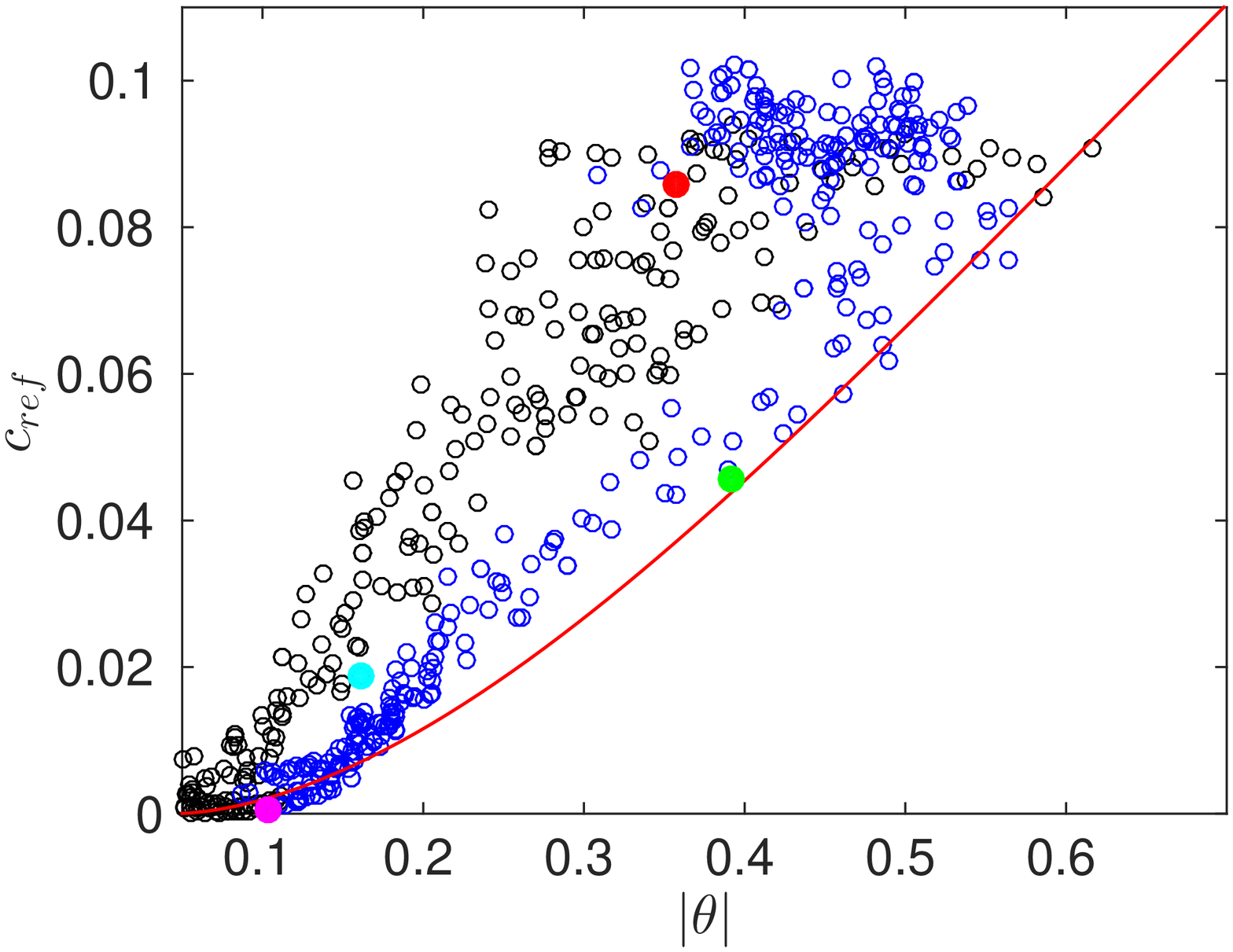}}
  \put(75,25){\vector(1,1){30}}
  \put(55,110){\vector(-1,-4){10}}
  \put(173,-5){\includegraphics[trim=0cm 0cm 0cm 0cm, clip, width=.33\textwidth]{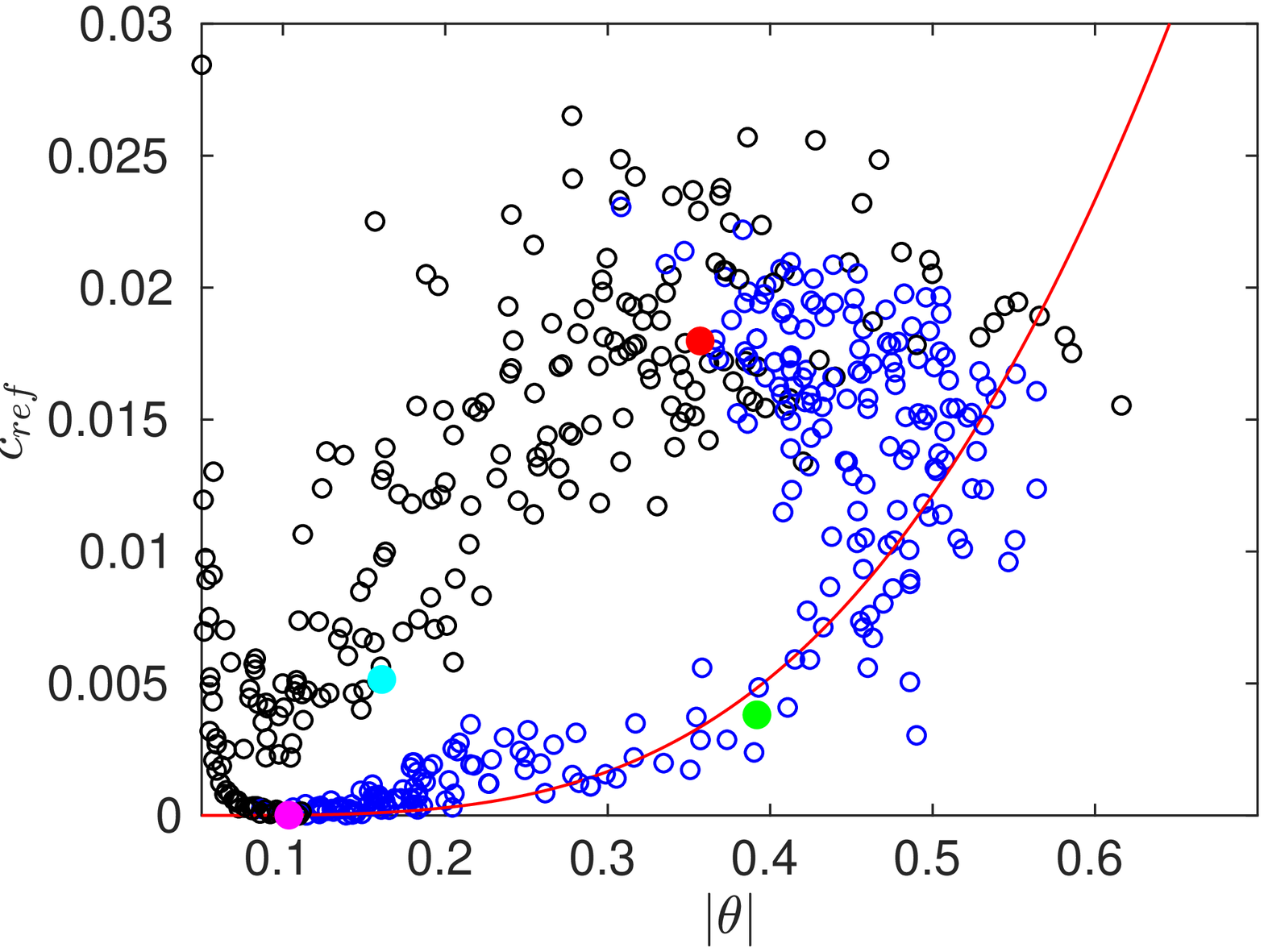}}
  \put(355,-5){\includegraphics[trim=0cm 0cm 0cm 0cm, clip, width=.33\textwidth]{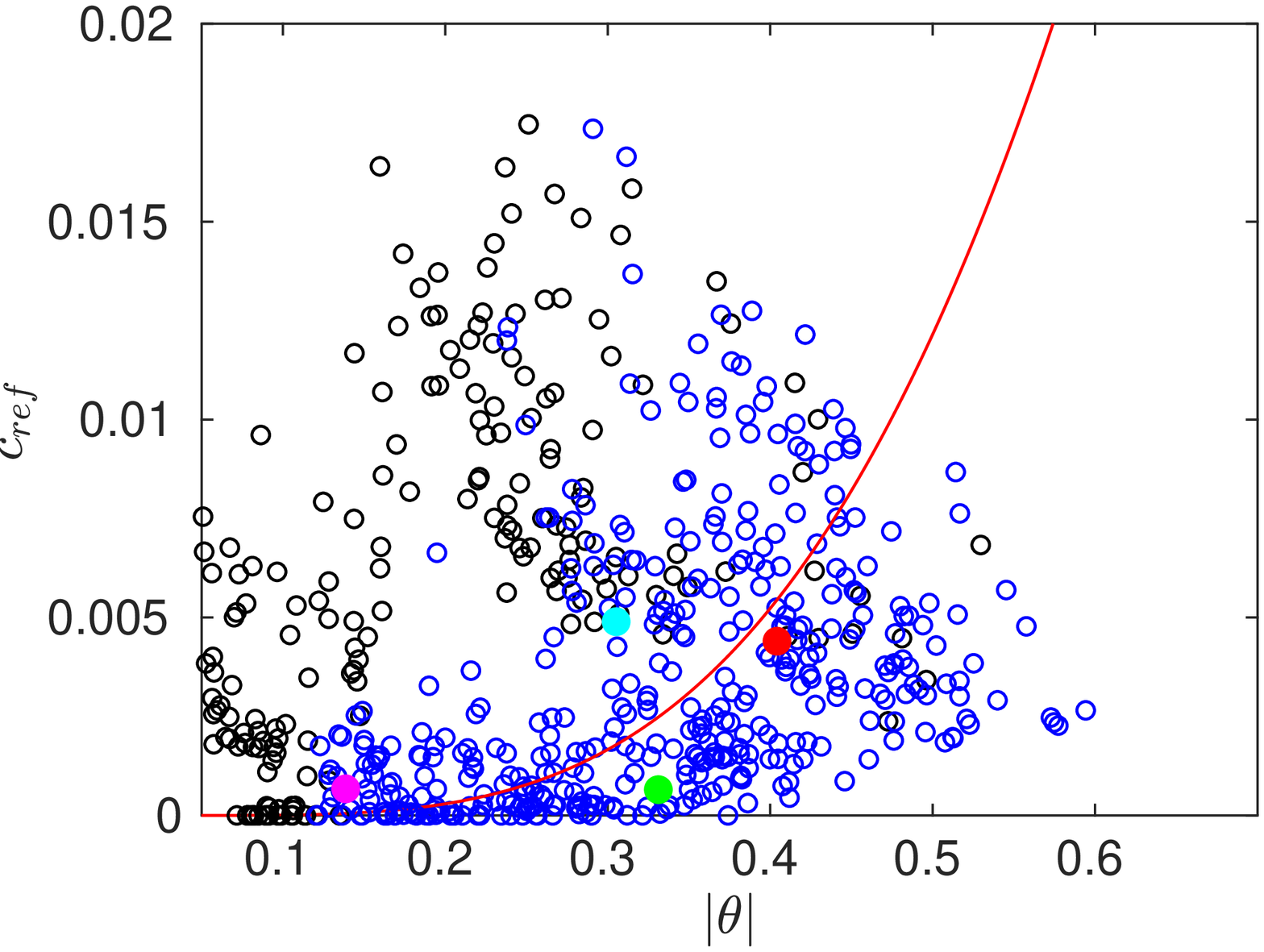}}
  \put(32,108){$(a)$}
  \put(205,108){$(b)$}
  \put(390,108){$(c)$}
\end{picture}
\caption{
\rev{}{%
Volumetric concentration of sediment provided by the DNS, at distances $(a)$ $2\,\ds^*$ above $x_{2,0}^*$ and $(b,c)$ $\salt^*$ above $\beds^*$, as a function of the Shields parameter for $(a,b)$ run~$3$ and $(c)$ run~$5$. %
Blue and black circles refer to accelerating and decelerating phases, respectively. %
The solid red line is the empirical value predicted by \eqref{c1} for $A=0.311$ and $\ell=1.75$ (panel~$a$) or for $A=0.2$ and $\ell=3.5$ (panels~$b,\,c$). %
The approximate critical value of the Shields parameter is equal to $0.05$. %
Arrows indicate the direction of the evolution during the wave-cycle. %
Coloured markers indicate values obtained at: %
$[${\protect\tikz \protect\draw[magenta,fill=magenta] (0,0) circle (.5ex);}$]$ $t=2.5\pi$ and 
$[${\protect\tikz \protect\draw[green,fill=green] (0,0) circle (.5ex);}$]$ $t=2.75\pi$; %
$[${\protect\tikz \protect\draw[red,fill=red] (0,0) circle (.5ex);}$]$ $t=3\pi$ and %
$[${\protect\tikz \protect\draw[cyan,fill=cyan] (0,0) circle (.5ex);}$]$ $t=3.25\pi$. %
}
}
\label{fig4a}
\end{figure}

\rev{}{%
For practical applications, when $\shields$ is large and in particular larger than a critical value $\shields_{cr,susp}$, the evaluation of the sediment transport rate requires the evaluation of the averaged volumetric sediment concentration $c$, hereafter simply referred to as ``concentration'', as a function of $\xf{2}^*$ and $t^*$. %
In other words, the numerical models, used to quantify the sediment transport rate, solve an advection-diffusion equation for $c(\xf{2}^*,t^*)$ where the sediment diffusivity is usually assumed to be proportional to the eddy viscosity and the eddy viscosity is evaluated by means of a turbulence model. %
First, the numerical solution of the momentum and continuity equations for the fluid is obtained at the time $t^*$ and the Shields parameter $\shields$ is evaluated. %
Then, the concentration is updated by imposing an appropriate bottom boundary condition. %
}%
\rev{%
For example, \citet{zyserman1994} proposed to fix $\xref^*$ equal to $2\,\ds^*$ above $x_{2,0}^*$ and to evaluate $c_{ref}$ by means of %
}{%
For example, \citet{zyserman1994} proposed to fix the concentration $c_{ref}$ %
\begin{equation}
\label{c1}
c_{ref}=\frac{A\left(\shields-0.045\right)^\ell}{1+0.72(\shields-0.045)^\ell}
\:\:,
\end{equation}
with $A=0.311$ and $\ell=1.75$, at the distance $2\,\ds^*$ from the bottom. %
}%
Figure~\ref{fig4a}a shows a comparison between the empirical relationship \eqref{c1} and the data provided by the present numerical simulations %
\rev{}{%
at the elevation $x_{2,ref}^*=x_{2,0}^*+2\,\ds^*$ %
}%
for $R_\delta$ is equal to $1000$ and $\ds^*/\del^*$ to $0.168$ (run~$3$), $x_{2,0}^*$ being the value of $\beds^*$ when the sediments do not move. %
It clearly appears that the values of $\cref$ provided by the present numerical simulations depend not only on $\shields$, since different values of $\cref$ are found for the same value of $\shields$ depending on the accelerating/decelerating phase of the cycle. %

Even though it is quite common to fix the reference distance from the bottom equal to $2 d^*$, %
\rev{%
\citep[see i.a.][]{zyserman1994} %
}{}%
it would be more appropriate to choose it equal to $\salt^*$ because the sediment grains, which move sliding, rolling and saltating in a bottom layer of thickness $\salt^*$ are usually assumed to contribute to the bed load. %
\rev{}{%
Figure~\ref{fig4a}b is similar to figure~\ref{fig4a}a, but it shows the volumetric concentration provided by the numerical simulation at $\beds^*+\salt^*$. %
The hysteresis observed in figure~\ref{fig4a}a is amplified, as $\salt^*$ is mostly larger than $2\,\ds^*$ and the values of $\cref$ obtained during the accelerating phases are significantly different from those obtained during the decelerating phases (see blue and black circles in figure~\ref{fig4a}b). %
}%
Indeed, as pointed out by \citet{fredsoe1992}, the boundary condition \eqref{c1} is questionable for an unsteady flow: when the bed shear stress rapidly increases, it is reasonable to assume that the reference concentration immediately adapts to the new conditions since the suspended sediments have to be picked-up from the bed and transported for a very small distance to reach the reference level which is of the order of the grain size. %
\rev{}{%
Therefore, %
during the phases characterised by increasing values of $\shields$ (i.e. for $d\vert\shields\vert/dt>0$), it is reasonable to use expression \eqref{c1} to estimate the concentration $c_{ref}$. %
Considering that $\salt^*$ significantly differs from $2\,\ds^*$, if $x_{2,ref}^*$ is assumed to be equal to $\salt^*$ above $\beds^*$, the coefficients appearing in the expression \eqref{c1} should be given different values. %
In fact, the expression \eqref{c1} fits the numerical results of run~$3$ for $A=0.2$ and $\ell=3.5$ (cf. figure~\ref{fig4a}b). %
For these values of the coefficients, expression \eqref{c1} adapts fairly to the results of run~$5$, where a significant amount of particles were observed to go into suspension (see figure~\ref{fig4a}c). %

When the bed shear stress decreases (i.e. $d\vert\shields\vert/dt<0$), the sediments in suspension cannot settle faster than the settling velocity and the concentration at the reference level cannot adapt instantaneously to the condition prescribed by the boundary condition. %
During these phases, \citet{fredsoe1992} suggest that the concentration at the reference level is approximately equal to the concentration $c(\xf{2}^*+w_s^*\Delta t^*,t^*-\Delta t^*)$, where $w_s^*$ denotes the settling velocity of sediment particles and $\Delta t^*$ is an arbitrary time interval such that $\Delta t^*\ll \omega^{*-1}$. %
\begin{figure}
\begin{picture}(0,120)(0,0)
  \put(-2,-10){\includegraphics[trim=0cm 0cm 0cm 0cm, clip, width=.32\textwidth]{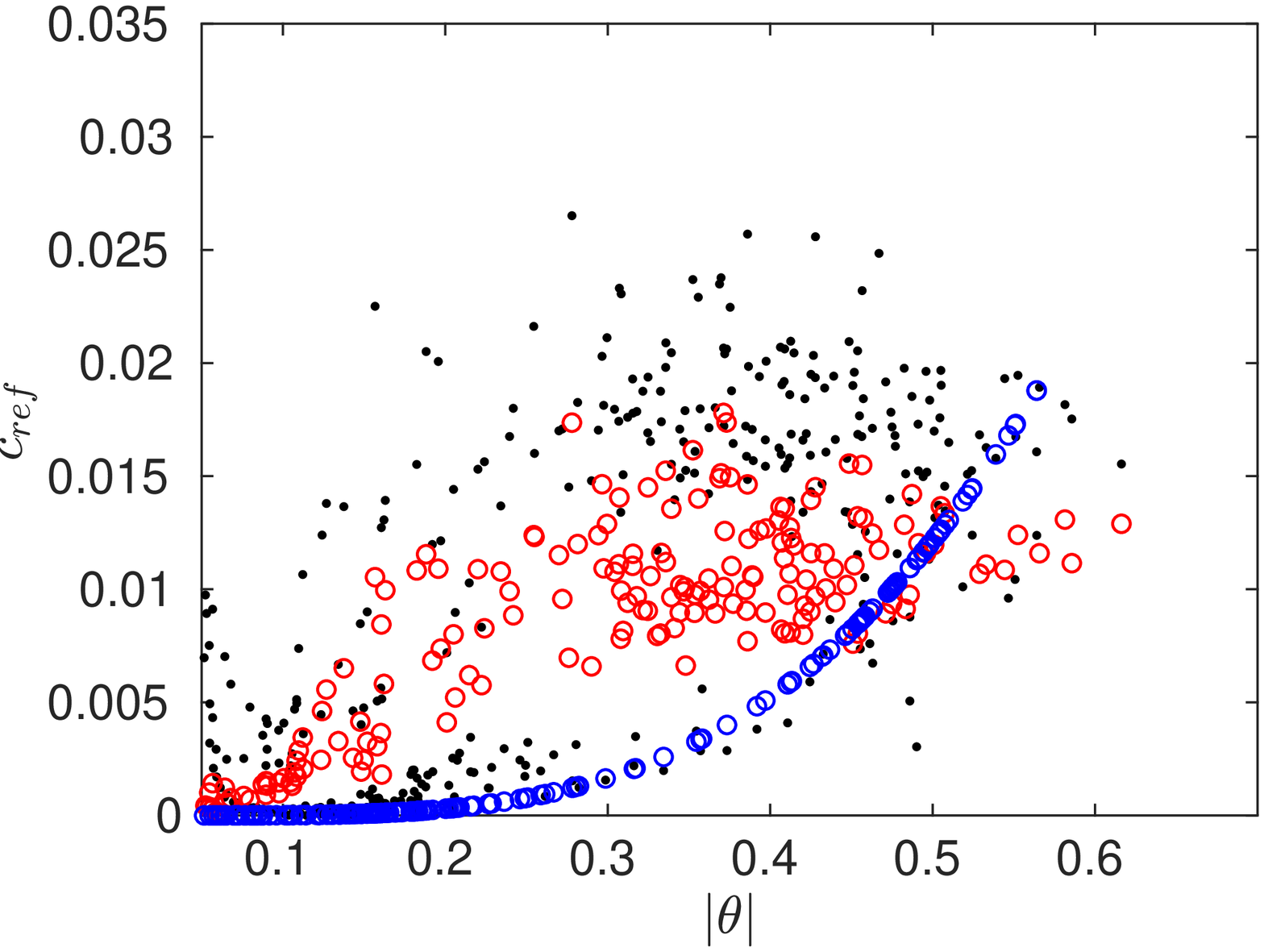}}
  \put(175,-10){\includegraphics[trim=0cm 0cm 0cm 0cm, clip, width=.32\textwidth]{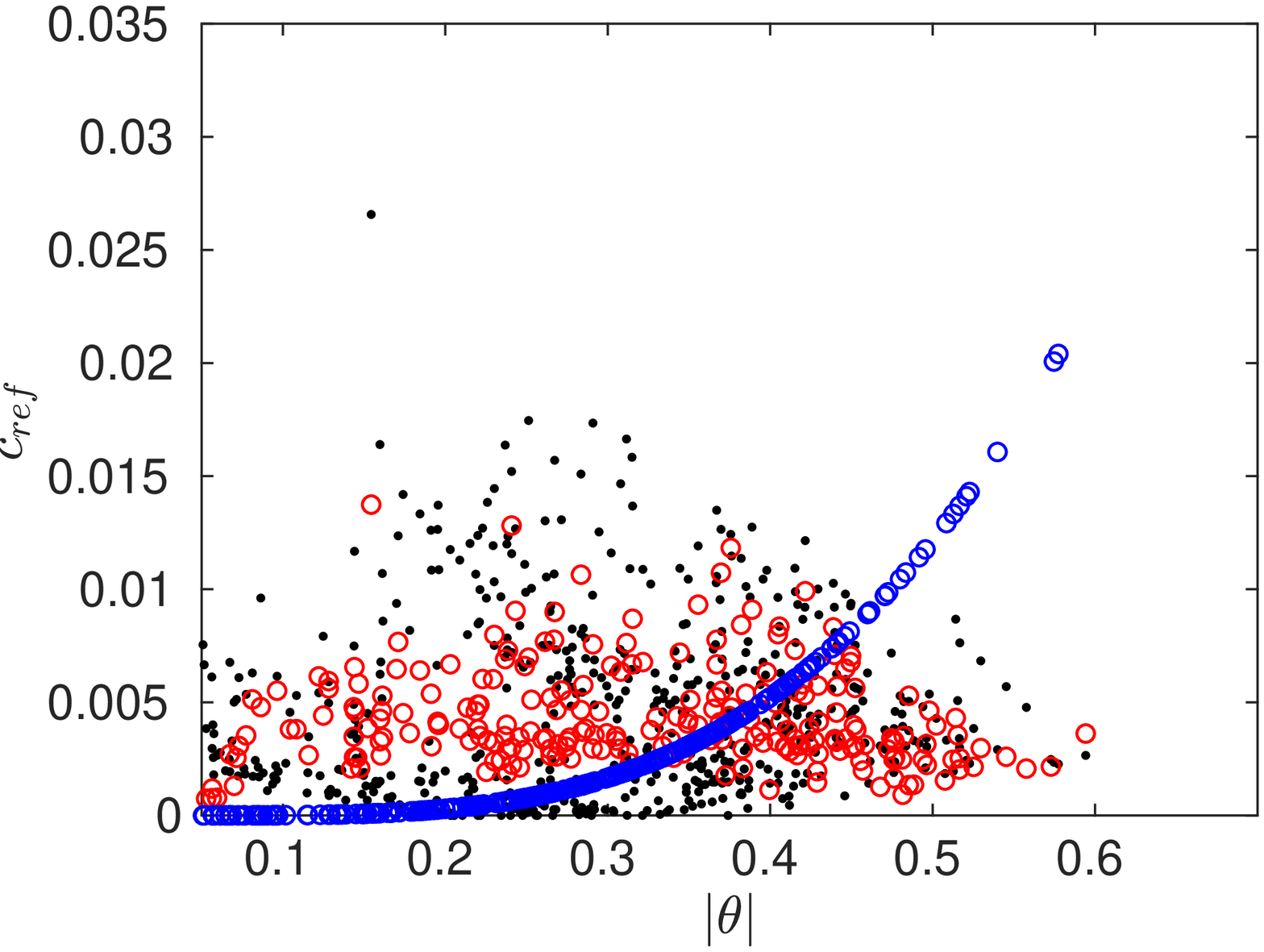}}
  \put(355,-10){\includegraphics[trim=0cm 0cm 0cm 0cm, clip, width=.33\textwidth]{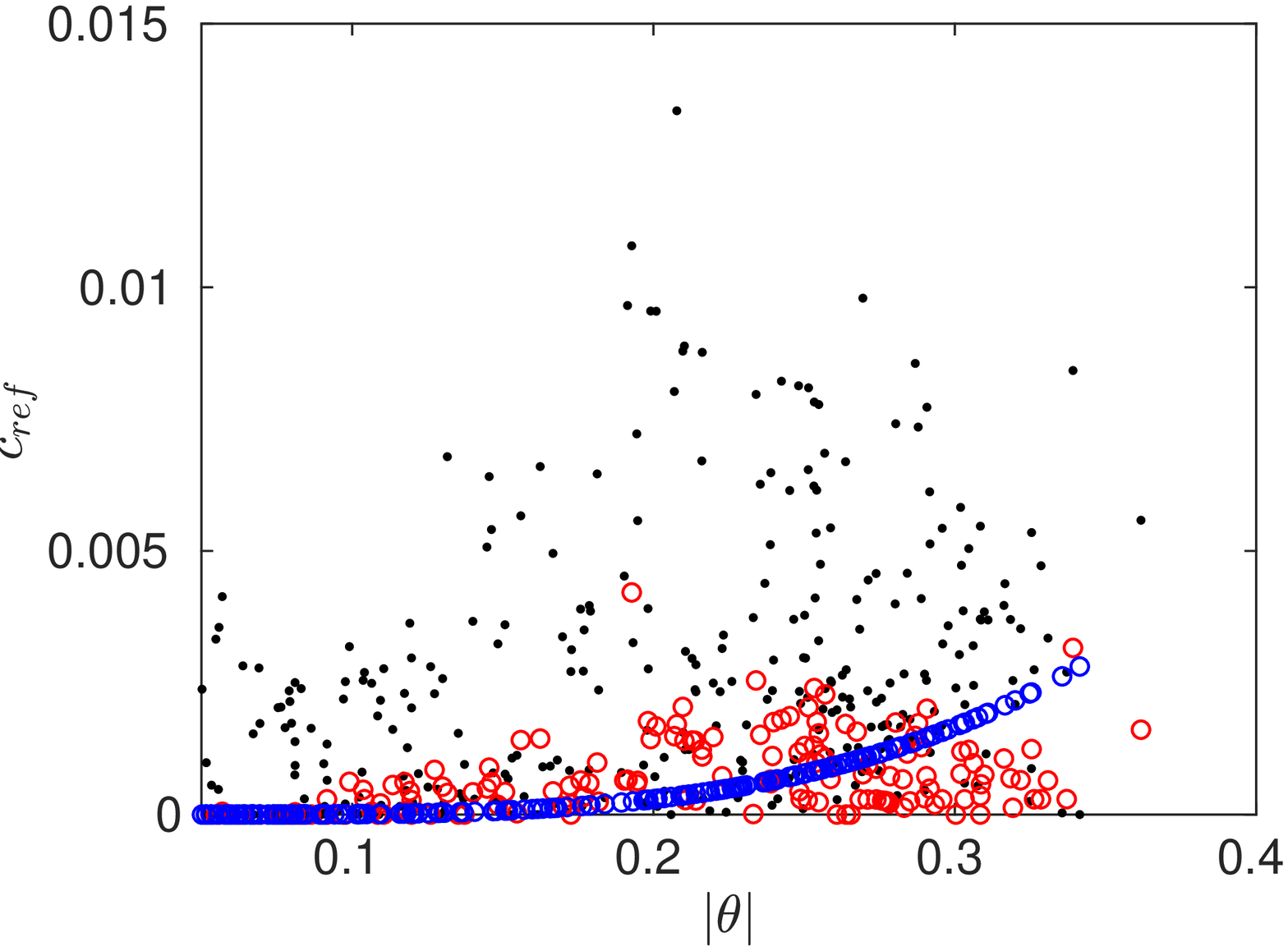}}
  \put(30,102){$(a)$}
  \put(205,102){$(b)$}
  \put(390,102){$(c)$}
\end{picture}
\caption{%
\rev{}{%
Volumetric concentration of sediment as function of the absolute value of the Shields parameter. %
Black points indicate the values obtained 
by means of the direct numerical simulations. %
The open circle indicate the values obtained following \citet{fredsoe1992}'s approach, i.e.
when $d\vert\shields\vert/dt>0$ the value of $\cref$ is computed by means of equation~\eqref{c1} (blue circles) %
whereas, when $d\vert\shields\vert/dt<0$, $\cref$ is assumed equal to the concentration 
at the elevation $\xf{2}^*=\beds^*+\salt^*(t^*-\Delta t^*)+w_s^*\Delta t^*$ at time $t^*-\Delta t^*$
(red circles). %
Panels~$(a)$, $(b)$ and $(c)$ refer to runs~$3$, $5$ and $2$, respectively. %
}
}%
\label{fig14f}
\end{figure}
Figure~\ref{fig14f} shows the volumetric sediment concentration $c$ at $x_{2,ref}^*=\beds^*+\salt^*$ (black dots), obtained from the results of runs~$3$, $5$ and $2$, as a function of $\vert\shields\vert$. %
Blue circles indicate the values of $\cref$ evaluated by means of \eqref{c1} during the phases characterized by growing values of $\vert\shields\vert$, with $A=0.2$ and $\ell=3.5$. %
Red circles are the values obtained following \citet{fredsoe1992}'s approach, i.e. $\cref=c(\beds^*+\salt^*+w_s^*\Delta t^*,t^*-\Delta t^*)$, with $w_s^*$ equal to the fall velocity of a single particle in still water. %
The agreement between the numerical results and the empirical boundary condition at $x_{2,ref}^*$ could be improved taking into account that the settling velocity in a turbulent flow differs from the fall velocity of a single particle in still water and is significantly smaller. %
This decrease of the fall velocity is due to the ``loitering effect'': a particle settling in a turbulent flow spends more time (loitering) in those regions of the flow field characterized by an upward velocity component than in the regions characterized by a downward velocity component \citep{nielsen1992}. %
Empirical formulas typical of coastal applications, that practically consist of multiplying $w_s^*$ by a constant smaller than $1$, allow for evaluation of the loitering effect. %
In principle, the settling velocity is also affected by the sediment concentration, but for the values of $c_{ref}$ the effect of particle interactions can be neglected. %
Figure~\ref{sketch}a shows that the time-averaged vertical particle velocity divided by its value $w_s^*$ far from the bottom and plotted against the time-averaged volumetric concentration $\overline{c}$ normalised by the concentration $c_0$ of the resting bed significantly deviates from $w_s^*$ for values of concentration larger than $0.1\,c_0$ (i.e. approximately for $\overline{c}>0.05$) which is well above the maximum value of $\cref$ (see figure~\ref{fig14f}). %
%
\begin{figure}
\begin{picture}(0,210)(0,10)
  \put(0,0){\includegraphics[trim=0cm 0cm 0cm 0cm, clip, width=.43\textwidth]{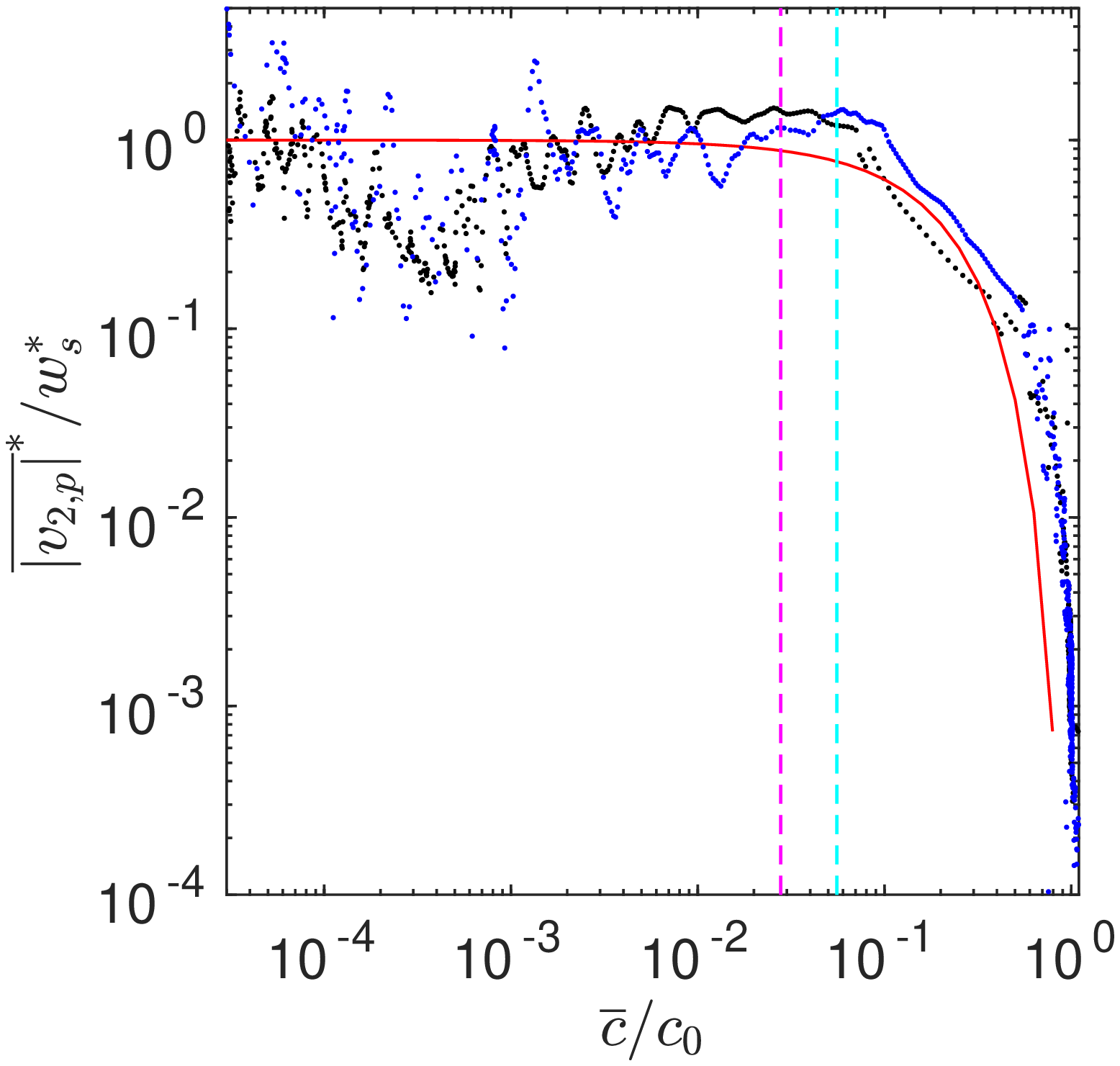}}
  \put(255,0){\includegraphics[trim=0cm 0cm 0cm 0cm, clip, width=.53\textwidth]{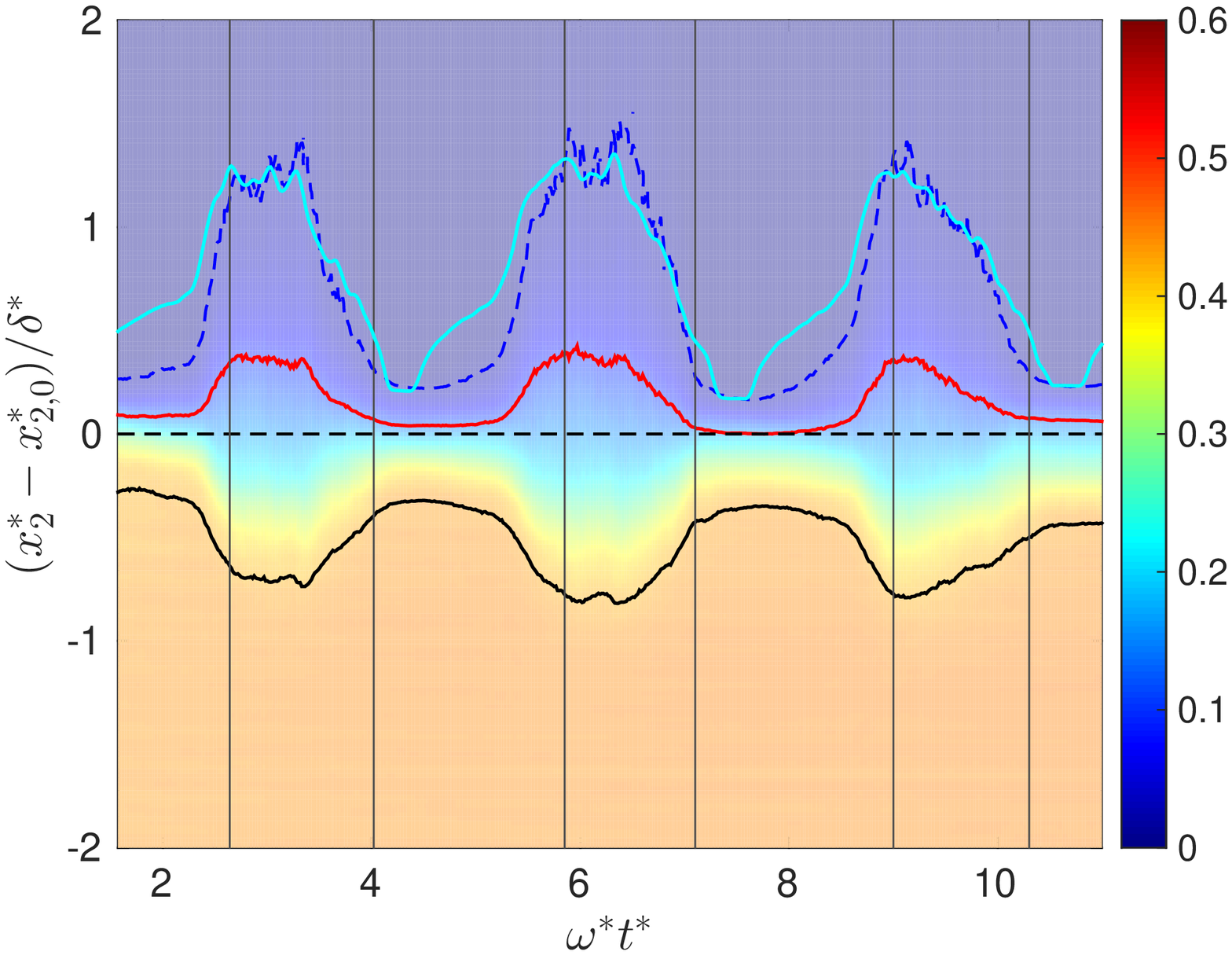}}
  \put(10,200){$(a)$}
  \put(250,200){$(b)$}
  \put(283,70){$\rbed$}
  \put(282,132){$\beds$}
  \put(282,192){$x_{2,ref}$}
  \put(316,150){$\salt$}
  \put(314,138){\vector(0,1){39}}
  \put(314,177){\vector(0,-1){39}}
  \put(296,77){\line(1,1){12}}
  \put(294,188){\line(1,-4){6}}
  \put(315,110){$x_{2,0}$}
  \put(318,28){\rotatebox{90}{\small$\frac{d\vert\shields\vert}{dt}<0$}}
  \put(392,28){\rotatebox{90}{\small$\frac{d\vert\shields\vert}{dt}<0$}}
  \put(467,28){\rotatebox{90}{\small$\frac{d\vert\shields\vert}{dt}<0$}}
\end{picture}
\caption{%
\rev{}{%
Panel~$(a)$ shows the time-averaged vertical velocity of particles, normalised by the value attained far from the bottom, as a function of the volumetric concentration of sediment, normalised by the value attained within the resting bed, for the runs~$3$ (black dots) and $5$ (blue dots). %
The red line indicates the value predicted by the formula proposed by \citet{richardson1954}. %
The broken lines indicate the maximum value of $\cref$ observed for run~$3$ (magenta) and run~$5$ (cyan). %
Panel~$(b)$ shows the time development of the bottom surface $\beds$ (red line), the resting bed elevation $\rbed$ (black line) and the top of the saltation layer which is denoted by $x_{2,ref}$ (cyan line), referred to the minimum bottom elevation $x_{2,0}$ (black broken line) for run~$3$. %
Wall-normal coordinates are non-dimensionalised by $\del^*$. %
Background colours are shaded according to the volumetric concentration of sediment particles. %
The contour line at $c=0.016$ (blue broken line) almost coincides with $x_{2,ref}$ during the phases characterised by decreasing values of $\frac{d\vert\shields\vert}{dt}$. %
}
}%
\label{sketch}
\end{figure}
%
\rev{}{%
The properties of the particle motion, like the particle velocity, which are associated with individual particles, are transferred to the computational (Eulerian) grid by using the procedure described by \citet{uhlmann2008}. %
}%

Besides the difficulties of estimating $w_s^*$, the position of the bottom is supposed to be fixed and known, whereas it is typically not. %
\rev{%
For the sake of clarity, figure~\ref{sketch}b shows the evolution of the bed elevation $\beds^*$ (red line) and of the reference elevation $\xref^*$ (cyan line) above $\beds^*$ for the run~$3$, the origin of the wall-normal coordinates being set at the constant level $x_{2,0}$ equal the minimum bed elevation $\min(\beds^*)$ (see the broken horizontal line in figure~\ref{sketch}). %
It is worth to remind that presently $\beds^*$ is defined as the maximum elevation at which $c=0.1$ and is therefore time dependent, but the elaborations that follow are independent of the definition of the bottom surface (which in most of practical cases is assumed constant) as long as the bottom shear stress is evaluated at $\beds^*$. %
}{%
For the sake of clarity, figure~\ref{sketch}b shows the evolution of the bed elevation $\beds^*$ (red line) and of the reference elevation $x_{2,ref}^*$ (cyan line) for the run~$3$, the origin of the wall-normal coordinates being set at the constant level $x_{2,0}$ equal the minimum bottom elevation $\min(\beds^*)$ (see the broken horizontal line in figure~\ref{sketch}b). %
It should be noted that the bottom surface $\beds^*$ is defined as the maximum elevation at which $c=0.1$ and that the bottom shear stress is evaluated at $\beds^*$. %
}%
When sediment particles deposit, mainly during the decelerating phases, the resting bed elevation increases and the bottom surface elevation decreases because of the compaction of the sediments. %
The numerical results show that the bottom  surface elevation $\beds^*$ is related to the \emph{deposition rate}, namely the dimensionless negative wall-normal component of the mean particle velocity, $v_{2,pd}^*/\vs^*$, $\vs^*$ denoting the characteristic particle velocity equal to $\sqrt{(s-1)\g^*\ds^*}$. %
In Appendix~\ref{appen2}, the following approximation $\beda^*$ of the distance of the bottom elevation $\beds^*$ above the level $x_{2,0}^*$ at the time $t^*$ is obtained %
\begin{equation}
\beda^* 
=
x_{2,0}^* 
+ 
\ds^*\,
\widehat{C}\,
k_D\left(\shields\right) 
\:\:,
\label{eqapx4}
\end{equation}
where $\widehat{C}=\frac{0.44}{0.1 + c_0}\left(\frac{\del^*}{\ds^*}\right)^{1.5}$ and $k_D=\frac{k_D^*}{\vs^*}$ is the dimensionless \emph{deposition constant} computed by means of~\citet{papavergos1984}'s formula (see \eqref{mccoy} in Appendix~\ref{appen2}). %
It should be noted that the expression~\eqref{eqapx4} was obtained on the basis of the present DNS data and could not be reliable for values of the parameters significantly different from those presently investigated. %

Therefore, the reference elevation $x_{2,ref}^*$ at which the volumetric concentration should be enforced equal to $\cref$ can be evaluated by %
\begin{equation}
x_{2,ref}^*(\shields) 
= 
\beda^*(\shields) 
+ 
\salt^*(\shields)
=
x_{2,0}^* 
+ 
d^*\left[1 + \widehat{{\cal C}}\,k_D\left(\shields\right) + A_b\left(\dfrac{\shields-\shields_{cs}}{\shields_{cs}}\right)^m\right]
\:\:.
\label{eqxref}
\end{equation}
Figure~\ref{fig14d} shows good agreement between the volumetric concentration evaluated at $\xf{2}^*=\beds^*+\salt^*$ (black dots) and that evaluated at $\xf{2}^*=\beda^* + \salt^*$. %
\begin{figure}[t]
\begin{picture}(0,130)(0,0)
  \put(0,-10){\includegraphics[trim=0cm 0cm 0cm 0cm, clip, width=.33\textwidth]{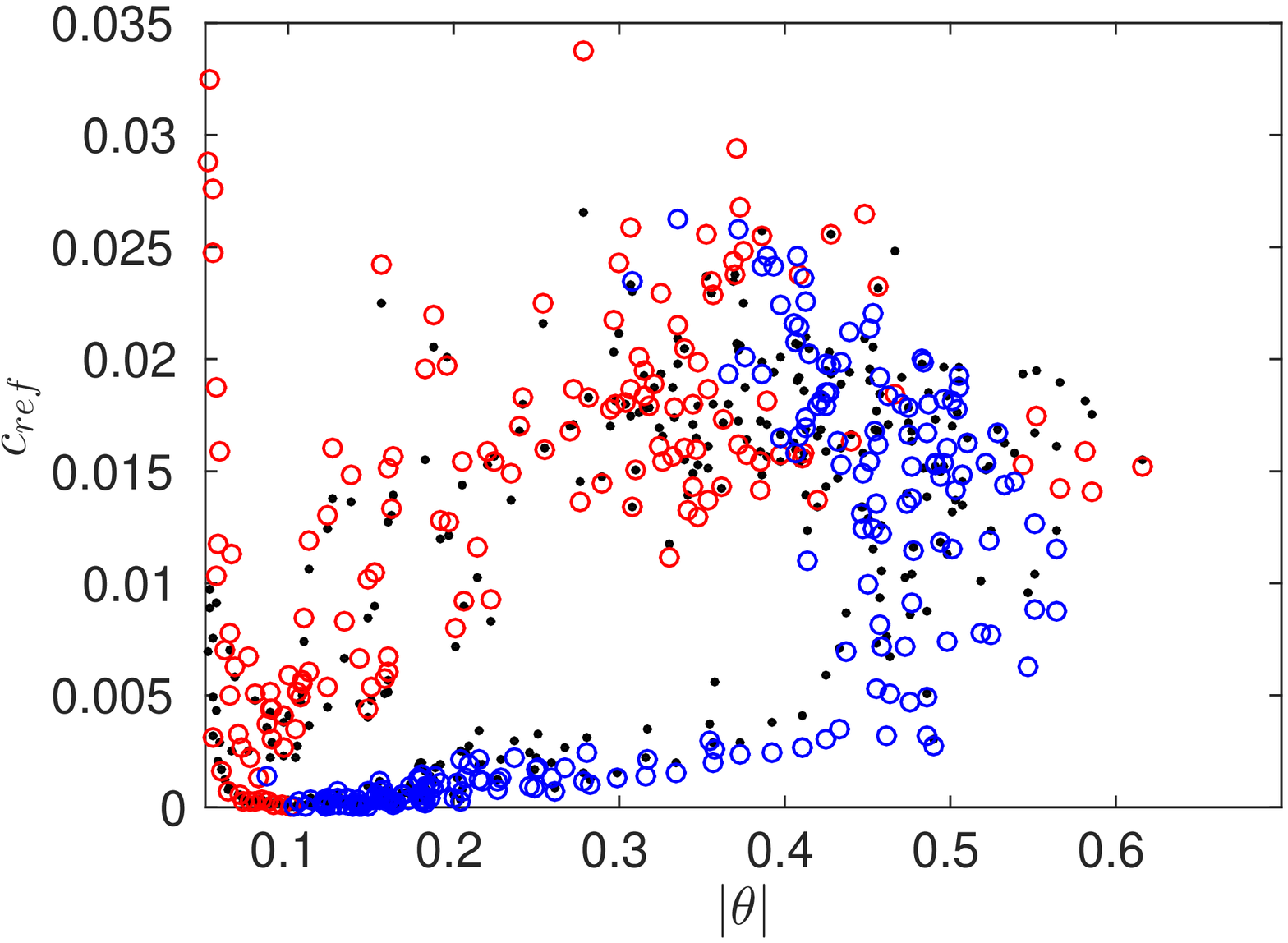}}
  \put(180,-10){\includegraphics[trim=0cm 0cm 0cm 0cm, clip, width=.32\textwidth]{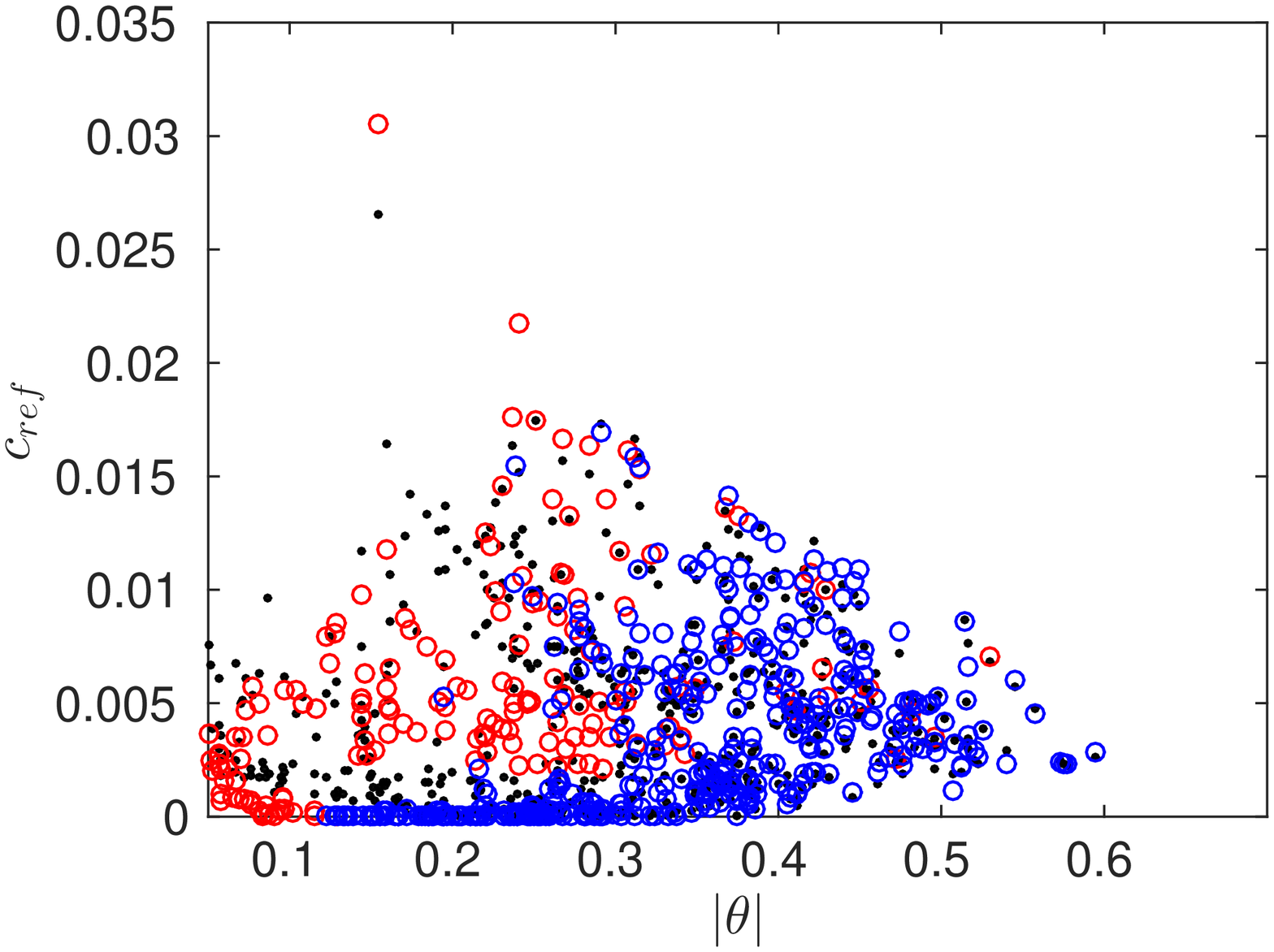}}
  \put(360,-10){\includegraphics[trim=0cm 0cm 0cm 0cm, clip, width=.33\textwidth]{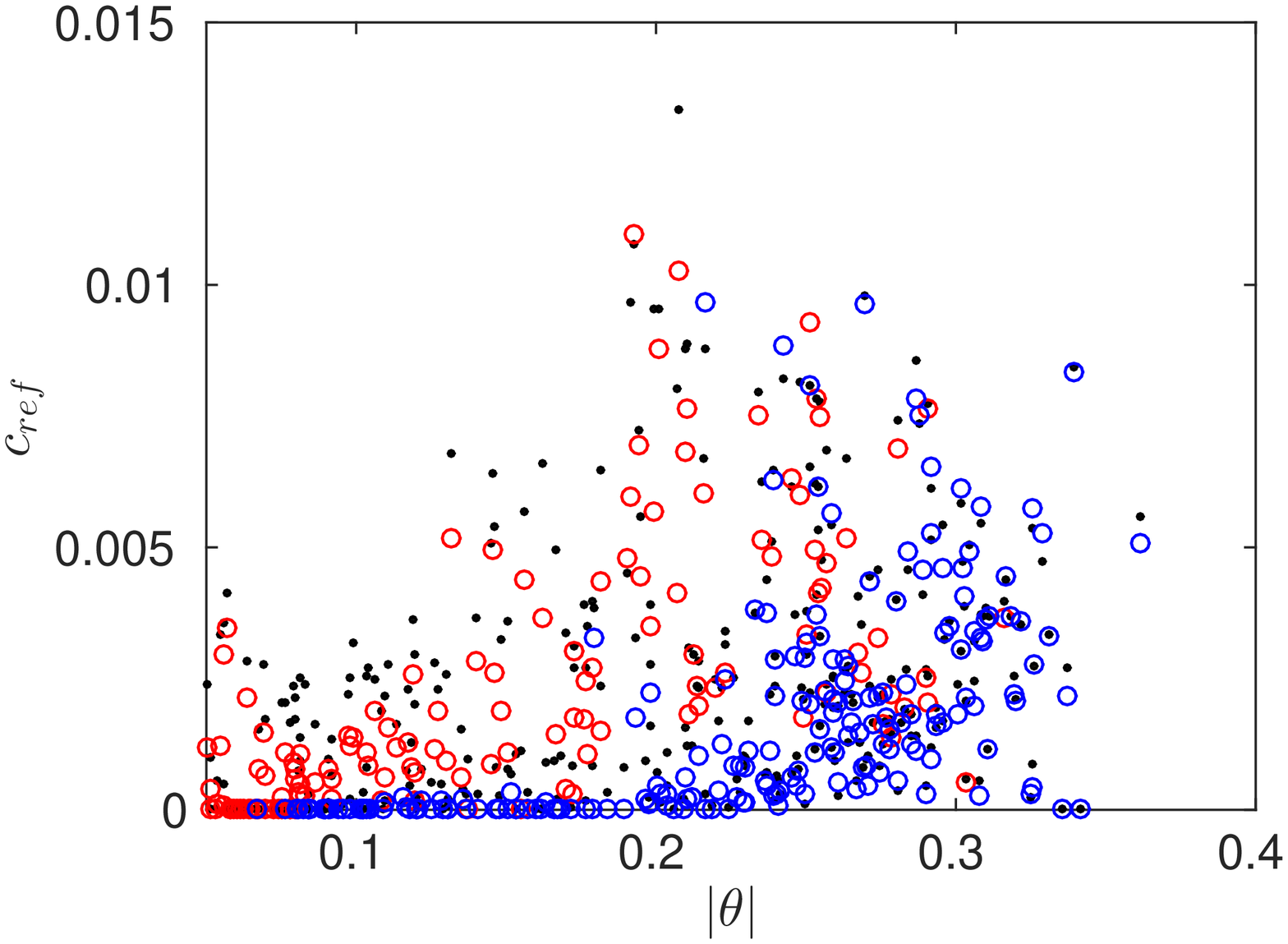}}
  \put(36,105){$(a)$}
  \put(216,105){$(b)$}
  \put(396,105){$(c)$}
\end{picture}
\caption{%
\rev{}{%
Volumetric concentration of sediment evaluated at $\xf{2}^*=\beds^*+\salt^*$ as a function of the absolute value of the Shields parameter. %
Black dots are the same as in figure~\ref{fig14f}, while empty circles indicate the values evaluated at $\beda^*+\salt^*$ (see \eqref{mu}, \eqref{eqxref} and \eqref{eqapx4}). %
Blue and red circles refer to the accelerating and decelerating phases, respectively. %
Panels~$(a)$, $(b)$ and $(c)$ refer to runs~$3$, $5$ and $2$, respectively. %
}
}%
\label{fig14d}
\end{figure}
\begin{figure}[t]
\begin{picture}(0,130)(0,0)
  \put(0,-10){\includegraphics[trim=0cm 0cm 0cm 0cm, clip, width=.33\textwidth]{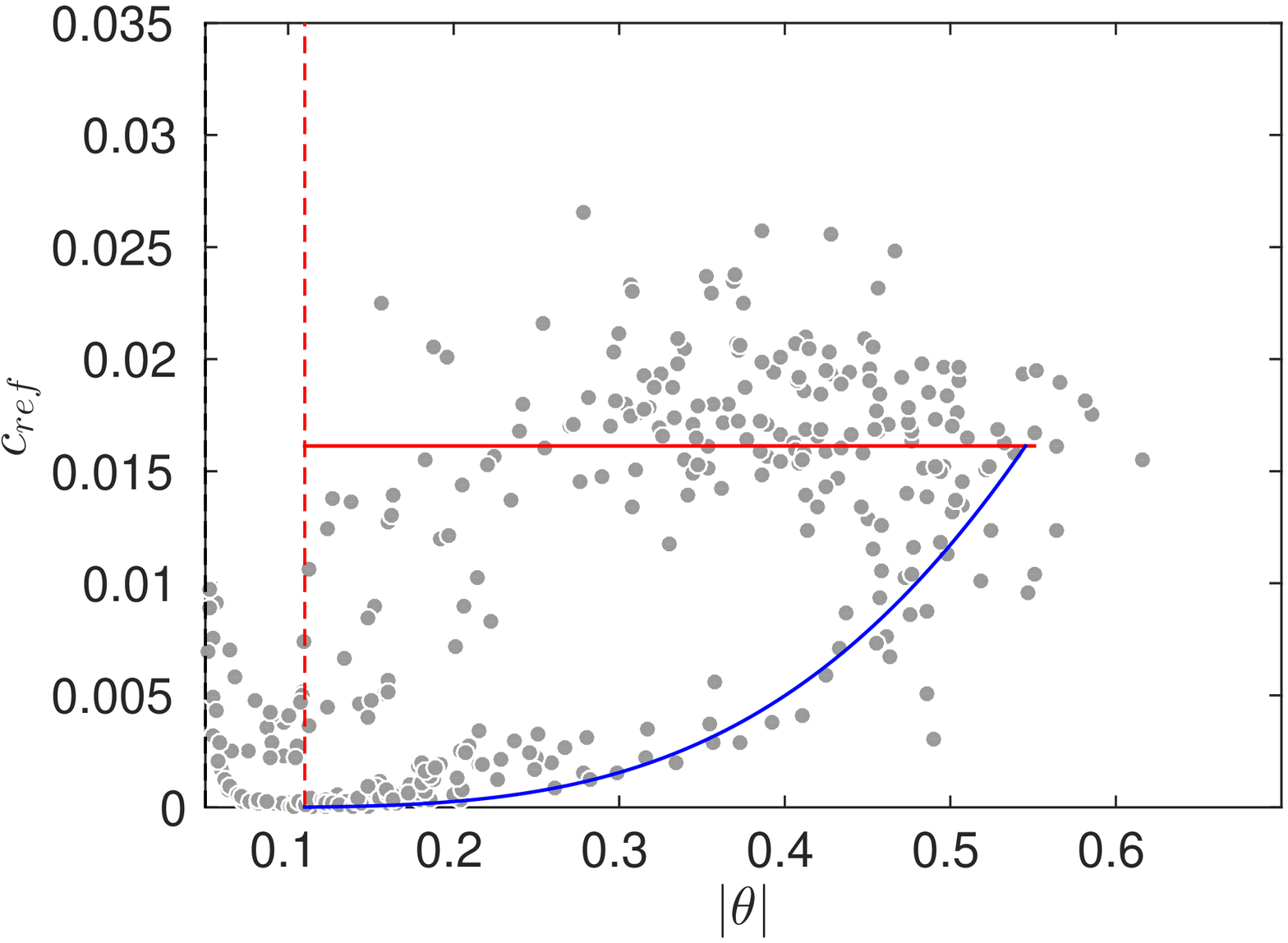}}
  \put(180,-10){\includegraphics[trim=0cm 0cm 0cm 0cm, clip, width=.33\textwidth]{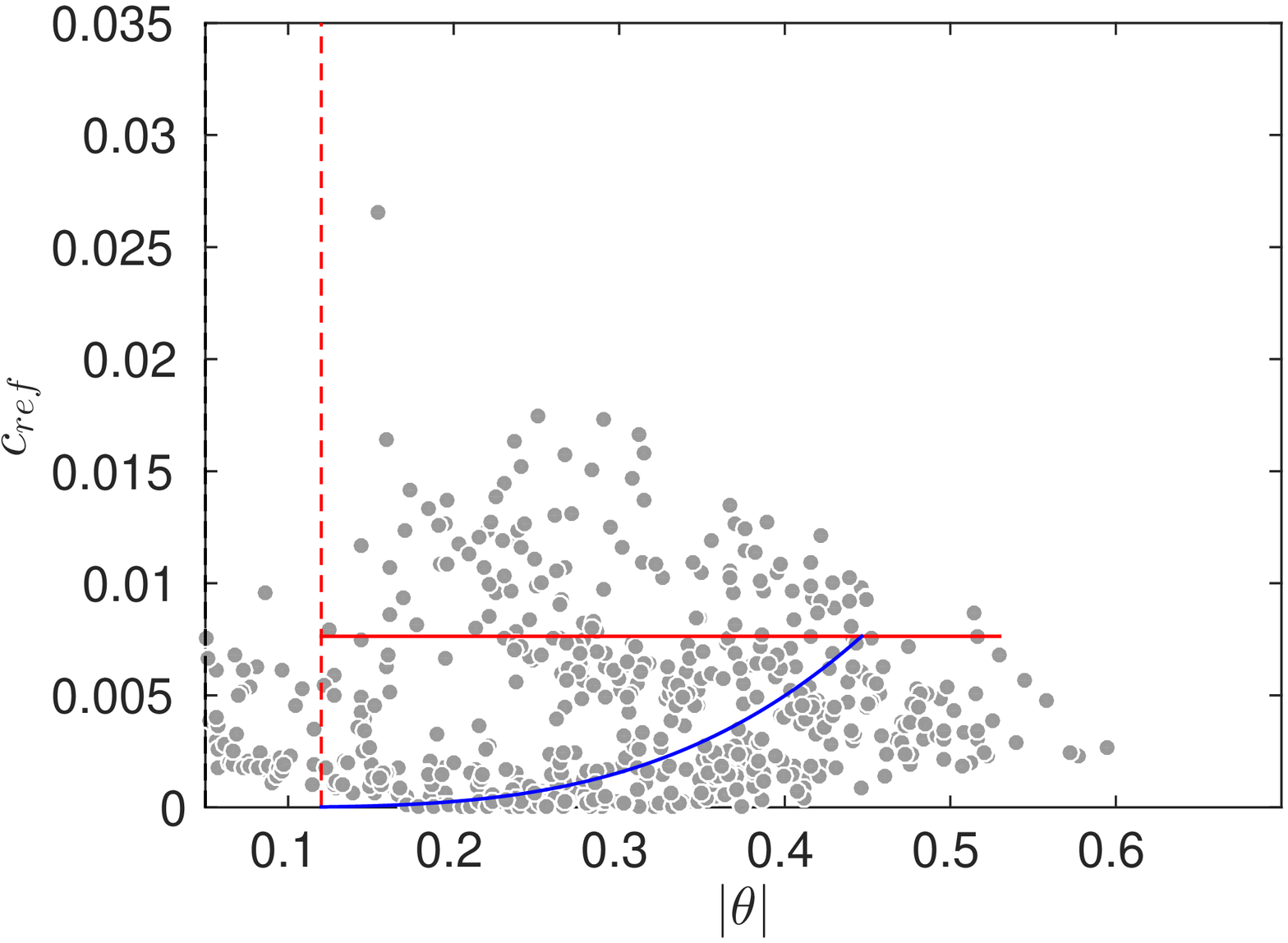}}
  \put(360,-10){\includegraphics[trim=0cm 0cm 0cm 0cm, clip, width=.33\textwidth]{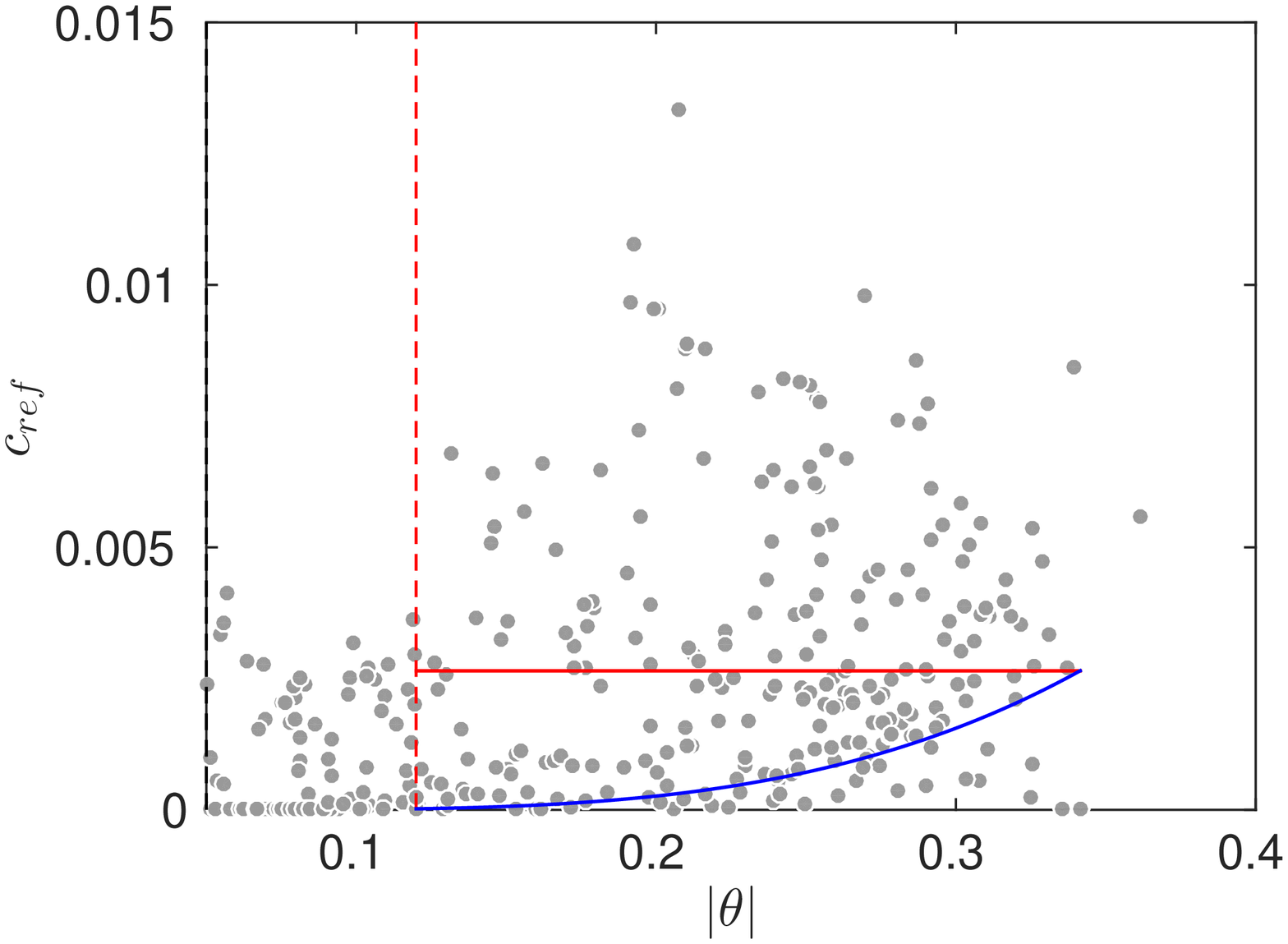}}
  \put(36,105){$(a)$}
  \put(216,105){$(b)$}
  \put(396,105){$(c)$}
\end{picture}
\caption{%
\rev{}{%
Volumetric concentration of sediment evaluated at $\xf{2}^*=\beds^*+\salt^*$ as a function of the absolute value of the Shields parameter. Gray dots indicate the values obtained by DNS, while the solid lines are the values of $\cref$ computed by means of \eqref{c1} and following the procedure described in the text. %
Blue and red lines refer to the phases characterised by $d\vert\shields\vert/dt>0$ (until $d\vert\shields\vert/dt$ becomes negative for the first time during the late acceleration phases) and $d\vert\shields\vert/dt<0$, respectively. %
The red broken line indicate the critical value $\shields_{cr,susp}$ of $\shields$ for sediment suspension which turns out $0.11$ for run~$3$ and to $0.12$ for runs~$2$ and $5$ \citep{rijn1984b}. %
Panels~$(a)$, $(b)$ and $(c)$ refer to runs~$3$, $5$ and $2$, respectively. %
}
}%
\label{fig14e}
\end{figure}

Finally, one is left with the problem of estimating the reference concentration $\cref$ to be applied at the reference elevation $x_{2,ref}^*$. %
During the phases characterised by increasing values of the Shields parameter ($d\vert\theta\vert/dt>0$), $\cref$ can be obtained by means of \eqref{c1} with $A=0.2$ and $\ell=3.5$. %
During the phases characterised by decreasing values of the Shields parameter ($d\vert\theta\vert/dt<0$), $\cref$ can be assumed constant and equal to the value provided by \eqref{c1} when $d\vert\theta\vert/dt$ vanishes (i.e. when $\theta$ decreases for the first time at the end of the accelerating phases). %
In fact, the reference elevation $x_{2,ref}^*$ follows the vertical excursion of the upper bound of the saltating layer where the concentration can be assumed constant (for example see the blue broken line in figure~\ref{sketch}b). %
For values of $\shields$ smaller than the threshold value $\shields_{cr,susp}$ for the suspension, the amount of suspended sediments can be considered negligible and set equal to zero. %
Figure~\ref{fig14e} shows a fair agreement between the values of $\cref$ computed with the aforementioned procedure and those obtained from runs~$2$, $3$ and $5$. %
The value of $\shields_{cr,susp}$ (red broken line in figure~\ref{fig14e}) was computed with the criterion proposed by \citet{rijn1984b}. %
}

\section{Conclusions}

The dynamics of sediment particles in the oscillatory boundary layer generated close to the sea bottom by propagating surface waves is evaluated by means of direct numerical simulations. %
The interaction of sediment grains with the turbulent vortex structures is explicitly computed using the immersed boundary approach. %
Values of the parameters typical of the shoaling region are considered. %
The results show that some of the empirical formulas used to quantify the bed load and the suspended load can be tuned to provide a more accurate evaluation of the sediment transport rate. %
\rev{%
For example, %
}{%
The present investigation focuses in particular on the sediment transport rate due to bed load, namely the transport of sediments in the region above the resting bed surface and below $x_{2,ref}^*$. %
The flux of sediment in the region closer to the bed in the engineering applications is related to the value of the Shields parameter. %
For small values of the Shields parameter, when the sediment transport rate is quite small, the inertia of sediment grains, the imposed pressure gradient and the turbulent vortex structures play a role too. %
The bedload sediment transport rate observed for negative values of $d\vert\shields\vert/dt$ is larger than that observed for positive values of $d\vert\shields\vert/dt$. %
However, the differences are significant only when $\shields$ is quite small and $\Phi_b$ assumes negligible values. %
Hence, the differences can be safely neglected. %
The maximum excursion of the resting bed and thickness of the layer between the resting bed and the bottom surface (defined as the surface where the volumetric concentration is equal to $0.1$) are found to be in line with the experimental results and the predictions that can be obtained with empirical formulae. %
Above the bottom surface, saltating particles characterise a layer the thickness of which depends on the Shields parameter and can be fairly predicted by using an empirical formula proposed for steady flows. %
The upper boundary of the saltation layer is assumed to be the reference level $x_{2,ref}^*$ separating the bed load from the suspended load. %
}%
\rev{%
The fall velocity should evaluated taking into account the presence %
\rev{%
and t%
}{%
of turbulent eddies and of other settling particles that cause a significant reduction of the fall velocity. %
Indeed, at high sediment concentration, the settling velocity reduces to a small fraction of the clear water value. %
The numerical values of the hindered settling velocity cannot be quantitatively compared with the relationship proposed by \citet{richardson1954} because in our simulations the hindering effects are superimposed to turbulent effects and, to our knowledge, there is no way to separate the two effects. %
On the other hand t%
}%
he bottom boundary condition for sediment concentration should consider the acceleration/deceleration of the forcing flow. %
}{}%
\rev{}{%
A practical approach is suggested to evaluate the reference concentration at $x_{2,ref}^*$ to improve the prediction of the suspended load. %
In particular, an expression is proposed to predict the excursion of the bottom elevation during the wave cycle, which allows us to evaluate $x_{2,ref}^*$ with a fair accuracy. %
The volumetric sediment concentration at $x_{2,ref}^*$ increases during the phases characterised by increasing values of the Shields parameter and can be predicted using an empirical formula. %
When the Shields parameter decreases, the concentration at $x_{2,ref}$ remains approximately constant since the suspended sediments predominantly settle. %
}%

\rev{%
However, the numerical results should be supported 
}{%
The predictor of the bottom surface excursion was developed only on the basis of the present DNS data and might be different when values of the parameters far from those presently simulated are considered. %
Therefore, a systematic exploration of the parameter space carried out by laboratory measurements and further direct numerical simulations would be beneficial. %
\rev{%
extend the validity of the procedure. %
Finally, it would be useful to provide an accurate description of the motion of sediments both in the bedload region and in suspension, a challenge that could be tackled more easily by numerical means. %
}{}
}%

\vspace{.3cm}

\paragraph*{Acknowledgements}\ %
\rev{}{The authors are grateful to Prof. Markus Uhlmann for the code that was used to make the present simulations.} %
This study has been partially supported by Ministero dell'Istruzione dell'Universit\`a e della Ricerca - MIUR (under grant FUNBREAK-PRIN2017 no.~20172B7MY9) and by the Office of Naval Research (U.S.A.) (under grant no.~12292911). %
The Authors also acknowledge the support of CINECA, which provided computational resources on Marconi under the PRACE project MOST SEA (Proposal ID: 2017174199) and the ISCRA (class B) project MOSTSEAP (application no. HP10BEHEPJ). J.S. and J.C. were supported under base funding to the U.S. Naval Research Laboratory from the Office of Naval Research. %

\appendix
\section{Appendix}
\label{appen1}
\begin{figure}
\begin{picture}(0,180)(0,0)
  \put(0,0){\includegraphics[trim=0cm 0cm 0cm 0cm, clip, width=.49\textwidth]{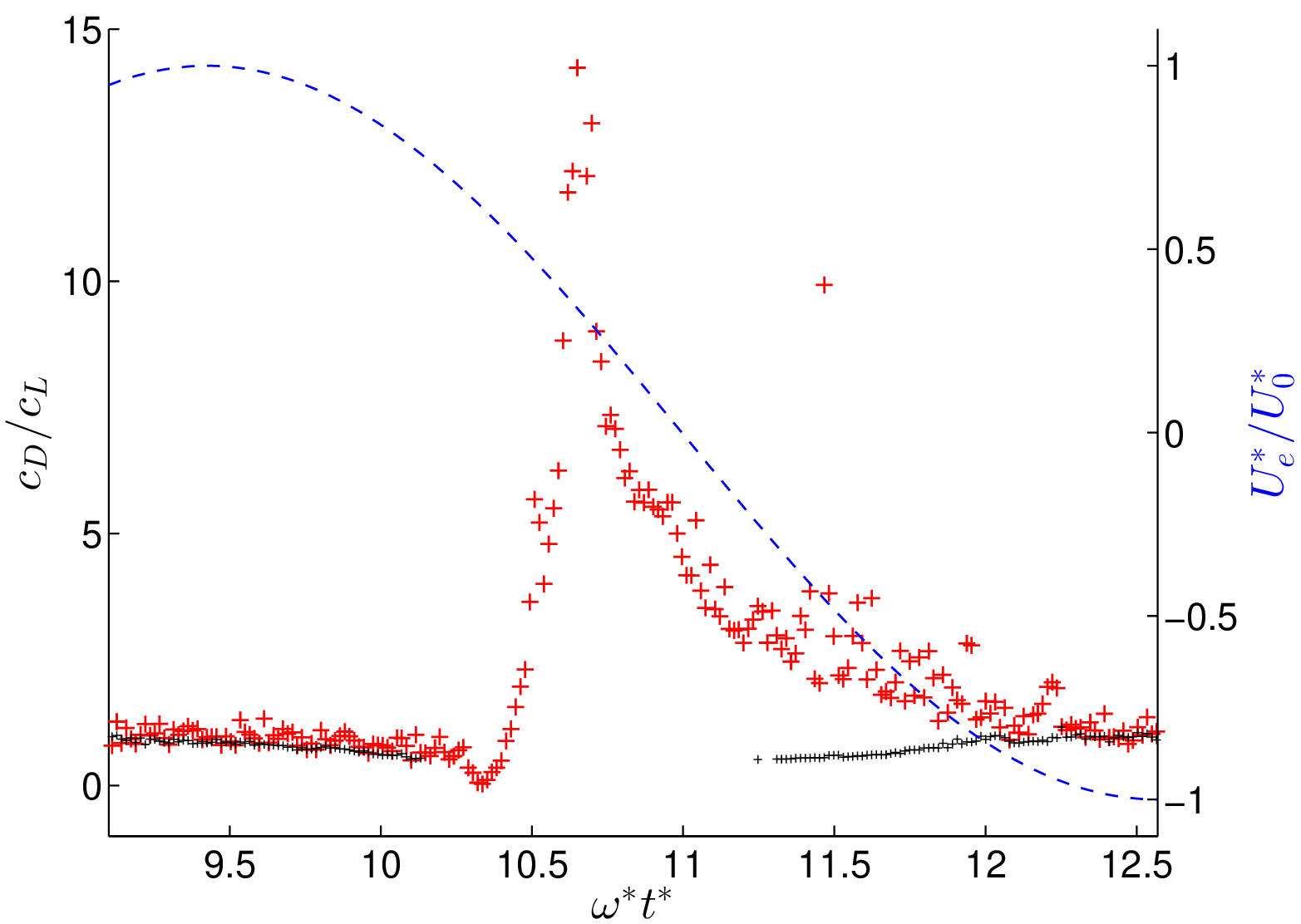}}
  \put(270,0){\includegraphics[trim=0cm 0cm 0cm 0cm, clip, width=.49\textwidth]{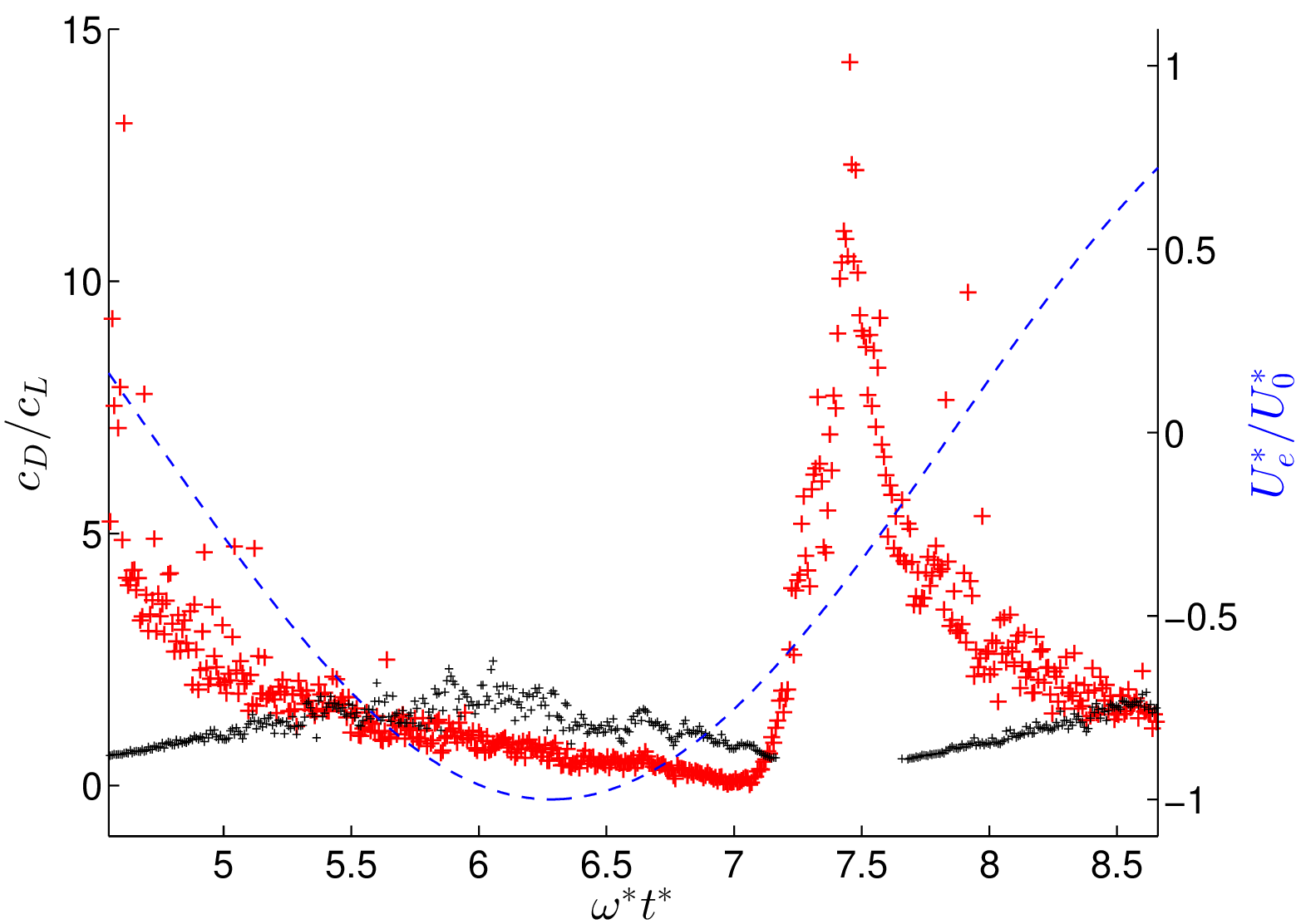}}
  \put(-2,160){$(a)$}
  \put(268,160){$(b)$}
\end{picture}
\caption{%
Time development of the values of $c_D/c_L$ obtained with the present model (black crosses) and of $F_D/F_L$ (which is equal to $c_D/c_L$) obtained with the DNS (red crosses) for $\ds=0.335$ and $(a)$ $\Rdel=775$, $(b)$ $\Rdel=1500$. %
The broken blue lines indicate the streamwise velocity far from the bottom. %
}
\label{figapp1}
\end{figure}

Equations~\eqref{drag}, \eqref{resi} and \eqref{v_p} show that the modelled mean drag force acting on moving particles is equal to \[F_D^*=\dfrac{\mu_d W_{sph}^*}{1+\mu_d\dfrac{c_L}{c_D}}\] and the lift force is equal to \[F_L^*=\dfrac{\mu_d W_{sph}^*}{\dfrac{c_D}{c_L} + \mu_d}\:,\] where $W_{sph}^*=\densf^*(s-1)\g^*\frac{\pi\ds^{*3}}{6}$ denotes the immersed weight of an individual sediment particle. %
Therefore, the ratio $c_D/c_L$, which depends on the Reynolds number $Re$ (defined by~\eqref{Re}), is the only parameter which affects the fluid-particle interaction and control sediment dynamics. %
Figure~\ref{figapp1} shows the values of the ratio $c_D/c_L$ predicted by the present model for $\ds=0.335, \Rdel=750$ and $\ds=0.335, \Rdel=1500$ along with the values computed by the DNS of \citet{mazzuoli2019b}. %
The results plotted in figure~\ref{figapp1} show that the simplified approach provides reasonable values of $c_D/c_L$ but for the phases of the cycle close to flow inversion, i.e. when the sediments do not move and the sediment transport rate vanishes. %
\rev{}{%
When the sediment particles saltate, taking into account that the number of particles which do long and high jumps is significantly smaller than that of the particles doing short and low jumps, it is reasonable to consider values of $L^*$ significantly smaller than $0.5~\salt^*$. %
}%
\rev{%
As pointed out in the text, such %
}{%
The %
}%
good agreement between the model predictions and the results of the DNS is obtained by fixing $L^*=0.08~\salt^*$, which is approximately constant and equal to $0.3~\ds^*$ during the phases where sediment particles are saltating for all the cases presently considered. %

\rev{}{%
\section{Appendix}
\label{appen2}

Figure~\ref{sketch}b shows the time-development of the three interfaces that are conventionally identified between the clear water and the resting sediments: $(i)$ the resting bottom surface, above which the sediment volumetric concentration is smaller than the mean concentration of the underlying bed, $(ii)$ the bottom surface where the bottom shear stress is evaluated and the momentum exchange between the bulk flow and the sediment is maximum (this surface is identified by the maximum elevation where the concentration is equal to $0.1$) and $(iii)$ the top of the saltating layer, which delimits the bed load layer. %
In the absence of bedforms, as in the present cases, these interfaces can be identified with horizontal planes. %
Let us focus on the layer bounded below by the resting bed elevation $\rbed^*$ and above by the bottom surface elevation $\beds^*$. %
During an oscillation, the sediments flow through both the upper and lower interfaces. %
The positive and negative sediment fluxes through the resting bed elevation are both smaller than those through the bottom surface elevation. %
At the bottom surface elevation, we refer to the positive and negative vertical fluxes of sediments as erosion and deposition. %
In particular, since the volumetric concentration at $\beds^*$ is equal to $0.1$ by definition, the \emph{erosion rate} and the \emph{deposition rate} are identified with mean values of the positive and negative contributions of the vertical particle velocity $\vpy^*$, that are denoted by $v_{2,pe}^*$ and $v_{2,pd}^*$, respectively. %
The deposition and erosion rates at $\beds^*$ are interrelated by the continuity equation for the sediment %
\begin{equation}
v_{2,p}^* 
= 
\dfrac{c_d v_{2,pd}^*+c_e v_{2,pe}^*}{0.1}\:\:,
\label{cpart}
\end{equation}
where $c_d$ and $c_e$ are the concentrations of the particles directed downwards and upwards, respectively, with $c_d+c_e=0.1$ at $\beds^*$. %
By assuming that $c_d$ and $c_e$ increases for increasing values of $v_{2,pd}^*$ and $v_{2,pe}^*$, respectively, namely $c_d=-0.1\,v_{2,pd}^*/(v_{2,pe}^*-v_{2,pd}^*)$ and $c_e=0.1\,v_{2,pe}^*/(v_{2,pe}^*-v_{2,pd}^*)$, equation~\eqref{cpart} gives %
\begin{equation}
v_{2,p}^* 
= 
v_{2,pd}^* + v_{2,pe}^*\:\:.
\label{cpart1}
\end{equation}
At the equilibrium, e.g. in steady conditions, $\beds^*$ is constant and $v_{2,pd}^* = -v_{2,pe}^*$. %
However, in an oscillatory flow, 
the sum of deposition and erosion rates does not cancel out. %
It is reasonable to assume that the deposition and erosion rates are somehow correlated because the turbulent vortices simultaneously affect the deposition and erosion rates. %

The balance of the sediment in the region between the resting bed and bottom elevations can be expressed in terms of volumetric concentration %
\begin{equation}
\underbrace{\dfrac{\partial}{\partial t^*}\displaystyle\int_{\rbed^*}^{\beds^*}\svf\, d\xf{2}^*}_{(I)} 
= 
\underbrace{\svf(\beds^*)\dfrac{d\beds^*}{dt^*}}_{(II)} 
- 
\underbrace{\svf(\rbed^*)\dfrac{d\rbed^*}{dt^*}}_{(III)} 
+ 
\underbrace{\svf(\rbed^*)v_{2,p}^*(\rbed^*)}_{(IV)} 
- 
\underbrace{\svf(\beds^*)v_{2,p}^*(\beds^*)}_{(V)} 
\:\:,
\label{exn1}
\end{equation}
i.e. the positive time-variation of the sediments  (term $I$) is equal to the velocity of expansion of the boundaries (terms $II$ and $III$) plus the inward flux of sediments (terms $IV$ and $V$) \citep[see also][for a detailed description of the generalised Exner equation]{paola2005}. %

As aforementioned, according to the present DNS data, the contribution of the term $V$ is smaller than the terms $I-III$ while $IV$ is negligible. %
Figure~\ref{sketch}b shows that the layer between the resting bed surface and the bottom surface experiences expansions/contractions, when the flow is accelerating/decelerating, that are almost symmetric with respect to the mid horizontal plane. %
It follows that $\frac{d\rbed^*}{dt^*}=-\frac{d\beds^*}{dt^*}$ and equation~\eqref{exn1} becomes %
\begin{equation}
\underbrace{\dfrac{\partial}{\partial t^*}\displaystyle\int_{-\beds^*}^{\beds^*}\svf\, d\xf{2}^*}_{(I)} 
+ 
\underbrace{\svf(\beds^*)v_{2,p}^*(\beds^*)}_{(V)} 
\simeq 
\underbrace{(0.1+c_0)\dfrac{d\beds^*}{dt^*}}_{(II+III)} 
\:\:,
\label{exn2}
\end{equation}
where the values of the volumetric concentration at the resting bed and bottom elevations were replaced by $\svf(\beds^*)=0.1$ and $\svf(\rbed^*)=c_0$, respectively. %
\begin{figure}
\begin{picture}(0,180)(0,0)
  \put(100,-10){\includegraphics[trim=0cm 0cm 0cm 0cm, clip, width=.5\textwidth]{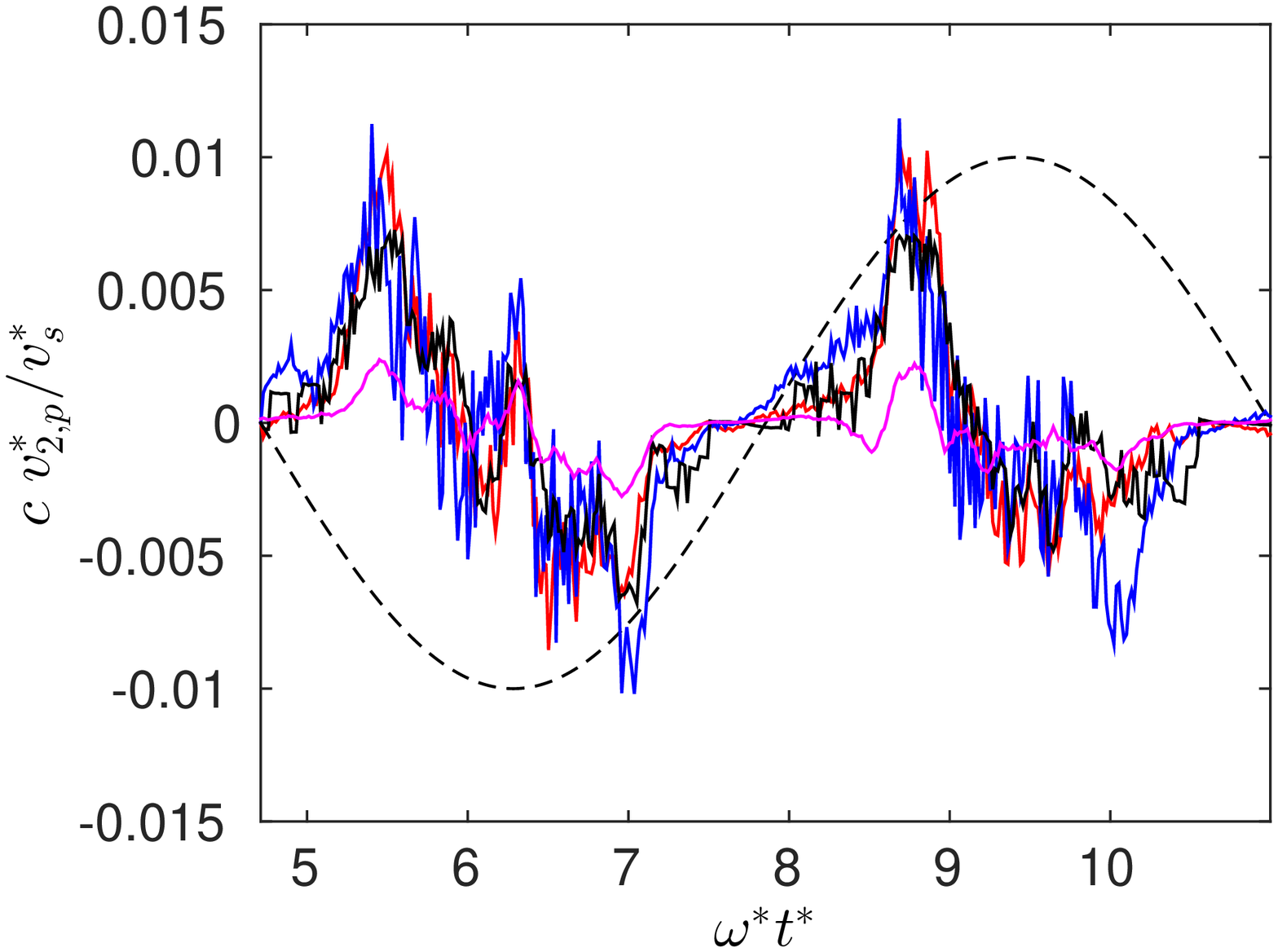}}
\end{picture}
\caption{%
\rev{}{%
Time development of the terms of equation~\eqref{exn2} for run~$3$ obtained by means of DNS: $I$ (black solid line), $II+III$ (red line) and $V$ (magenta line). %
The blue line is the value of the right hand side of equation~\eqref{exn3}. %
Each term is divided by the characteristic particle velocity $v_s^*=\sqrt{(s-1)\g^*\ds^*}$. %
The broken line qualitatively describes the fluid velocity far from the bottom. %
}
}
\label{apfig1}
\end{figure}
The values of each term of equation~\eqref{exn2} for run~$3$ are shown in figure~\ref{apfig1} (similar diagrams were obtained also for the other runs). %
The oscillations of the bottom elevation are essentially caused by the rearrangement of sediment particles under the action of the hydrodynamic forces. %
Since $v_{2,p}^*(\beds^*)$ is small, in particular during the decelerating phases, the erosion and deposition rates have nearly the same magnitude $v_{2,pe}^* \sim -v_{2,pd}^*$ and it is reasonable to look for a relationship between the bottom elevation and such magnitude, for instance $\vert v_{2,pd}^*\vert$ (note that $v_{2,pd}^*$ is always negative by definition). %

Indeed, on the basis of the present simulations, the left hand side of equation~\eqref{exn2} was found to be fairly correlated with the time derivative of the deposition rate and the following empirical relationship was obtained: %
\begin{equation}
(0.1+c_0)\dfrac{d\beds^*}{dt^*} 
\simeq 
{\cal C}_1\svf(\beds^*)\sqrt{\dfrac{\ds^*}{\del^*}}
\dfrac{1}{\om^*}\dfrac{\partial \vert v_{2,pd}^*\vert_{\xf{2}^*=\beds^*}}{\partial t^*}
\label{exn3}
\end{equation}
where ${\cal C}_1$ is a constant equal to $0.29$ and $\svf(\beds^*)=0.1$. %

Hence, by integrating equation~\eqref{exn3} between $t_0^*$, at which $\beds^*=x_{2,0}^*$ (see figure~\ref{sketch}) and $v_{2,p}^*\simeq 0$, and the generic time $t^*$, the following expression is readily obtained %
\begin{equation}
\beds^*\left(t^*\right)
\simeq 
x_{2,0}^* 
+ 
\ds^*
{\cal C}_2
\dfrac{\left\vert v_{2,pd}^*\left(t^*\right)\right\vert}{\vs^*} 
\:\:,
\label{x2apx2}
\end{equation}
where ${\cal C}_2$ is a constant equal to $\frac{0.1\,{\cal C}_1\,Ga}{2(0.1 + c_0)}\left(\frac{\del^*}{\ds^*}\right)^{1.5}$, $Ga$ denotes the Galileo number defined by \eqref{gal} and $\vs^*=\sqrt{(s-1)\g^*\ds^*}$ is the characteristic velocity of the sediment particles. %

Figure~\ref{apfig2} shows the comparison between the values of $\beds^*$ provided by the numerical simulations and equation~\eqref{x2apx2} for runs~$2$, $3$ and $5$. %
\begin{figure}[t]
\begin{picture}(0,130)(0,0)
  \put(0,-10){\includegraphics[trim=0cm 0cm 0cm 0cm, clip, width=.33\textwidth]{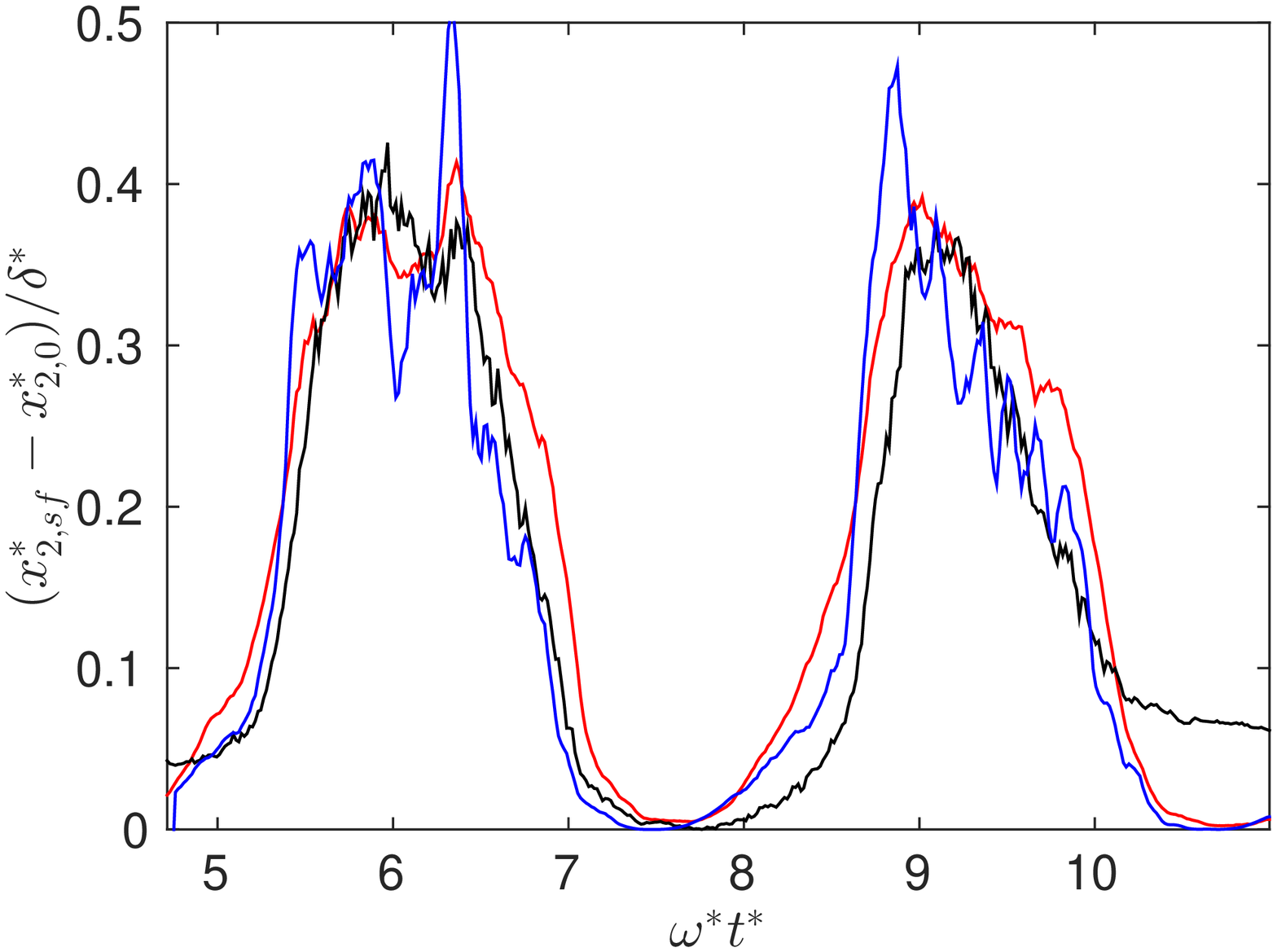}}
  \put(180,-10){\includegraphics[trim=0cm 0cm 0cm 0cm, clip, width=.33\textwidth]{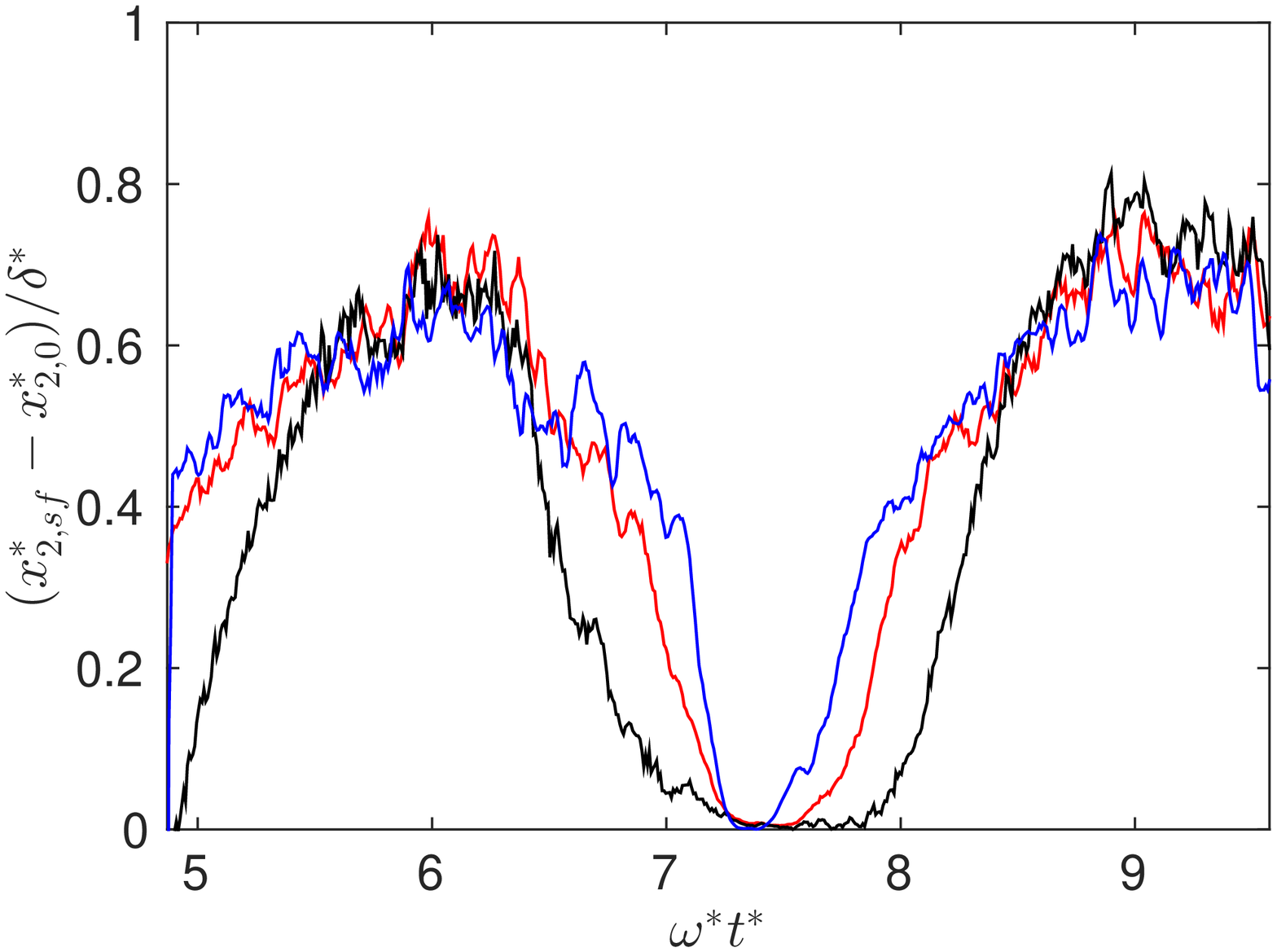}}
  \put(360,-10){\includegraphics[trim=0cm 0cm 0cm 0cm, clip, width=.335\textwidth]{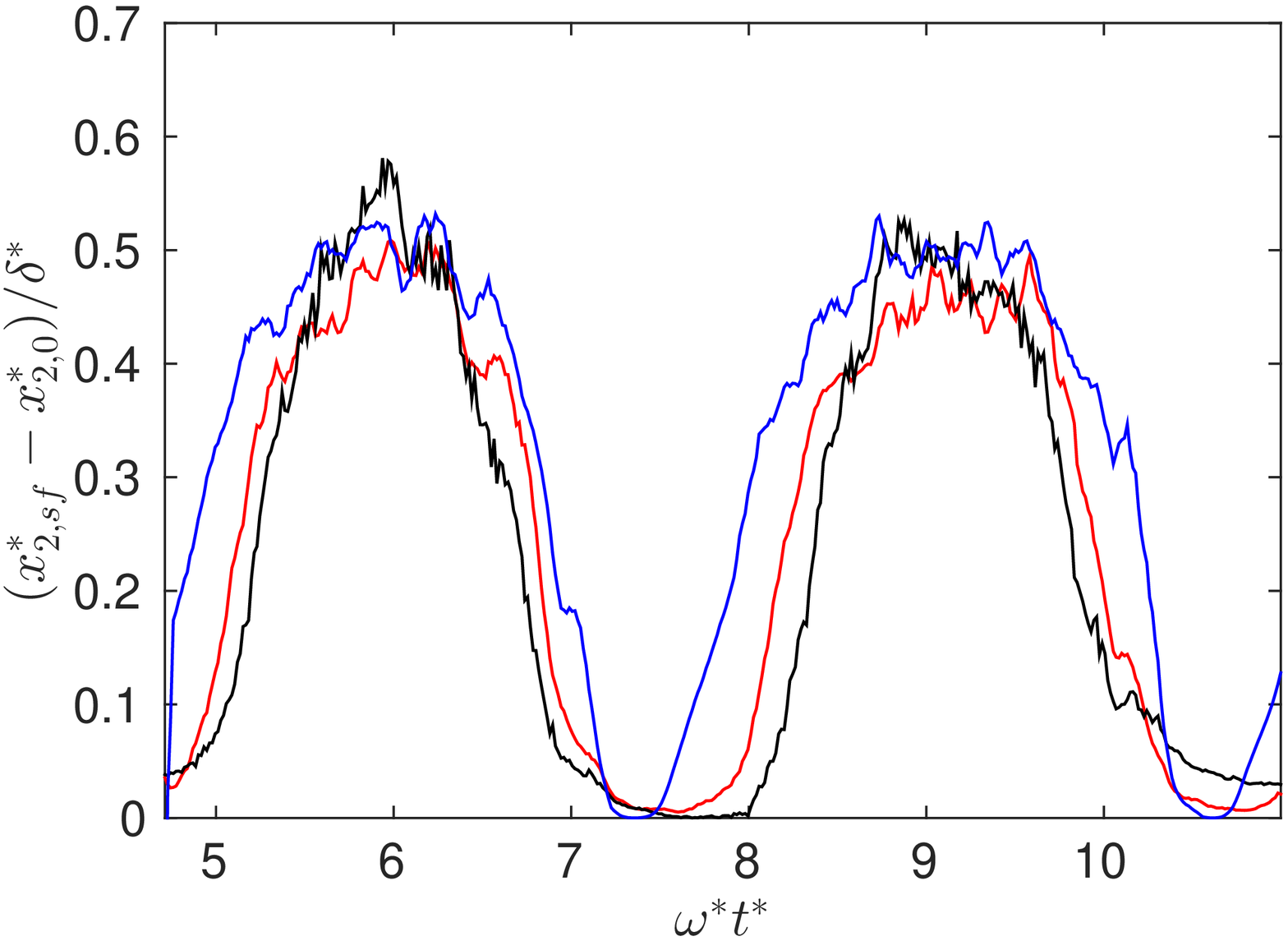}}
  \put(35,105){$(a)$}
  \put(215,105){$(b)$}
  \put(395,105){$(c)$}
\end{picture}
\caption{%
\rev{}{%
Time development of the dimensionless excursion the bottom elevation $\beds^*$ above the elevation $x_{2,0}^*$, obtained by means of DNS (black lines) and approximated by
equation~\eqref{x2apx2} (red lines) and by equation~\eqref{eqapx3} (blue lines). %
Panels~$(a)$, $(b)$ and $(c)$ refer to runs~$3$, $5$ and $2$, respectively. %
}
}
\label{apfig2}
\end{figure}
The magnitude of the deposition rate can be estimated by using the expression of the \textit{deposition constant} $k_D$ that was initially proposed by \citet{mccoy1977} for turbulent vertical channel flows and then adapted by \citet{papavergos1984} to the horizontal configuration %
\begin{equation}
\left\lbrace
\begin{array}{lll}
k_D^*
\simeq -u_{\tau}^* k_1 & \text{if} & t_p<0.2 \\
k_D^*= -u_{\tau}^* k_2 t_p^2 & \text{if} & 0.2<t_p<20 \\
k_D^*= -u_{\tau}^* k_3 & \text{if} & t_p>20
\:\:,
\end{array}
\right.
\label{mccoy}
\end{equation}
where $k_1=8\times 10^{-5}$, $k_2=2\times 10^{-3}$ and $k_3=0.8$, $u_\tau^*$ is the friction velocity which is equal to $v_s^*\sqrt{\shields}$ and $t_p$ is the dimensionless relaxation time of sediment particles $t_p=t_p^*u_\tau^{*2}/\nu^*$, with $t_p^*=s\,\ds^{*2}/(18\Phi\nu^*)$, which is equal to %
\begin{equation}
t_p = \dfrac{s\,Ga^2}{18}\dfrac{\vert\shields\vert}{\phi}\:\:. %
\label{tp}
\end{equation}
The function $\phi$ depends on the particle Reynolds number based on the vertical component of the relative particle velocity and corrects the value of $t_p^*$ to take into account the effects associated with the disturbances of the flow field caused by a spherical solid particle \citep[see][]{balachandar2010}. %
For the present simulations, $\phi$ takes values ranging about $1$. %
The prediction of the deposition rate, denoted by $\widehat{v}_{2,pd}^*$, is found to be equal to the deposition constant $k_D^*$ multiplied by $2.0$ for run~$3$ and to scale linearly with $Ga^{-1}$ for runs~$2$ and $5$, viz.
\begin{equation}
\widehat{v}_{2,pd}^*=-{\cal C}_3\,Ga^{-1}\,k_D^*
\label{eqkd}
\end{equation}
where ${\cal C}_3=30.35$, which is equal to the product between $2.0$ and the value of $Ga$ for run~$3$. %
\begin{figure}[t]
\begin{picture}(0,130)(0,0)
  \put(0,-10){\includegraphics[trim=0cm 0cm 0cm 0cm, clip, width=.33\textwidth]{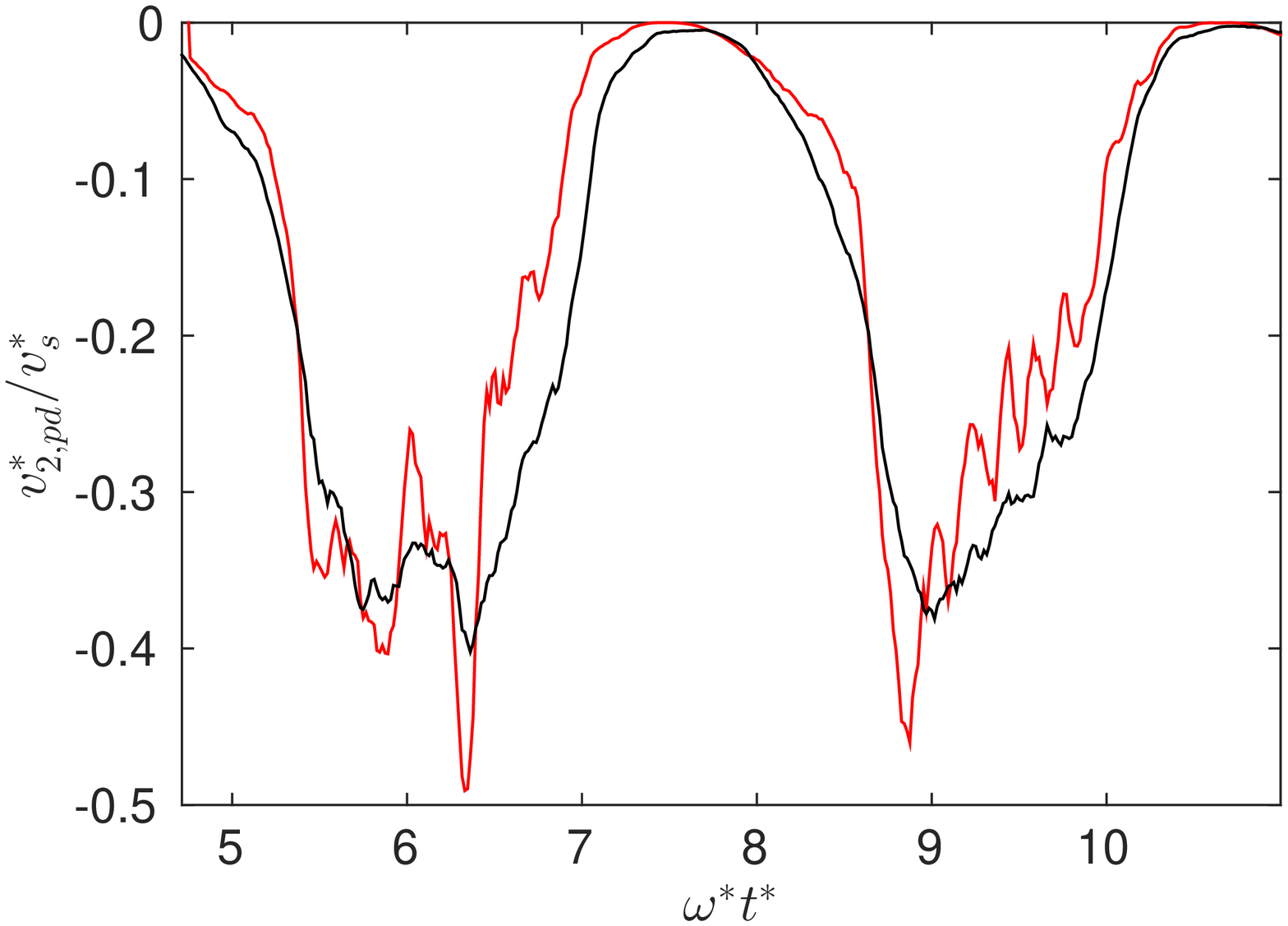}}
  \put(180,-10){\includegraphics[trim=0cm 0cm 0cm 0cm, clip, width=.33\textwidth]{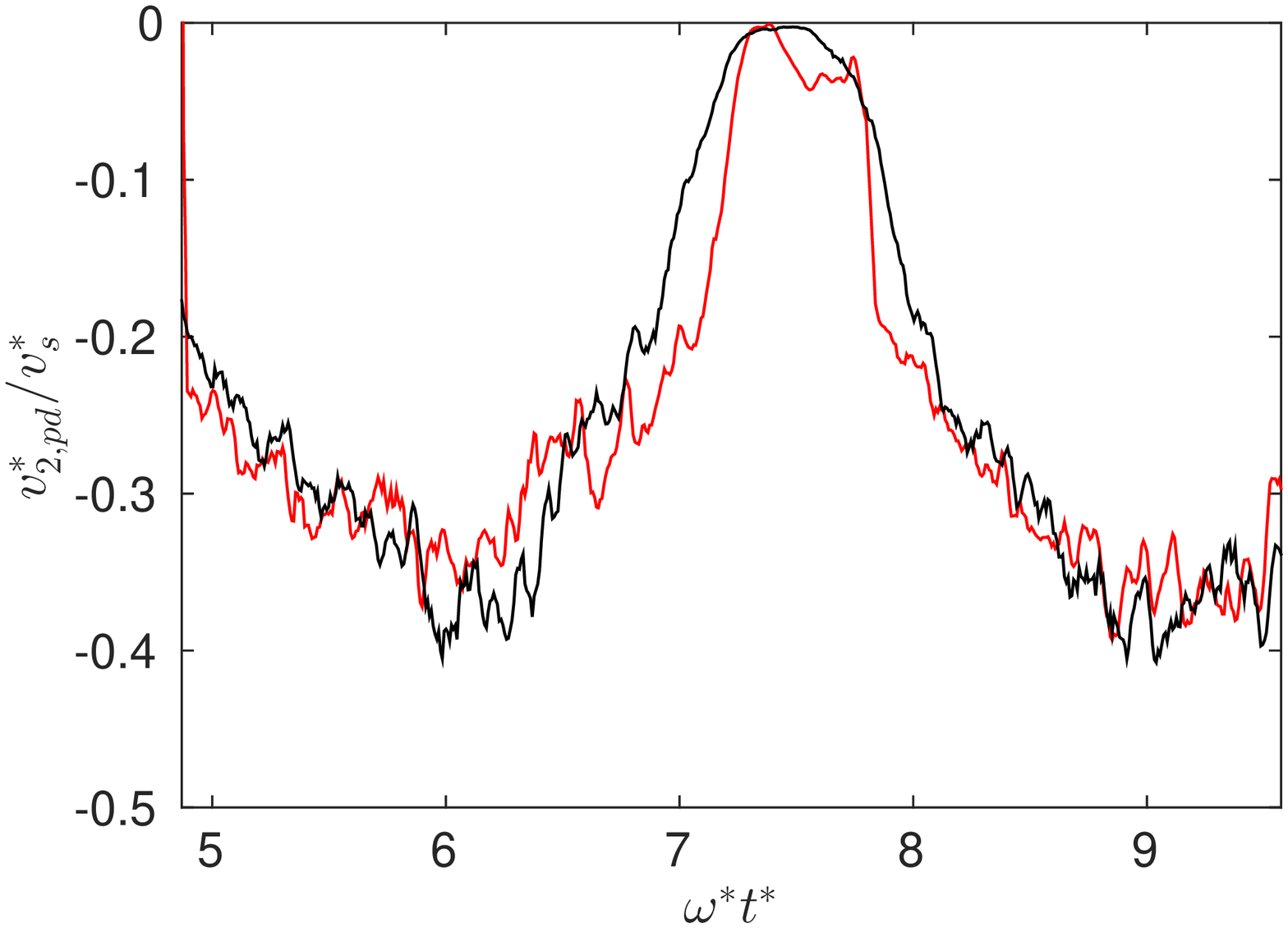}}
  \put(360,-10){\includegraphics[trim=0cm 0cm 0cm 0cm, clip, width=.33\textwidth]{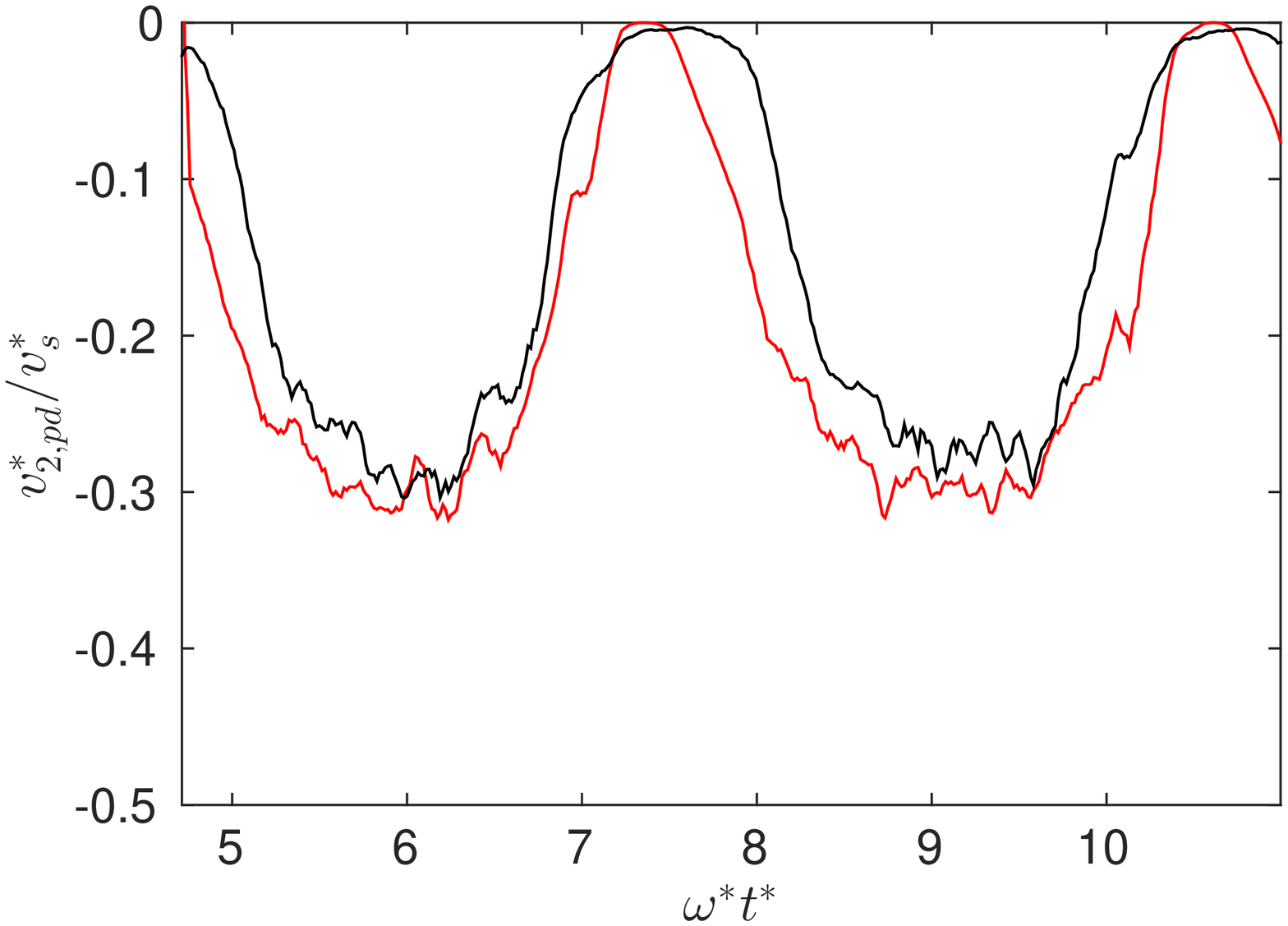}}
  \put(35,105){$(a)$}
  \put(215,105){$(b)$}
  \put(395,105){$(c)$}
\end{picture}
\caption{%
\rev{}{%
Time development of the ``deposition rate'' $v_{2,pd}^*$ obtained by DNS (black lines) and of its approximation $\widehat{v}_{2,pd}^*$ obtained by equation~\eqref{eqkd} (red lines), divided by the characteristic particle velocity $v_s^*=\sqrt{(s-1)\g^*\ds^*}$. %
Panels~$(a)$, $(b)$ and $(c)$ refer to runs~$3$, $5$ and $2$, respectively. %
}
}
\label{apfig3}
\end{figure}
Figure~\ref{apfig3} show that a fair agreement between $\widehat{v}_{2,pd}^*$ and $v_{2,pd}^*$ is obtained for the present runs. %
Hence, by replacing $v_{2,pd}^*$ with $\widehat{v}_{2,pd}^*$ in equation~\eqref{x2apx2}, the following approximation of the bottom elevation $\beda^*$ at time $t^*$ is obtained %
\begin{equation}
\beda^*
= 
x_{2,0}^* 
+ 
\ds^*\,
\widehat{C}\,
k_D\left(\shields\right) 
\:\:,
\label{eqapx3}
\end{equation}
with $\widehat{C}={\cal C}_2 {\cal C}_3=\frac{0.44}{0.1 + c_0}\left(\frac{\del^*}{\ds^*}\right)^{1.5}$ and $k_D=\frac{k_D^*}{\vs^*}$ provided by \eqref{mccoy} (see the blue lines in figure~\ref{apfig2}). %
}%

\end{document}